\pdfoutput=1
\documentclass[showpacs,preprintnumbers,amsmath,amssymb]{revtex4}

\usepackage{graphicx}
\usepackage{dcolumn}
\usepackage{bm}
\usepackage{epsfig}

\newcommand{\dhd}{{\textstyle d}
\lower.03ex\hbox{\kern-0.40em$^{\scriptstyle-}$}\kern-0.08em{}}

\newcommand{\bu}{{\bullet}}

\begin{document}

\preprint{JLAB-THY-07-741}

\title{
NLO evolution of color
dipoles}

\author{Ian Balitsky and Giovanni A. Chirilli}
\affiliation{
Physics Dept., ODU, Norfolk VA 23529, \\
and \\
Theory Group, Jlab, 12000 Jefferson Ave, Newport News, VA 23606
}
\email{balitsky@jlab.org, chirilli@jlab.org}

\date{\today}

\begin{abstract}
The small-$x$ deep inelastic scattering in the saturation region is governed by 
the non-linear evolution of Wilson-line operators. 
In the leading logarithmic approximation it is given by the BK
equation for the evolution of color dipoles. In the next-to-leading order the BK equation gets contributions from quark and gluon loops as well as from the tree gluon diagrams with quadratic and cubic nonlinearities. 
We calculate the gluon contribution to small-x evolution of 
Wilson lines (the quark part was obtained earlier).

\end{abstract}

\pacs{12.38.Bx,  12.38.Cy}

\maketitle

\section{\label{sec:in}Introduction }

A general feature of high-energy scattering is that a fast particle moves along its straight-line classical trajectory and the only quantum effect is the eikonal phase factor acquired along this propagation path. In QCD, for the fast quark or gluon scattering off some target, this eikonal phase factor is a Wilson line - the infinite gauge link ordered along the straight line collinear to particle's velocity $n^\mu$:
\begin{equation}
U^\eta(x_\perp)={\rm Pexp}\Big\{ig\int_{-\infty}^\infty\!\!  du ~n_\mu 
~A^\mu(un+x_\perp)\Big\},~~~~
\label{defU}
\end{equation}
Here $A_\mu$ is the gluon field of the target, $x_\perp$ is the transverse
position of the particle which remains unchanged throughout the collision, and the 
index $\eta$ labels the rapidity of the particle. Repeating the above argument for the target (moving fast in the spectator's frame) we see that 
particles with very different rapidities perceive each other as Wilson lines and
therefore these Wilson-line operators form
the convenient effective degrees of freedom in high-energy QCD (for a review, see ref. \cite{mobzor}).

Let us consider the deep inelastic scattering from a hadron at small 
$x_B=Q^2/(2p\cdot q)$.  The virtual photon decomposes into a pair of fast quarks 
 moving along straight lines separated by some transverse distance.
The propagation of this quark-antiquark pair reduces  to the 
``propagator of the color dipole''  $U(x_\perp)U^\dagger(y_\perp)$ - two Wilson lines ordered along the direction collinear to quarks' velocity. The structure function of a hadron is proportional to a matrix element of this color dipole operator
\begin{equation}
\hat{\cal U}^\eta(x_\perp,y_\perp)=1-{1\over N_c}
{\rm Tr}\{\hat{U}^\eta(x_\perp)\hat{U}^{\dagger\eta}(y_\perp)\}
\label{fla1}
\end{equation}
switched between the target's states ($N_c=3$ for QCD).  The gluon parton density is 
approximately
\begin{equation}
x_BG(x_B,\mu^2=Q^2)~
\simeq ~\left.\langle p|~\hat{\cal U}^\eta(x_\perp,0)|p\rangle
\right|_{x_\perp^2=Q^{-2}}
\label{fla2}
\end{equation}
where $\eta=\ln{1\over x_B}$. (As usual, we denote operators by ``hat'').
The energy dependence of the structure function is translated then into the dependence of the color dipole on the slope of the Wilson lines determined by the rapidity $\eta$.

Thus, the  small-x behavior of the structure functions is  governed by the 
rapidity evolution of color dipoles \cite{mu94,nnn}. 
At relatively high energies and for sufficiently small dipoles we can use the leading logarithmic approximation (LLA)
where  $ \alpha_s\ll 1,~ \alpha_s\ln x_B\sim 1$ and get the non-linear BK evolution equation for the color
dipoles \cite{npb96,yura}:
\begin{eqnarray}
&&\hspace{-1mm}
{d\over d\eta}~\hat{\cal U}(x,y)~=~
{\alpha_sN_c\over 2\pi^2}\!\int\!d^2z~ {(x-y)^2\over(x-z)^2(z-y)^2}
[\hat{\cal U}(x,z)+\hat{\cal U}(y,z)-\hat{\cal U}(x,y)-\hat{\cal U}(x,z)\hat{\cal U}(z,y)]
\label{bk}
\end{eqnarray}
The first three terms correspond to the linear BFKL evolution \cite{bfkl} and describe the parton emission while the last term is responsible for the parton annihilation. For sufficiently high $x_B$ the parton emission balances the parton annihilation so the partons reach the state of saturation\cite{saturation} with
the characteristic transverse momentum $Q_s$ growing with energy $1/x_B$
(for a review, see \cite{satreviews})

As usual, to get the region of application of the leading-order evolution equation one needs to find the next-to-leading order (NLO) corrections. In the case of the small-x evolution equation (\ref{bk}) there is another reason why NLO corrections are important.  Unlike the DGLAP evolution, the argument of the coupling constant in Eq. (\ref{bk}) is left undetermined in 
the LLA, and usually it is set by hand to be $Q_s$. Careful analysis of this argument is very important  from both theoretical and experimental points of view. 
From the theoretical viewpoint, we need to know whether the
coupling constant is determined by the size of the original dipole $|x-y|$ or of the size of the produced dipoles $|x-z|$ and/or 
$|z-y|$ since we may get a very different behavior
of the solutions of the equation (\ref{bk}). On the experimental side, the cross section is  proportional to some power of the coupling constant so  the argument determines 
how big (or how small) is the cross section. The typical argument of $\alpha_s$ is 
the characteristic transverse momenta of the process. For high enough energies, they are of order of the saturation scale $Q_s$ which is $\sim 2\div 3$ GeV for the LHC collider, so even the difference between $\alpha(Q_s)$ and 
$\alpha(2Q_s)$ can make a substantial impact on the cross section. The precise form of the argument of $\alpha_s$  should come from the solution of the BK equation with the running coupling constant, and 
the starting point of the analysis of the argument of $\alpha_s$ in Eq.  (\ref{bk}) is the calculation of the NLO evolution.

Let us present our result for the  NLO evolution of the color dipole
(hereafter we use notations $X\equiv x-z$,  $X'\equiv x-z'$, $Y\equiv y-z$, 
and  $Y'\equiv y-z'$)
\begin{eqnarray}
&&\hspace{-2mm}
{d\over d\eta}{\rm Tr}\{\hat{U}_x \hat{U}^{\dagger}_y\}~
\label{nlobk}\\
&&\hspace{-2mm}
=~{\alpha_s\over 2\pi^2}
\!\int\!d^2z~
{(x-y)^2\over X^2 Y^2}\Big\{1+{\alpha_s\over 4\pi}\Big[b\ln(x-y)^2\mu^2
-b{X^2-Y^2\over (x-y)^2}\ln{X^2\over Y^2}+
({67\over 9}-{\pi^2\over 3})N_c-{10\over 9}n_f
\nonumber\\
&&\hspace{62mm} 
-~
2N_c\ln{X^2\over(x-y)^2}\ln{Y^2\over(x-y)^2}\Big]\Big\}
~[{\rm Tr}\{\hat{U}_x \hat{U}^{\dagger}_z\}{\rm Tr}\{\hat{U}_z \hat{U}^{\dagger}_y\}
-N_c{\rm Tr}\{\hat{U}_x \hat{U}^{\dagger}_y\}]   
\nonumber\\
&&\hspace{-2mm} 
+~{\alpha_s^2\over 16\pi^4}
\int \!d^2 zd^2 z'
\Bigg[
\Big(-{4\over (z-z')^4}+\Big\{2{X^2{Y'}^2+{X'}^2Y^2-4(x-y)^2(z-z')^2\over  (z-z')^4[X^2{Y'}^2-{X'}^2Y^2]}\nonumber\\ 
&&\hspace{-2mm}
+~{(x-y)^4\over X^2{Y'}^2-{X'}^2Y^2}\Big[
{1\over X^2{Y'}^2}+{1\over Y^2{X'}^2}\Big]
+{(x-y)^2\over (z-z')^2}\Big[{1\over X^2{Y'}^2}-{1\over {X'}^2Y^2}\Big]\Big\}
\ln{X^2{Y'}^2\over {X'}^2Y^2}\Big)
\nonumber\\ 
&&\hspace{52mm}
\times~[{\rm Tr}\{\hat{U}_x\hat{U}^\dagger_z\}{\rm Tr}\{\hat{U}_z\hat{U}^\dagger_{z'}\}{\rm Tr}\{\hat{U}_{z'}\hat{U}^\dagger_y\}
-{\rm Tr}\{\hat{U}_x\hat{U}^\dagger_z \hat{U}_{z'}U^\dagger_y\hat{U}_z\hat{U}^\dagger_{z'}\}-(z'\rightarrow z)]
\nonumber\\ 
&&\hspace{-2mm}
+~\Big\{{(x-y)^2\over (z-z')^2 }\Big[{1\over X^2{Y'}^2}+{1\over Y^2{X'}^2}\Big]
-{(x-y)^4\over  X^2{Y'}^2{X'}^2Y^2}\Big\}\ln{X^2{Y'}^2\over {X'}^2Y^2}
~{\rm Tr}\{\hat{U}_x \hat{U}^\dagger_z\}{\rm Tr}\{\hat{U}_z\hat{U}^\dagger_{z'}\}{\rm Tr}\{\hat{U}_{z'}\hat{U}^\dagger_y\}
\nonumber\\
&&\hspace{-2mm}
+~4n_f
\Big\{{4\over(z-z')^4}
-2{{X'}^2Y^2+{Y'}^2X^2-(x-y)^2(z-z')^2\over (z-z')^4(X^2{Y'}^2-{X'}^2Y^2-)}
\ln{X^2{Y'}^2\over {X'}^2Y^2}\Big\}{\rm Tr}\{t^a\hat{U}_xt^b\hat{U}^{\dagger}_y\}
[{\rm Tr}\{t^a
\hat{U}_zt^b \hat{U}^\dagger_{z'}\}-(z'\rightarrow z)]\Bigg]
\nonumber
\end{eqnarray}
Here $\mu$ is the normalization point in the $\overline{MS}$ scheme and
$b={11\over 3}N_c-{2\over 3}n_f$ is the first coefficient of the $\beta$-function. 
The result of this paper is the gluon part of the evolution, the quark part of Eq. (\ref{nlobk}) proportional to $n_f$ was found earlier \cite{prd75,kw1}.  Also, the terms with cubic nonlinearities were previously found in the large-$N_c$ approximation in Ref. \cite{balbel}. The NLO kernel is a sum of the running-coupling part (proportional to $b$), the non-conformal  double-log 
term $\sim \ln{(x-y)^2\over (x-z)^2} \ln{(x-y)^2\over (x-z)^2}$ and the three conformal terms which depend on the two four-point conformal ratios ${X^2{Y'}^2\over {X'}^2Y^2}$ 
and ${(x-y)^2(z-z')^2\over X^2 {Y'}^2}$. Note that the logarithm of the second conformal ratio 
$\ln{(x-y)^2(z-z')^2\over X^2 {Y'}^2}$ is absent.

It should be emphasized that the  NLO result itself does not lead automatically to
the argument of coupling constant $\alpha_s$ in Eq. \ref{bk}. 
 In order to get this argument one can use the 
 renormalon-based approach\cite{renormalons}: first  get  the quark part 
 of the running coupling constant  coming from the bubble chain of quark loops and then make a conjecture 
 that the gluon part 
 of the $\beta$-function will follow that pattern. The Eq. (\ref{nlobk}) proves this conjecture in the first nontrivial order:
the quark part of the $\beta$ - function ${2\over 3}n_f$  calculated earlier
 gets promoted to full $b$. 
The analysis of the argument of the coupling constant was performed in Refs. \cite{prd75,kw1} and we briefly review it  in Sect. 7 for completeness. Roughly speaking, the argument of $\alpha_s$ is determined by the size of the smallest dipole $min(|x-y|,|x-z|,|y-z|)$.

The paper is organized as follows. In  Sect. II we remind the derivation of the 
BK equation in the leading order in $\alpha_s$. In Sect. III and IV, which are central to the paper,
we calculate the gluon contribution to the NLO kernel of the small-$x$ evolution of color dipoles:
in Sect. III we calculate the part of the NLO kernel corresponding to one-to-three dipoles transition and
in Sect. IV  the one-to-two dipoles part.
In Sect. V  we assemble the NLO BK kernel and in Sect. VI we compare the forward NLO BK kernel to 
the NLO BFKL results \cite{nlobfkl}. The results of the analysis of the argument of coupling constant are
briefly reviewed in Sect. VII.
Appendix A is devoted to the calculation of the UV-divergent part of the one-to-three dipole kernel
and in  Appendix B we discuss the dependence of the NLO kernel on the cutoff in 
the longitudinal momenta.

\section{Derivation of the BK equation}
Before discussing the small-x evolution of color dipole in the next-to-leading approximation it is instructive to recall the derivation of the leading-order (BK)
evolution equation. 
As discussed in the Introduction, the dependence of the structure functions 
on $x_B$ comes from the dependence of Wilson-line operators   
\begin{equation}
\hat{U}^\eta(x_\perp)={\rm Pexp}\Big\{ig\int_{-\infty}^\infty\!\!  du ~n_\mu ~\hat{A}^\mu(un+x_\perp)\Big\},~~~~
n\equiv p_1+e^{-2\eta}p_2
\label{defy}
\end{equation}
on the slope of the supporting line.  
The momenta $p_1$ and $p_2$ are the light-like
vectors such that $q=p_1-x_B p_2$ and  $p=p_2+{m^2\over s}p_1$ where
$p$ is the momentum of the target and $m$ is the mass. Throughout the paper, we use the 
Sudakov variables $p=\alpha p_1+\beta p_2 +p_\perp$ and the notations 
$x_\bullet\equiv x_\mu p_1^\mu$ and $x_\ast\equiv x_\mu p_2^\mu$ related to
the light-cone coordinates: $x_\ast=x^+\sqrt{s/2},~x_\bullet=x^-\sqrt{s/2}$. 

\begin{figure}

\vspace{-40mm}
\includegraphics[width=1.1\textwidth]{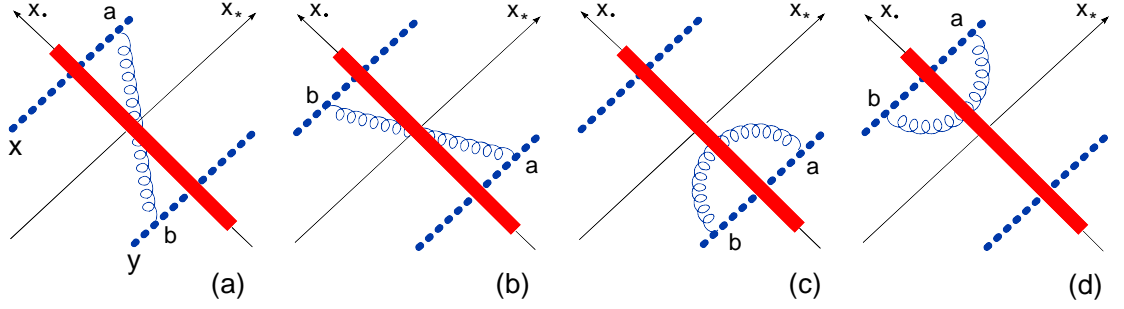}

\vspace{-170mm}
\caption{Leading-order diagrams for the small-$x$ evolution of color dipole\label{bkevol}. Gauge links are denoted by dotted lines.}
\end{figure}

To find the evolution of the color dipole (\ref{fla1}) with respect to the slope of the 
Wilson lines in the leading log approximation
we consider the matrix element of the color dipole between (arbitrary) target states and integrate over the gluons with rapidities $\eta_1>\eta>\eta_2=\eta_1-\Delta\eta$ leaving the gluons with $\eta<\eta_2$ as
a background field (to be integrated over later).
In the frame of gluons with $\eta\sim\eta_1$ the fields with
$\eta<\eta_2$ shrink to a pancake and we obtain the four diagrams shown in Fig. 
\ref{bkevol}. Technically,  to find the kernel in the leading-ordrer approximation we 
write down the general form of the operator equation for the evolution of the color dipole 
\begin{eqnarray}
&&\hspace{-6mm}
{\partial\over\partial\eta}{\rm Tr}\{\hat{U}_x\hat{U}^\dagger_y\}=
K_{\rm LO}{\rm Tr}\{\hat{U}_x\hat{U}^\dagger_y\}+...
\label{eveq}
\end{eqnarray}
(where dots stand for the higher orders of the expansion) 
and calculate the l.h.s. of Eq. (\ref{eveq}) in the shock-wave background
\begin{eqnarray}
&&\hspace{-2mm}
{\partial\over\partial\eta}\langle{\rm Tr}\{\hat{U}_x\hat{U}^\dagger_y\}\rangle_{\rm shockwave}=
\langle K_{\rm LO}{\rm Tr}\{\hat{U}_x\hat{U}^\dagger_y\}\rangle_{\rm shockwave}
\label{eveqmaels}
\end{eqnarray}
In what follows we replace $\langle ...\rangle_{\rm shockwave}$ 
by $\langle ...\rangle$ for brevity.
With future NLO computation in view, we will perform the leading-order calculation in the lightcone gauge $p_2^\mu A_\mu=0$. The  gluon propagator in a shock-wave external field has the form\cite{prd99,balbel} 
\begin{eqnarray}
&&\hspace{-2mm}\langle \hat{A}^a_{\mu}(x)\hat{A}^b_{\nu}(y)\rangle ~=~
\theta(x_\ast y_\ast)\delta^{ab}
{s\over 2}\!\int\! \dhd\alpha\dhd\beta(x_\perp|{d_{\mu\nu}\over i(\alpha\beta s-p_\perp^2+i\epsilon)}|y_\perp)~
\label{gluprop}\\
&&\hspace{-2mm}
-~\theta(x_\ast)\theta(-y_\ast)
\!\int_0^\infty\!\dhd\alpha~ {e^{-i\alpha(x-y)_\bullet}\over 2\alpha}
(x_\perp|e^{-i{p_\perp^2\over\alpha s}x_\ast}
\Big[g^\perp_{\mu\xi}-{2\over \alpha s}(p^\perp_\mu p_{2\xi}+p_{2\mu}p^\perp_\xi)\Big]U^{ab} 
\Big[g^{\perp\xi}_\nu -{2\over \alpha s}(p_2^\xi p^\perp_\nu+p_{2\nu}p_\perp^\xi) \Big]
e^{i{p_\perp^2\over\alpha s}y_\ast}|y_\perp)
\nonumber \\
&&\hspace{-2mm}-~\theta(-x_\ast)\theta(y_\ast)
\!\int_0^\infty\!\dhd\alpha~ {e^{i\alpha(x-y)_\bullet}\over 2\alpha}
(x_\perp|e^{i{p_\perp^2\over\alpha s}x_\ast}
\Big[g^\perp_{\mu\xi}-{2\over \alpha s}(p^\perp_\mu p_{2\xi}+p_{2\mu}p^\perp_\xi)\Big]
U^{\dagger ab}
\Big[ g^{\perp\xi}_{~\nu}-{2\over \alpha s}(p_2^\xi p^\perp_\nu+p_{2\nu}p_\perp^\xi)\Big]
e^{-i{p_\perp^2\over\alpha s}y_\ast}|y_\perp)
\nonumber
\end{eqnarray}
where 
\begin{equation}
d_{\mu\nu}(k)\equiv g^{\perp}_{\mu\nu}-
{2\over s\alpha}(k^{\perp}_{\mu}p_{2\nu}+k^{\perp}_{\nu}p_{2\mu})
-{4\beta\over s\alpha}p_{2\mu}p_{2\nu}
\label{demunu}
\end{equation}
Hereafter we use Schwinger's notations 
$(x_\perp|F(p_\perp)|y_\perp)\equiv \int\!\dhd p~e^{i(p,x-y)_\perp}F(p_\perp)$ 
(the scalar product of the four-dimensional vectors in our notations is 
$x\cdot y={2\over s}(x_\ast y_\bullet+x_\ast y_\bullet)-(x,y)_\perp$). 
Note that the interaction with the shock wave does not change the $\alpha$-component
of the gluon momentum.

\begin{figure}

\vspace{-38mm}
\includegraphics[width=\textwidth]{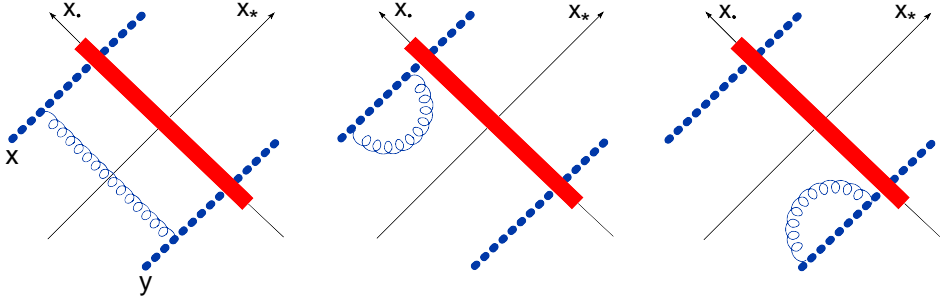}

\vspace{-150mm}
\caption{Leading-order diagrams proportional to the original dipole. \label{virdiagrams}}
\end{figure}

We obtain
\begin{equation}
\hspace{-0mm}\!\int_0^\infty \! du\! \int^0_{-\infty} \! 
dv~\langle\hat{A}^a_\bu(un+x_\perp)
\hat{A}^b_\bu(vn+y_\perp)\rangle_{\rm Fig. \ref{bkevol}a}
~=~-4\alpha_s
\int_{e^{-\eta_2}}^\infty\!{d\alpha\over\alpha}
(x_\perp|{p_i\over p_\perp^2+\alpha^2e^{-2\eta_1}s}
U^{ab}{p_i\over p_\perp^2+\alpha^2 e^{-2\eta_1}s}|y_\perp)
\label{bk1}
\end{equation}
Formally, the integral over $\alpha$ diverges at the lower limit, but since we integrate over the rapidities $\eta>\eta_2$ we get (in the LLA) 
\begin{eqnarray}
&&\hspace{-26mm}\!\int_0^\infty \! du \int^0_{-\infty} \! dv~\langle \hat{A}^a_\bu(un+x_\perp)
\hat{A}^b_\bu(vn+y_\perp)\rangle_{\rm Fig. \ref{bkevol}a}
~=~-4\alpha_s\Delta\eta
(x_\perp|{p_i\over p_\perp^2}U^{ab}{p_i\over p_\perp^2}|y_\perp)
\label{bk2}
\end{eqnarray}
and therefore
\begin{eqnarray}
&&\hspace{-2mm}\langle \hat{U}_x\otimes \hat{U}^\dagger_y\rangle_{\rm Fig. \ref{bkevol}a}^{\eta_1}
~=~-{\alpha_s\over \pi^2}\Delta\eta~ \{t^aU_x\otimes t^bU^\dagger_y\}^{\eta_2}
\!\int\! d^2z_\perp {(x-z,y-z)_\perp\over (x-z)_\perp^2(y-z)_\perp^2}
(U_z^{\eta_2})^{ab}
\label{bk3}
\end{eqnarray}
The contribution of the diagram in Fig.  \ref{bkevol}b is obtained from Eq. (\ref{bk3})
by the replacement $t^aU_x\otimes t^bU^\dagger_y \rightarrow U_x t^b\otimes U^\dagger_y t^a$, $x\leftrightarrow y$ and the two remaining diagrams are obtained from
Eq. \ref{bk2} by taking $y=x$ (Fig. \ref{bkevol}c) and  $x=y$ (Fig. \ref{bkevol}d).
Finally, one obtains
\begin{eqnarray}
&&\hspace{-2mm}\langle \hat{U}_x\otimes \hat{U}^\dagger_y\rangle_{\rm Fig. \ref{bkevol}}^{\eta_1}
~=~-{\alpha_s\Delta\eta\over \pi^2} \{t^aU_x\otimes t^bU^\dagger_y
+ U_x t^b\otimes U^\dagger_y t^a\}^{\eta_2}
\!\int\! d^2z_\perp {(x-z,y-z)_\perp\over (x-z)_\perp^2(y-z)_\perp^2}
(U_z^{\eta_2})^{ab}
\nonumber\\
&&\hspace{-2mm}+~{\alpha_s\Delta\eta\over \pi^2}
 \{t^aU_xt^b\otimes U^\dagger_y\}^{\eta_2}
\!\int\! {d^2z_\perp \over (x-z)_\perp^2}
(U_z^{\eta_2})^{ab}
+{\alpha_s\Delta\eta\over \pi^2} \{U_x\otimes t^bU^\dagger_yt^a\}^{\eta_2}
\!\int\! {d^2z_\perp\over (y-z)_\perp^2}
(U_z^{\eta_2})^{ab}
\label{bk4}
\end{eqnarray}
so
\begin{eqnarray}
&&\hspace{-2mm}
\langle{\rm Tr}\{\hat{U}_x \hat{U}^\dagger_y\}\rangle_{\rm Fig. \ref{bkevol}}^{\eta_1}
~=~{\alpha_s\Delta\eta\over 2\pi^2} 
\!\int\! d^2z_\perp {(x-y)^2_\perp\over (x-z)_\perp^2(y-z)_\perp^2}
[{\rm Tr}\{U_x U^\dagger_z\}{\rm Tr}\{U_z U^\dagger_y\}
-{1\over N_c}{\rm Tr}\{U_x U^\dagger_y\}]^{\eta_2}
\label{bk5}
\end{eqnarray}
There are also contributions coming from the diagrams shown in Fig. \ref{virdiagrams} (plus graphs obtained by reflection with respect to the shock wave).
These diagrams are proportional to the original dipole ${\rm Tr}\{U_x U^\dagger_y\}$ 
and therefore the corresponding term can be derived from the contribution
of Fig. 1 graphs using the requirement that the r.h.s. of the evolution equation
should vanish for $x=y$ since 
$\lim_{x\rightarrow y}{\partial\over\partial\eta}{\rm Tr}\{U_x U^\dagger_y\}=0$).
It is easy to see that this requirement leads to 
\begin{eqnarray}
&&\hspace{-2mm}
\langle{\rm Tr}\{\hat{U}_x \hat{U}^\dagger_y\}\rangle^{\eta_1}
~=~{\alpha_s\Delta\eta\over 2\pi^2} 
\!\int\! d^2z_\perp {(x-y)^2_\perp\over (x-z)_\perp^2(y-z)_\perp^2}
[{\rm Tr}\{U_x U^\dagger_z\}{\rm Tr}\{U_z U^\dagger_y\}
-N_c{\rm Tr}\{U_x U^\dagger_y\}]^{\eta_2}
\label{bk6}
\end{eqnarray}
which is equivalent to the BK equation for the evolution of the color dipole (\ref{bk}).

\section{Diagrams with two gluon-shockwave intersections}

\subsection{``Cut self-energy'' diagrams}

In the next-to-leading order there are three types of diagrams. Diagrams of the first type  have two intersections
of the emitted gluons with the shock wave, diagrams of the second type have one intersection, and finally there are 
diagrams of the third type without intersections.
In principle, there could have been contributions coming from the gluon loop which lies entirely in the shock wave, but we will demonstrate below that
such terms are absent (see the discussion at the end of Sect. \ref{sect:compare}).
 
 For the NLO calculation we use the lightcone gauge $p_2^\mu A_\mu=0$. Also, we find it convenient to change the prescription  for the cutoff in the longitudinal direction.  
 We consider the light-like dipoles (in the $p_1$ direction) and impose the cutoff
 on the maximal $\alpha$ emitted by any gluon from the Wilson lines so
\begin{eqnarray}
&&\hspace{-2mm} 
 U^\eta_x~=~{\rm Pexp}\Big[ig\!\int_{-\infty}^\infty\!\! du p_1^\mu A^\eta_\mu(up_1+x_\perp)\Big]
 \nonumber\\ 
&&\hspace{-2mm}
A^\eta_\mu(x)~=~\int\!\dhd^4 k \theta(e^\eta-|\alpha_k|)e^{-ik\cdot x} A_\mu(k)
\label{cutoff}
\end{eqnarray}
 As we will see below, the (almost) conformal result (\ref{nlobk}) comes from the
 regularization (\ref{cutoff}). In Appendix B we will present the NLO kernel
 for the cutoff with the slope (\ref{defy}). 
\begin{figure}

\vspace{-22mm}
\includegraphics[width=\textwidth]{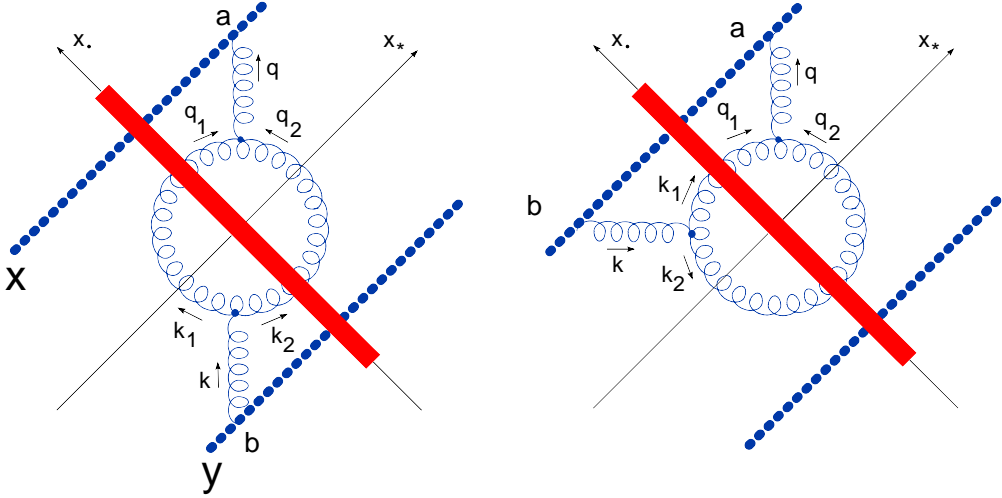}

\vspace{-145mm}
\caption{``Cut self-energy'' diagram\label{nlobk2}.}
\end{figure}
We start with the calculation of the Fig. \ref{nlobk2}a diagram.  Multiplying two propagators (\ref{gluprop}),  two 3-gluon vertices and two bare propagators we obtain

\begin{eqnarray}
&&\hspace{-6mm}g^2\int_0^{\infty}du\int^0_{-\infty}dv 
\langle \hat{A}^a_{\bullet}(up_1+x_{\perp})\hat{A}^b_{\bullet}(vp_1+y_{\perp})\rangle
\label{selfenergy1}\\
&&\hspace{-6mm}
=~{1\over 2}
g^4{s^2\over 4}f^{anl}f^{bn'l'}\int
\dhd\alpha\dhd\alpha_1\dhd\beta\dhd\beta' \dhd\beta_1\dhd\beta'_1
\dhd\beta_2\dhd\beta'_2\!\int d^2z d^2z'
\int \dhd^2q_1\dhd^2q_2\dhd^2k_1\dhd^2k_2~e^{i(q_1+q_2,x)_{\perp}
-i(k_1+k_2,y)_{\perp}}
\nonumber\\ 
&&\hspace{-6mm}
{4\alpha_1(\alpha-\alpha_1) 
U^{nn'}_zU^{ll'}_{z'}e^{-i(q_1-k_1,z)_\perp-i(q_2-k_2,z')_\perp}
\over(\beta-\beta_1-\beta_2+i\epsilon)
(\beta'-\beta'_1-\beta'_2+i\epsilon)(\beta-i\epsilon)(\beta'-i\epsilon)}
~{d_{\bullet\lambda}(\alpha p_1+\beta p_2+q_{1\perp}+k_{1\perp})\over \alpha\beta
s-(q_1+q_2)_\perp^2+i\epsilon}
{d_{\lambda'\bullet}(\alpha p_1+\beta' p_2+q_{2\perp}+k_{2\perp})\over \alpha\beta'
s-(k_1+k_2)_\perp^2+i\epsilon}
\nonumber\\ 
&&\hspace{-6mm}
{d_{\mu\xi}(\alpha_1 p_1+\beta_1 p_2+q_{1\perp})
\over \alpha_1\beta_1 s-q_{1\perp}^2+i\epsilon}
{d^\xi_{~\mu'}(\alpha_1 p_1+\beta'_1 p_2+k_{1\perp})
\over \alpha_1\beta'_1 s-k_{1\perp}^2+i\epsilon}
{d_{\nu\eta}((\alpha-\alpha_1) p_1+\beta_2 p_2+q_{2\perp})
\over (\alpha-\alpha_1)\beta_2 s-q_{2\perp}^2+i\epsilon}
~{d^\eta_{~\nu'}((\alpha-\alpha_1)p_1+\beta'_2 p_2+k_{2\perp})
\over (\alpha-\alpha_1)\beta'_2 s-k_{2\perp}^2+i\epsilon}
\nonumber\\ 
&&\hspace{-6mm}
\Gamma^{\mu\nu\lambda}
(\alpha p_1+q_{1\perp},(\alpha-\alpha_1)p_1+q_{2\perp},-\alpha p_1-q_{1\perp}
-q_{2\perp})~
\Gamma^{\mu'\nu'\lambda'}
(\alpha p_1+k_{1\perp},(\alpha-\alpha_1)p_1
+k_{2\perp},-\alpha p_1-k_{1\perp}-k_{2\perp})
\nonumber
\end{eqnarray}
where
\begin{equation}
\Gamma_{\mu\nu\lambda}(p,k,-p-k)=(p-k)_\lambda g_{\mu\nu}
+(2k+p)_\mu g_{\nu\lambda}+(-2p-k)_\nu g_{\lambda\mu}
\end{equation}
In this formula
${1\over \beta-i\epsilon}$ comes from the integration over $u$ parameter in the l.h.s. and  ${1\over \beta-\beta_1-\beta_2+i\epsilon}$  from 
the integration of the left three-gluon vertex over the half-space $x_\ast>0$. 
Similarly, we get ${1\over \beta' -i\epsilon}$ from the integration over $v$ parameter and 
${1\over \beta'-\beta'_1-\beta'_2+i\epsilon}$  from 
the integration of the right three-gluon vertex over the half-space $x_\ast<0$. 
The factor ${1\over 2}$ in the r.h.s. is combinatorial. Note that in the light-cone gauge 
one can always neglect the 
$\beta p_{2\xi}$ components of the momenta in the three-gluon vertex since they are always multiplied by 
some $d_{\xi\eta}$.
 
Taking residues at $\beta=\beta'=0$ and $\beta_2=-\beta_1$, $\beta'_2=-\beta'_1$ 
we obtain

\begin{eqnarray}
&&\hspace{-6mm}g^2\!\int_0^{\infty}du\int^0_{-\infty}dv 
\langle \hat{A}^a_{\bullet}(up_1+x_{\perp})\hat{A}^b_{\bullet}(vp_1+y_{\perp})\rangle
\label{selfenergy2}\\
&&\hspace{-6mm}=~{1\over 2}
g^4{s^2\over 4}f^{anl}f^{bn'l'}\!\int\!
\dhd\alpha\dhd\alpha_1\dhd\beta_1\dhd\beta'_1\!\int d^2z d^2z'
\!\int \!\dhd^2q_1\dhd^2q_2\dhd^2k_1\dhd^2k_2
~e^{i(q_1+q_2,x)_{\perp}
-i(k_1+k_2,y)_{\perp}}
\nonumber\\ 
&&\hspace{-6mm}
4{\alpha_1(\alpha-\alpha_1)\over\alpha^2}
 U^{nn'}_zU^{ll'}_{z'}e^{-i(q_1-k_1)z-i(q_2-k_2)z'}
~{(q_{1\perp}+q_{2\perp})_\lambda\over (q_1+q_2)_\perp^2}
{(k_{1\perp}+k_{2\perp})_{\lambda'}\over (k_1+k_2)_\perp^2}
\nonumber\\ 
&&\hspace{-6mm}
{d_\mu^{~\xi}(\alpha_1 p_1+q_{1\perp})
\over \alpha_1\beta_1 s-q_{1\perp}^2+i\epsilon}
{d_{\xi\mu'}(\alpha_1 p_1+k_{1\perp})
\over \alpha_1\beta'_1 s-k_{1\perp}^2+i\epsilon}
{d_\nu^{~\eta}((\alpha-\alpha_1) p_1+q_{2\perp})
\over -(\alpha-\alpha_1)\beta_1 s-q_{2\perp}^2+i\epsilon}
~{d_{\eta\nu'}((\alpha-\alpha_1)p_1+k_{2\perp})
\over -(\alpha-\alpha_1)\beta'_1 s-k_{2\perp}^2+i\epsilon}
\nonumber\\ 
&&\hspace{-6mm}
\Gamma^{\mu\nu\lambda}
(\alpha_1 p_1+q_{1\perp},(\alpha-\alpha_1)p_1+q_{2\perp},-\alpha p_1-q_{1\perp}
-q_{2\perp})~
\Gamma^{\mu'\nu'\lambda'}
(\alpha_1 p_1+k_{1\perp},(\alpha-\alpha_1)p_1
+k_{2\perp},-\alpha p_1-k_{1\perp}-k_{2\perp})
\nonumber
\end{eqnarray}
We have omitted terms $\sim \beta p_2$ in the arguments of $d_{\xi\eta}$ since
they do not contribute to $d_{\mu\xi}d^{\xi\mu'}$,  see Eq. (\ref{demunu}).
Introducing the variable $u=\alpha_1/\alpha$ and taking residues at $\beta_1={q_1^2\over\alpha_1 s}$ 
and $\beta'_1={k_1^2\over\alpha_1 s}$ we obtain
\begin{eqnarray}
&&\hspace{-6mm}
-~{g^4\over 8\pi^2}
f^{anl}f^{bn'l'}\int_0^\sigma
{d\alpha\over\alpha}\!\int_0^1\! du~\bar{u} u\!\int d^2z d^2z'
\!\int \!\dhd^2q_1\dhd^2q_2\dhd^2k_1\dhd^2k_2~e^{i(q_1+q_2,x)_{\perp}
-i(k_1+k_2,y)_{\perp}-i(q_1-k_1,z)-i(q_2-k_2,z')}
\nonumber\\ 
&&\hspace{-6mm}
\times~ U^{nn'}_zU^{ll'}_{z'}
{(q_{1\perp}+q_{2\perp})_\lambda(k_{1\perp}+k_{2\perp})_{\lambda'}
\over (q_1+q_2)_\perp^2(k_1+k_2)_\perp^2}
~{d_{\mu\xi}(u\alpha p_1+q_{1\perp})d^\xi_{~\mu'}(u\alpha p_1+k_{1\perp})
\over q_{1\perp}^2\bar{u}+q_{2\perp}^2 u}~
{d_{\nu\eta}(\bar{u} \alpha p_1+q_{2\perp})
d^\eta_{~\nu'}(\bar{u}\alpha p_1+k_{2\perp})
\over  k_{1\perp}^2\bar{u}+k_{2\perp}^2 u}
\nonumber\\ 
&&\hspace{-6mm}
\times~\Gamma^{\mu\nu\lambda}
(u\alpha p_1+q_{1\perp},\bar{u}\alpha p_1+q_{2\perp},-\alpha p_1-q_{1\perp}
-q_{2\perp})~
\Gamma^{\mu'\nu'\lambda'}
(u\alpha p_1+k_{1\perp},\bar{u}\alpha p_1
+k_{2\perp},-\alpha p_1-k_{1\perp}-k_{2\perp})
\label{selfenergy3}
\end{eqnarray}
where we have imposed a cutoff $\alpha<\sigma$ in accordance with 
Eq. (\ref{cutoff}).

Using formulas 
\begin{eqnarray}
&&\hspace{-5mm}
d_{\mu\xi}(u\alpha p_1+q_{\perp})d^{\xi\mu'}(u\alpha p_1+k_{\perp})
~=~\Big(g^\perp_{\mu\xi}-{2\over s\alpha u}p_{2\mu}q^\perp_{\xi}\Big)
\Big(g_\perp^{\xi\mu'}-{2\over s\alpha u}k^\xi_{\perp}\Big),
\nonumber\\
&&\hspace{-5mm}
-(q_1+q_2)_\perp^\lambda\Gamma_{\mu\nu\lambda}(\alpha up_1+q_{1\perp},
\alpha \bar{u} p_1+q_{2\perp},
-\alpha p_1-(q_1+q_2)_\perp)(g_\perp^{\mu i}-{2\over s\alpha u}p_2^\mu q_1^i)
(g_\perp^{\nu j}-{2\over s\alpha \bar{u} }p_2^\nu q_2^j)
\nonumber\\
&&\hspace{-5mm}=~(q_{1\perp}^2-q_{2\perp}^2)g_\perp^{ij}
+{2\over u} q_1^i(q_1+q_2)^j-{2\over  \bar{u}}(q_1+q_2)^iq_2^j
\label{nice}
\end{eqnarray}
we can represent the contribution of diagram in Fig. \ref{nlobk2}a in the form

\begin{eqnarray}
&&\hspace{-2mm}
\langle{\rm Tr}\{\hat{U}_x \hat{U}^\dagger_y\}\rangle_{\rm Fig. \ref{nlobk2}a}
~=~-{g^4\over 8\pi^2}{\rm Tr}\{t^aU_xt^bU^\dagger_y\}f^{anl}f^{bn'l'}
\!\int\! d^2z d^2z'U_z^{nn'}U_{z'}^{ll'}
\label{selfenergy4}\\
&&\hspace{-2mm}
\times
\int_0^{\sigma}\!
{d\alpha\over\alpha}\!\int_0^1\! du~ \bar{u}u
\!\int\! \dhd^2 q_1\dhd^2q_2\dhd^2k_1\dhd^2k_2~
{e^{i(q_1,x-z)_\perp+i(q_2,x-z')_\perp
-i(k_1,y-z)_\perp-i(k_2,y-z')_\perp}
\over (q_1+q_2)^2(k_1+k_2)^2(q_1^2\bar{u}+q_2^2u)(k_1^2\bar{u}+k_2^2u)}
\nonumber\\ 
&&\hspace{-2mm}\times~
\Big[(q_1^2-q_2^2)\delta_{ij}-{2\over u}q_{1i}(q_1+q_2)_j+
{2\over \bar{u}}(q_1+q_2)_iq_{2j}\Big]
\Big[(k_1^2-k_2^2)\delta_{ij}-{2\over u}k_{1i}(k_1+k_2)_j+
{2\over \bar{u}}(k_1+k_2)_ik_{2j}\Big]
\nonumber
\end{eqnarray}
Throughout the paper we use Greek letters for indices $\mu=0,1,2,3$ (with $g^{\mu\nu}$=(1,-1,-1,-1)) and Latin letters
for transverse indices $i=1,2$.

The diagram shown in Fig. \ref{nlobk2}b is obtained by the substitution
$e^{-i(k_1+k_2,y_\perp)}\rightarrow -e^{-i(k_1+k_2,x)_\perp}$ (the different sign comes from 
the replacement $[-\infty p_1,0]_y$ by $[0,-\infty p_1]_x$).  We get
\begin{eqnarray}
&&\hspace{-2mm}
\langle{\rm Tr}\{\hat{U}_x \hat{U}^\dagger_y\}\rangle_{\rm Fig. \ref{nlobk2}a+b}
~=~{g^4\over 8\pi^2}{\rm Tr}\{t^aU_xt^bU^\dagger_y\}f^{anl}f^{bn'l'}
\!\int\! d^2z d^2z'U_z^{nn'}U_{z'}^{ll'}
\label{vkladnlobk2}\\
&&\hspace{-2mm}
\times
\int_0^{\sigma}\!
{d\alpha\over\alpha}\!\int_0^1\! du~ \bar{u}u
\!\int\! \dhd^2 q_1\dhd^2q_2\dhd^2k_1\dhd^2k_2~
{e^{i(q_1+q_2,x)_{\perp}
-i(q_1-k_1,z)-i(q_2-k_2,z')}[e^{-i(k_1+k_2,x)_\perp}-e^{-i(k_1+k_2,y)_\perp}]
\over (q_1+q_2)^2(k_1+k_2)^2(q_1^2\bar{u}+q_2^2u)(k_1^2\bar{u}+k_2^2u)}
\nonumber\\ 
&&\hspace{-2mm}\times~
\Big[(q_1^2-q_2^2)\delta_{ij}-{2\over u}q_{1i}(q_1+q_2)_j+
{2\over \bar{u}}(q_1+q_2)_iq_{2j}\Big]
\Big[(k_1^2-k_2^2)\delta_{ij}-{2\over u}k_{1i}(k_1+k_2)_j+
{2\over \bar{u}}(k_1+k_2)_ik_{2j}\Big]
\nonumber
\end{eqnarray}
%

\subsection{``Cut vertex'' diagrams}
\begin{figure}

\vspace{-22mm}
\includegraphics[width=\textwidth]{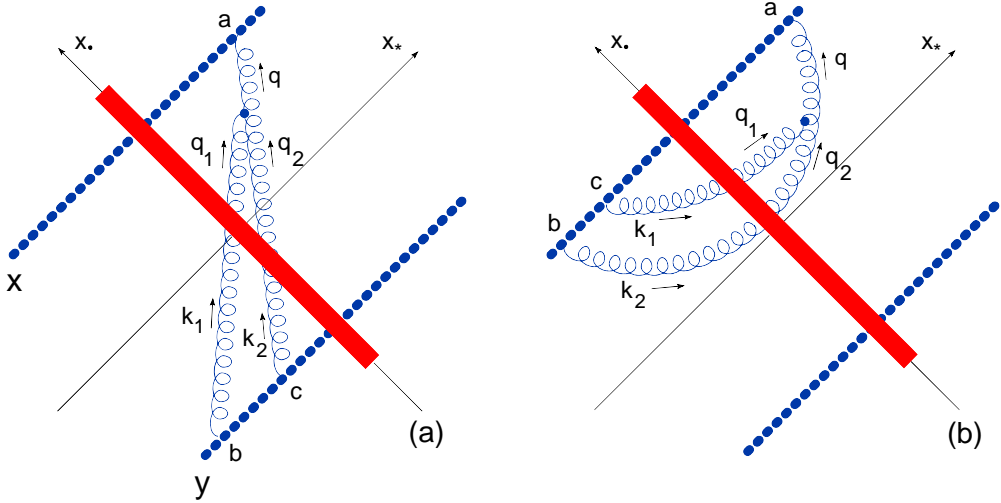}

\vspace{-145mm}
\caption{``Cut vertex'' diagrams\label{nlobk3}.}
\end{figure}
Next, consider the ``cut vertex'' diagram in Fig. \ref{nlobk3}a. The analog of Eq. (\ref{selfenergy2}) has the form:

\begin{eqnarray}
&&\hspace{-2mm}
g^3\!\int_0^{\infty}\!\!dt \!\int^0_{-\infty}\!\!du
\!\int^0_{u}\!\!dv\langle \hat{A}^a_{\bullet}(tp_1+x_\perp)
\hat{A}^b_{\bullet}(up_1+y_\perp)
\hat{A}^c_{\bullet}(vp_1+y_\perp)\rangle
\nonumber\\
&&\hspace{-2mm}
=~2g^4sf^{mna}\!\int\!\dhd\alpha_1\dhd\beta_1\dhd\alpha_2\dhd\beta_2
\!\int\! \dhd^2q_1\dhd^2q_2\dhd^2 k_1\dhd^2 k_2~
\Gamma^{\mu\nu\lambda}(\alpha_1p_1+q_{1\perp},\alpha_2p_1+q_{2\perp},
-(\alpha_1+\alpha_2)p_1-(q_1+q_2)_\perp)
\nonumber\\
&&\hspace{-2mm}
\times~
{\alpha_1\theta(\alpha_1)\theta(\alpha_2)
e^{i(q_1+q_2,x)_{\perp}}\Big((q_1+q_2)_{\perp}+2(\beta_1+\beta_2
) p_2\Big)_{\lambda}\over (\alpha_1+\alpha_2)(\beta_1+\beta_2-i\epsilon)}~
{\Big(k_1+{2(k_1,q_1)_{\perp}\over\alpha_1s}p_2\Big)_{\mu}
\Big(k_2+{2(k_2,q_2)_{\perp}\over\alpha_2s}p_2\Big)_{\nu}
\over  [(\alpha_1+\alpha_2)(\beta_1+\beta_2) s-(q_1+q_2)_{\perp}^2+i\epsilon]}
\nonumber\\
&&\hspace{42mm}
\times \int\!d^2z d^2z'
{e^{-i(k_1+k_2,y)_{\perp}}U_z^{mb}
U_{z'}^{nc}e^{-i(q_1-k_1,z)_\perp-i(q_2-k_2,z')_\perp}
\over
k_1^2(k_1^2\alpha_2+k_2^2\alpha_1)
(\alpha_1\beta_1 s-q_{1\perp}^2+i\epsilon)
(\alpha_2\beta_2 s-q_{2\perp}^2+i\epsilon)}
\label{vertex1}
\end{eqnarray}
Going to variables $\alpha=\alpha_1+\alpha_2$, $u=\alpha_1/\alpha$ and
taking residues at $\beta_1+\beta_2=0$ and $\beta_1={q_1^2\over\alpha_1}$  we get

\begin{eqnarray}
&&\hspace{-3mm}
g^3\!\int_0^{\infty}\!\!dt \!\int^0_{-\infty}\!\!du
\!\int^0_{u}\!\!dv\langle \hat{A}^a_{\bullet}(tp_1+x_\perp)
\hat{A}^b_{\bullet}(up_1+y_\perp)
\hat{A}^c_{\bullet}(vp_1+y_\perp)\rangle
\nonumber\\
&&\hspace{-3mm}
=~{g^4\over 2\pi^2}f^{mna}\!\int_0^\sigma\!{d\alpha\over\alpha^2}\!\int_0^1\! du~u
\!\int\! \dhd^2q_1\dhd^2q_2\dhd^2 k_1\dhd^2 k_2\!\int\!d^2z d^2z'
U_z^{mb}
U_{z'}^{nc}
{e^{i(q_1+q_2,x)_{\perp}-i(k_1+k_2,y)_{\perp}-i(q_1-k_1,z)_\perp-i(q_2-k_2,z')_\perp}
\over (q_1+q_2)_{\perp}^2(q_1^2\bar{u}+q_2^2u)
k_1^2(k_1^2\bar{u}+k_2^2u)}
\nonumber\\
&&\hspace{-3mm}
\times~\Big(k_1+{2(k_1,q_1)_{\perp}\over\alpha us}p_2\Big)_{\mu}
\Big(k_2+{2(k_2,q_2)_{\perp}\over\alpha \bar{u}s}p_2\Big)_{\nu}
(q_1+q_2)_{\perp\lambda}
\Gamma^{\mu\nu\lambda}(\alpha up_1+q_{1\perp},\alpha\bar{u}p_1+q_{2\perp},
-\alpha p_1-(q_1+q_2)_\perp)
\label{vertex2}
\end{eqnarray}
and therefore
\begin{eqnarray}
&&\hspace{-2mm}
\langle{\rm Tr}\{\hat{U}_x \hat{U}^\dagger_y\}\rangle_{\rm Fig. \ref{nlobk3}a}~
\label{vertex3}\\
&&\hspace{-3mm}=~
-i{g^4\over2\pi^2}f^{mna}~{\rm Tr}\{t^aU_x t^bt^cU^\dagger_y\}
\!\int_0^\sigma \!{d\alpha\over\alpha}\!\int_0^1\! du~\bar{u}u
\!\int \!\dhd^2 k_1\dhd^2k_2\dhd^2 q_1\dhd^2 q_2 
\!\int\!d^2z d^2z'U_z^{mb}U_{z'}^{nc}
~
\nonumber\\
&&\hspace{-2mm}
\times~           
 {(q_1^2-q_2^2)\delta_{ij}-
{2\over u}q_{1i}(q_1+q_2)_j+
{2\over \bar{u}}(q_1+q_2)_iq_{2j}
\over (q_1+q_2)^2(q_1^2\bar{u}+q_2^2u)}  ~
{k_{1i}k_{2j}\over \bar{u}k_1^2(k_1^2\bar{u}+k_2^2u)}  
e^{i(q_1+q_2,x)_{\perp}-i(k_1+k_2,y)_{\perp}-i(q_1-k_1,z)_\perp-i(q_2-k_2,z')_\perp}   
\nonumber     
\end{eqnarray}
where we have used the formula 
\begin{eqnarray}
&&\hspace{-2mm}
\Big(k_1+{2(k_1,q_1)_{\perp}\over\alpha us}p_2\Big)_{\mu}
\Big(k_2+{2(k_2,q_2)_{\perp}\over\alpha \bar{u}s}p_2\Big)_{\nu}
(q_1+q_2)_{\perp\lambda}
\Gamma^{\mu\nu\lambda}(\alpha up_1+q_{1\perp},\alpha\bar{u}p_1+q_{2\perp},
-\alpha p_1-(q_1+q_2)_\perp)
\nonumber\\
&&\hspace{-2mm}
=~k_{1i}k_{2j}\Big[(q_1^2-q_2^2)\delta_{ij}-
{2\over u}q_{1i}(q_1+q_2)_j+
{2\over \bar{u}}(q_1+q_2)_iq_{2j}\Big]
\label{vertex4}
\end{eqnarray}
following from Eq. (\ref{nice}).

The contribution of the diagram shown in Fig. \ref{nlobk3}b differs from Eq. (\ref{vertex3}) by the 
substituion $e^{-i(k_1+k_2,y)_\perp}\rightarrow e^{-i(k_1+k_2,x)_\perp}$ and changing the 
order of $t^b$, $t^c$ matrices. (Similarly to the case of the Fig. (\ref{nlobk2})b diagram, this 
prescription  follows from the replacement $[-\infty p_1,0]_y$ by $[0,\infty p_1]_x$ but now we consider the second term of the expansion in the gauge field). We get
\begin{eqnarray}
&&\hspace{-2mm}
\langle{\rm Tr}\{\hat{U}_x \hat{U}^\dagger_y\}\rangle_{\rm Fig. \ref{nlobk3}a+b}~
\label{vkladnlobk3}\\
&&\hspace{-3mm}=~
-i{g^4\over2\pi^2}
\!\int_0^\sigma \!{d\alpha\over\alpha}\!\int_0^1\! du~\bar{u}u
\!\int \!\dhd^2 k_1\dhd^2k_2\dhd^2 q_1\dhd^2 q_2 
\!\int\!d^2z d^2z'U_z^{mb}U_{z'}^{nc}
~ {(q_1^2-q_2^2)\delta_{ij}-
{2\over u}q_{1i}(q_1+q_2)_j+
{2\over \bar{u}}(q_1+q_2)_iq_{2j}
\over (q_1+q_2)^2(q_1^2\bar{u}+q_2^2u)} 
\nonumber\\
&&\hspace{-2mm}
\times~           
 ~
{k_{1i}k_{2j}\over \bar{u}k_1^2(k_1^2\bar{u}+k_2^2u)}  
e^{i(q_1+q_2,x)_{\perp}-i(q_1-k_1,z)_\perp-i(q_2-k_2,z')_\perp}   
~f^{mna}{\rm Tr}\{t^aU_x[  t^ct^be^{-i(k_1+k_2,x)}+
t^bt^ce^{-i(k_1+k_2,y)}]U^\dagger_y\}
\nonumber     
\end{eqnarray}

\begin{figure}

\vspace{-25mm}
\includegraphics[width=\textwidth]{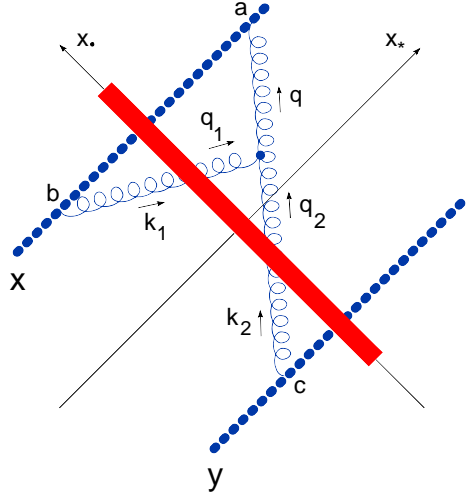}
\vspace{-142mm}
\caption{Another type of diagrams with two gluon-shockwave intersections \label{nlobk4}.}
\end{figure}

There is another type of  diagrams with two gluon-shockwave intersections shown in Fig. \ref{nlobk4}

%
\begin{eqnarray}
&&\hspace{-2mm}
g^3\!\int_0^{\infty}\!\!dt \!\int^0_{-\infty}\!\!du
\!\int^0_{-\infty}\!\!dv~\langle \hat{A}^a_{\bullet}(tp_1+x_\perp)
\hat{A}^b_{\bullet}(up_1+x_\perp)
\hat{A}^c_{\bullet}(vp_1+y_\perp)\rangle
\label{vertex5}\\
&&\hspace{-2mm}
=~2g^4s\!\int\!\dhd\alpha_1\dhd\alpha_2\dhd\beta_1\dhd\beta_2
\!\int\! d^2zd^2z'U_z^{mb}U_{z'}^{nc}
\!\int\! {\dhd^2k_1\dhd^2k_2\over k_1^2k_2^2}~
\int\! \dhd^2 q_1\dhd^2 q_2~\theta(\alpha_1)\theta(\alpha_2)
\nonumber\\
&&\hspace{-2mm}
\times~f^{mna}
\Gamma^{\mu\nu\lambda}(\alpha_1 p_1+q_1,\alpha_2 p_1+q_2,
-(\alpha_1+ \alpha_2)p_1-(q_1+q_2)_\perp) 
~e^{i(q_1,x-z)_{\perp}+i(q_2,x-z')_{\perp}-i(k_1,x-z)_{\perp}-i(k_2,y-z')_\perp}
\nonumber\\
&&\hspace{-2mm}
\times~
{[(q_1+q_2)_\perp+2(\beta_1+\beta_2)p_2]_\lambda
\over
(\alpha_1+\alpha_2)(\beta_1+\beta_2-i\epsilon)
[(\alpha_1+\alpha_2)(\beta_1+\beta_2) s-(q_1+q_2)^2_{\perp}+i\epsilon]}~
{k_{1\mu}+{2(k_1,q_1)_{\perp}\over\alpha_1 s}p_{2\mu}
\over  \alpha_1\beta_1 s-q_{1\perp}^2+i\epsilon}~
{k_{2\nu}+{2(k_2,q_2)_{\perp}\over\alpha_2 s}p_{2\nu}
\over \alpha_2\beta_2 s-q_{2\perp}^2+i\epsilon}
\Bigg\}
\nonumber
\end{eqnarray}
Taking residues at $\beta_1+\beta_2=0$ and at $\beta_1={q_1^2\over\alpha_1s}$ 
and going to variables $\alpha=\alpha_1+\alpha_2$ and $u=\alpha_1/\alpha$ we get

\begin{eqnarray}
&&\hspace{-2mm}
g^3\!\int_0^{\infty}\!\!dt \!\int^0_{-\infty}\!\!du
\!\int^0_{-\infty}\!\!dv~\langle \hat{A}^a_{\bullet}(tp_1+x_\perp)
\hat{A}^b_{\bullet}(up_1+x_\perp)
\hat{A}^c_{\bullet}(vp_1+y_\perp)\rangle
\label{vertex6}\\
&&\hspace{-2mm}=~{g^4\over 2\pi^2}f^{amn}\!\int_0^{\sigma}\!{d\alpha\over \alpha^2}
\!\int_0^1\! du
\!\int \!\dhd^2 q_1\dhd^2 q_2\dhd^2 k_1\dhd^2 k_2~
U_z^{mb}U_{z'}^{nc}~
{e^{i(q_1,x-z)_{\perp}+i(q_2,x-z')_{\perp}-i(k_1,x-z)_{\perp}-i(k_2,y-z')_\perp}
\over k_1^2k_2^2(q_1+q_2)^2(\bar{u} q_1^2+u q_2^2)}
\nonumber\\
&&\hspace{-2mm}
\Big(k_1+{2(q_1,k_1)_{\perp}p_2\over s\alpha u}\Big)^{\mu}
\Big(k_2+{2(q_2,k_2)_{\perp}p_2\over s\alpha\bar{u}}\Big)^{\nu}(q_1+q_2)^{\lambda}
\Gamma_{\mu\nu\lambda}(\alpha u p_1+q_1,\alpha\bar{u} p_1+q_2,
-\alpha p_1-(q_1+q_2)_\perp) 
\nonumber\\
&&\hspace{-2mm}
=~{g^4\over 2\pi^2}f^{amn}\!\int_0^{\sigma}\!{d\alpha\over \alpha^2}
\!\int_0^1\! du
\!\int \!\dhd^2 q_1\dhd^2 q_2\dhd^2 k_1\dhd^2 k_2~
U_z^{mb}U_{z'}^{nc}~
{e^{i(q_1,x-z)_{\perp}+i(q_2,x-z')_{\perp}-i(k_1,x-z)_{\perp}-i(k_2,y-z')_\perp}
\over k_1^2k_2^2(q_1+q_2)^2(\bar{u} q_1^2+u q_2^2)}
\nonumber\\
&&\hspace{-2mm}
\times~\Big[(q_{1\perp}^2-q_{2\perp}^2)\delta^{ij}-{2\over u} q_1^i(q_1+q_2)^j+
{2\over \bar{u}}(q_1+q_2)^iq_2^j\Big]k_{1i}k_{2j}
\nonumber
\end{eqnarray}
and therefore 
\begin{eqnarray}
&&\hspace{-12mm}
\langle{\rm Tr}\{\hat{U}_x \hat{U}^\dagger_y\}\rangle_{\rm Fig. \ref{nlobk4}}
\label{vertex7}\\
&&\hspace{-12mm}
=~
i{g^4\over 2\pi^2}f^{amn}
\!\int_0^\sigma\!{d\alpha\over\alpha}\!\int_0^1\!du\!\int\! d^2zd^2z'
\int\! \dhd^2q_1\dhd^2q_2\int\! \dhd^2k_1\dhd^2k_2
~e^{-i(q_1-k_1,z)_\perp-i(q_2-k_2,z')_\perp}U_z^{mm'}U_{z'}^{nn'}
\nonumber\\
&&\hspace{-12mm}
\times~
{e^{i(q_1+q_2,x)_\perp-i(k_1,x)_\perp-i(k_2,y)_\perp}\over (q_1+q_2)^2(q_1^2\bar{u}+q_2^2{u})}
\Big[(q_1^2-q_2^2)\delta_{ij}-{2\over u} q_{1i}(q_1+q_2)_j+
{2\over \bar{u}}(q_1+q_2)_iq_{2j}\Big]
{k_{1i}k_{2j}\over k_1^2k_2^2}{\rm Tr}\{t^aU_xt^{m'}t^{n'}
U^\dagger_y\}
\nonumber
\end{eqnarray}
where again we have used formula (\ref{vertex4}).

\begin{figure}

\vspace{-17mm}
\includegraphics[width=1.1\textwidth]{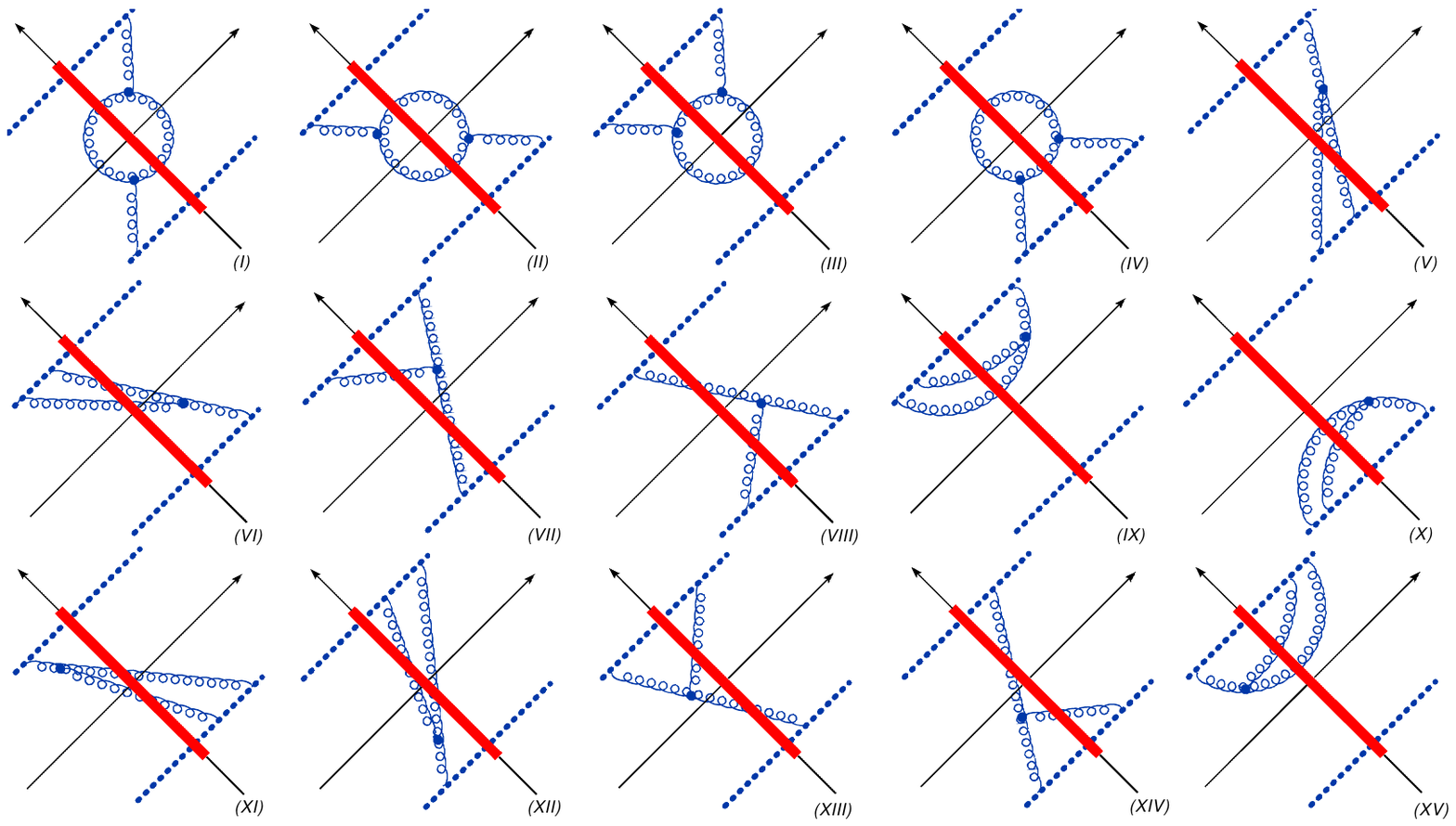}

\vspace{-170mm}
\includegraphics[width=1.1\textwidth]{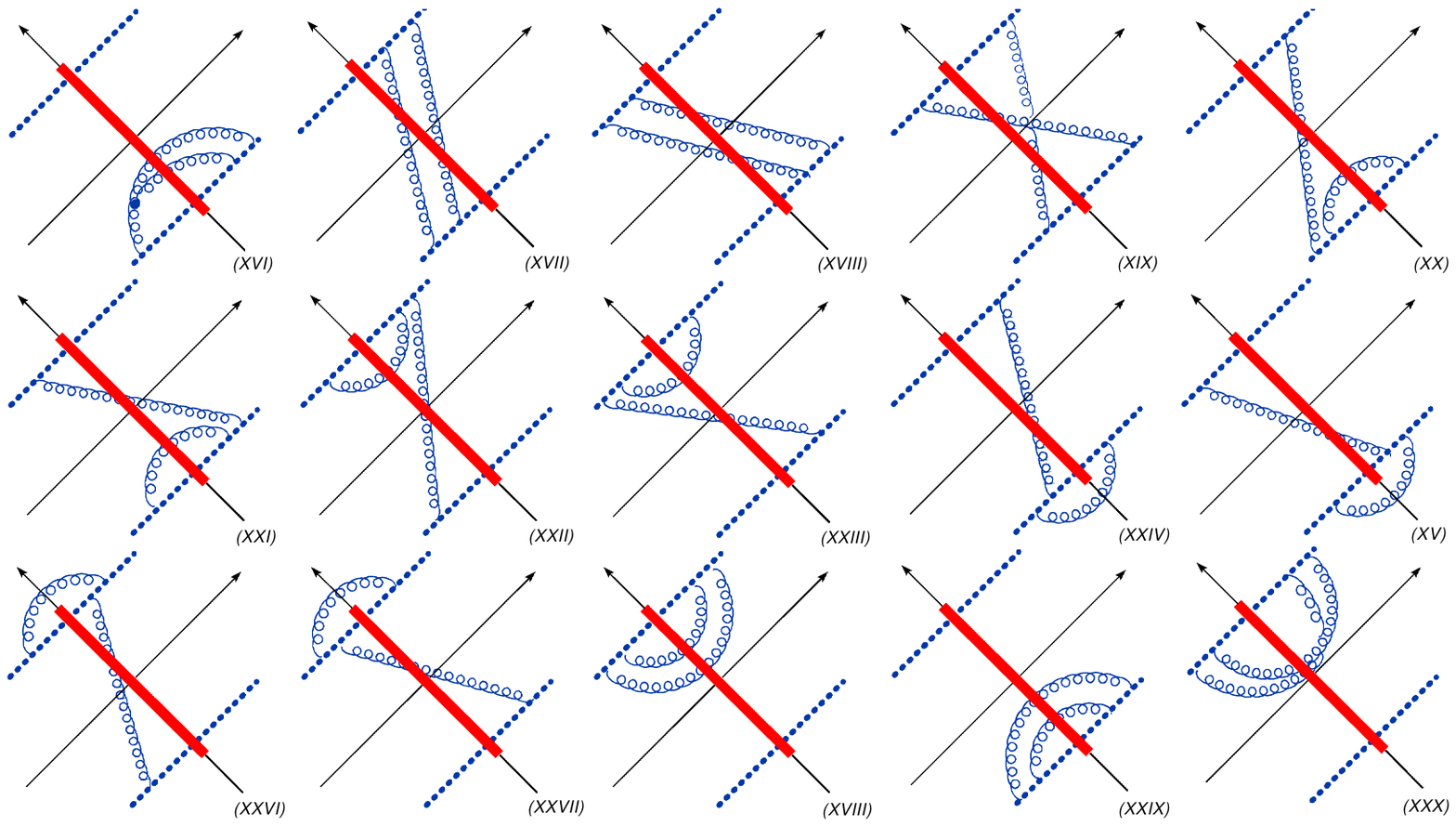}

\vspace{-170mm}
\includegraphics[width=1.1\textwidth]{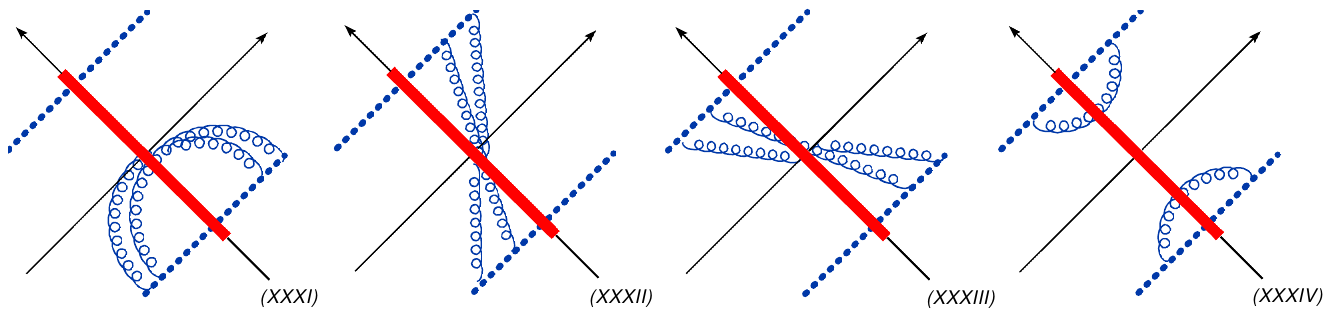}

\vspace{-199mm}
\caption{Diagrams with two cuts. \label{2cutdms}}
\end{figure}

The sum of the contributions (\ref{vkladnlobk2}), (\ref{vkladnlobk3}) and (\ref{vertex7})  can be represented as follows
\begin{eqnarray}
&&\hspace{-12mm}
\langle{\rm Tr}\{\hat{U}_x \hat{U}^\dagger_y\}\rangle_{\rm Fig. \ref{nlobk2} 
+ Fig. \ref{nlobk3}+ Fig. \ref{nlobk4}}~\equiv~
\langle{\rm Tr}\{\hat{U}_x \hat{U}^\dagger_y\}\rangle_{\rm Fig. \ref{2cutdms} ~
I+III+V+VII+IX}
\label{partialsum}\\
&&\hspace{-12mm}
=~
{g^2\over 8\pi^2}
\!\int_0^\infty\!{d\alpha\over\alpha}\!\int_0^1\!du~\bar{u} u~\int d^2zd^2z'
\int\! \dhd^2q_1\dhd^2q_2\int\! \dhd^2k_1\dhd^2k_2
~e^{-i(q_1-k_1,z)_\perp-i(q_2-k_2,z')_\perp}U_z^{mm'}U_{z'}^{nn'}
\nonumber\\
&&\hspace{-12mm}
\times~f^{amn}
{\rm Tr}\Big\{t^a
{e^{i(q_1+q_2,x)_\perp}
\over (q_1+q_2)^2(q_1^2\bar{u}+q_2^2{u})}
\Big[(q_1^2-q_2^2)\delta_{ij}-{2\over u} q_{1i}(q_1+q_2)_j+
{2\over \bar{u}}(q_1+q_2)_iq_{2j}\Big]U_x
\nonumber\\
&&\hspace{-12mm}
\times~
\Bigg[t^bf^{bm'n'}
{(e^{-i(k_1+k_2,x)_\perp}-e^{-i(k_1+k_2,y)_\perp})
\over (k_1+k_2)^2(k_1^2\bar{u}+k_2^2u)}
\Big[(k_1^2-k_2^2)\delta_{ij}-{2\over u} k_{1i}(k_1+k_2)_j+
{2\over\bar{u}}(k_1+k_2)_ik_{2j}\Big]
\nonumber\\
&&\hspace{-12mm}
-~4ik_{1i}k_{2j}\Bigg(
{t^{n'}t^{m'}e^{-i(k_1+k_2,x)_\perp}\over \bar{u}k_1^2(k_1^2\bar{u}+k_2^2u)}
-{e^{-i(k_1,x)_\perp-i(k_2,y)_\perp}\over \bar{u}uk_1^2k_2^2}t^{m'}t^{n'}
+~{t^{m'}t^{n'}e^{-i(k_1+k_2,y)_\perp}\over \bar{u}k_1^2(k_1^2\bar{u}+k_2^2u)}
\Bigg)\Bigg]
U^\dagger_y\Big\}
\nonumber
\end{eqnarray}
If we add contribution of the diagrams with the gluon on the right side of the shock wave attached to the Wilson line at the point $y$ instead of $x$ (which differs from Eq. (\ref{partialsum}) by the substitution $e^{i(q_1+q_2,x)_\perp}\rightarrow-e^{i(q_1+q_2,y)_\perp}$) we obtain 
\begin{eqnarray}
&&\hspace{-12mm}
\langle{\rm Tr}\{\hat{U}_x \hat{U}^\dagger_y\}\rangle_{\rm Fig.\ref{2cutdms}~I+II+...+X}
\label{partialsum1}\\
&&\hspace{-12mm}
=~
{g^2\over 8\pi^2}
\!\int_0^\infty\!{d\alpha\over\alpha}\!\int_0^1\!du~\bar{u} u\!\!\int d^2zd^2z'\!
\int\! \dhd^2q_1\dhd^2q_2\int\! \dhd^2k_1\dhd^2k_2
~e^{-i(q_1-k_1,z)_\perp-i(q_2-k_2,z')_\perp}U_z^{mm'}U_{z'}^{nn'}
\nonumber\\
&&\hspace{-12mm}
\times~f^{amn}
{\rm Tr}\Big\{t^a
{(e^{i(q_1+q_2,x)_\perp}-e^{i(q_1+q_2,y)_\perp})
\over (q_1+q_2)^2(q_1^2\bar{u}+q_2^2{u})}
\Big[(q_1^2-q_2^2)\delta_{ij}-{2\over u} q_{1i}(q_1+q_2)_j+
{2\over \bar{u}}(q_1+q_2)_iq_{2j}\Big]U_x
\nonumber\\
&&\hspace{-12mm}
\times~
\Bigg[t^bf^{bm'n'}
{(e^{-i(k_1+k_2,x)_\perp}-e^{-i(k_1+k_2,y)_\perp})
\over (k_1+k_2)^2(k_1^2\bar{u}+k_2^2u)}
\Big[(k_1^2-k_2^2)\delta_{ij}-{2\over u} k_{1i}(k_1+k_2)_j+
{2\over\bar{u}}(k_1+k_2)_ik_{2j}\Big]
\nonumber\\
&&\hspace{-12mm}
-~4ik_{1i}k_{2j}\Bigg(
{t^{n'}t^{m'}e^{-i(k_1+k_2,x)_\perp}\over \bar{u}k_1^2(k_1^2\bar{u}+k_2^2u)}
-{e^{-i(k_1,x)_\perp-i(k_2,y)_\perp}\over \bar{u}uk_1^2k_2^2}t^{m'}t^{n'}
+~{t^{m'}t^{n'}e^{-i(k_1+k_2,y)_\perp}\over \bar{u}k_1^2(k_1^2\bar{u}+k_2^2u)}
\Bigg)\Bigg]
U^\dagger_y\Big\}
\nonumber
\end{eqnarray}
The result (\ref{partialsum1}) can be obtained from the self-energy contribution 
(\ref{vkladnlobk2}) by the replacement of the term corresponding to the 
emission of the two gluons via the three-gluon vertex
\begin{eqnarray}
&&\hspace{-12mm}
t^bf^{bm'n'}
{(e^{-i(k_1+k_2,x)_\perp}-e^{-i(k_1+k_2,y)_\perp})
\over (k_1+k_2)^2(k_1^2\bar{u}+k_2^2u)}
\Big[(k_1^2-k_2^2)\delta_{ij}-{2\over u} k_{1i}(k_1+k_2)_j+
{2\over\bar{u}}(k_1+k_2)_ik_{2j}\Big]
\nonumber
\end{eqnarray}
with similar contribution containing the ``effective vertex''
\begin{eqnarray}
&&\hspace{-12mm}
{\rm S}^{m'n'}(k_1,k_2;x,y)\equiv\nonumber\\
&&\hspace{-12mm}
t^bf^{bm'n'}
{(e^{-i(k_1+k_2,x)_\perp}-e^{-i(k_1+k_2,y)_\perp})
\over (k_1+k_2)^2(k_1^2\bar{u}+k_2^2u)}
\Big[(k_1^2-k_2^2)\delta_{ij}-{2\over u} k_{1i}(k_1+k_2)_j+
{2\over\bar{u}}(k_1+k_2)_ik_{2j}\Big]
\nonumber\\
&&\hspace{-12mm}
-~4ik_{1i}k_{2j}\Bigg(
{t^{n'}t^{m'}e^{-i(k_1+k_2,x)_\perp}\over \bar{u}k_1^2(k_1^2\bar{u}+k_2^2u)}
-{e^{-i(k_1,x)_\perp-i(k_2,y)_\perp}\over \bar{u}uk_1^2k_2^2}t^{m'}t^{n'}
+~{t^{m'}t^{n'}e^{-i(k_1+k_2,y)_\perp}\over \bar{u}k_1^2(k_1^2\bar{u}+k_2^2u)}
\Bigg)\label{effective1}
\end{eqnarray}
%

It can be demonstrated that the sum of the contributions of the diagrams shown 
in Fig. \ref{2cutdms} I,..., IV, XI,..., XVI can be obtained from the self-energy contribution 
(\ref{vkladnlobk2}) by replacing
the gluon vertex 
\begin{eqnarray}
&&\hspace{-12mm}
t^af^{amn}
{(e^{i(q_1+q_2,x)_\perp}-e^{i(q_1+q_2,y)_\perp})
\over (q_1+q_2)^2(q_1^2\bar{u}+q_2^2{u})}
\Big[(q_1^2-q_2^2)\delta_{ij}-{2\over u} q_{1i}(q_1+q_2)_j+
{2\over \bar{u}}(q_1+q_2)_iq_{2j}\Big]\nonumber\\
\end{eqnarray}
with similar ``effective vertex''
\begin{eqnarray}
&&\hspace{-12mm}
t^af^{amn}
{(e^{i(q_1+q_2,x)_\perp}-e^{i(q_1+q_2,y)_\perp})
\over (q_1+q_2)^2(q_1^2\bar{u}+q_2^2{u})}
\Big[(q_1^2-q_2^2)\delta_{ij}-{2\over u} q_{1i}(q_1+q_2)_j+
{2\over \bar{u}}(q_1+q_2)_iq_{2j}\Big]
\nonumber\\
&&\hspace{-12mm}
+~4iq_{1i}q_{2j}\Bigg(
{t^mt^ne^{i(q_1+q_2,x)_\perp}\over\bar{u} q_1^2(q_1^2\bar{u}+q_2^2u)}
-{e^{i(q_1,x)_\perp+i(q_2,y)_\perp}\over \bar{u}uq_1^2q_2^2}t^nt^m
+{t^nt^me^{i(q_1+q_2,y)_\perp}\over \bar{u}q_1^2(q_1^2\bar{u}+q_2^2u)}
\Bigg)\label{effective2}
\end{eqnarray}
Note that (\ref{effective2}) is equal to ${\rm S}^{\dagger mn}(q_1,q_2;x,y)$.
Let us consider now the box diagrams topology shown in fig. \ref{2cutdms} XVII-XXXIV. 
The calculation of these diagrams is similar to the above calculation of  ``cut self-energy'' 
and ``cut vertex'' diagrams so we present here only the final result  
\begin{eqnarray}
&&\hspace{-12mm}
\langle{\rm Tr}\{\hat{U}_x \hat{U}^\dagger_y\}\rangle_{\rm Fig. \ref{2cutdms}~XVII+...+XXXIV}
\label{boxsum}\\
&&\hspace{-12mm}
=~
{g^2\over 2\pi^2}
\!\int_0^\sigma\!{d\alpha\over\alpha}\!\int_0^1\!du~\bar{u} u\!\int d^2zd^2z'
\!\int\! \dhd^2q_1\dhd^2q_2\int\! \dhd^2k_1\dhd^2k_2
~e^{-i(q_1-k_1,z)_\perp-i(q_2-k_2,z')_\perp}U_z^{mm'}U_{z'}^{nn'}
\nonumber\\
&&\hspace{-12mm}
\times~
{\rm Tr}\Big\{\Big[
q_{1i}q_{2j}\Bigg(
{t^mt^ne^{i(q_1+q_2,x)_\perp}\over\bar{u} q_1^2(q_1^2\bar{u}+q_2^2u)}
+{t^nt^me^{i(q_1+q_2,x)_\perp}\over uq_2^2(q_1^2\bar{u}+q_2^2u)}
-{e^{i(q_1,x)_\perp+i(q_2,y)_\perp}\over \bar{u}uq_1^2q_2^2}t^nt^m
-{e^{i(q_1,y)_\perp+i(q_2,x)_\perp}\over \bar{u}uq_1^2q_2^2}t^mt^n
\nonumber\\
&&\hspace{62mm}
+~{t^nt^me^{i(q_1+q_2,y)_\perp}\over \bar{u}q_1^2(q_1^2\bar{u}+q_2^2u)}
+{t^mt^ne^{i(q_1+q_2,y)_\perp}\over uq_2^2(q_1^2\bar{u}+q_2^2u)}\Bigg)\Bigg]U_x
\nonumber\\
&&\hspace{-12mm}
\times~
\Big[k_{1i}k_{2j}\Bigg(
{t^{n'}t^{m'}e^{-i(k_1+k_2,x)_\perp}\over \bar{u}k_1^2(k_1^2\bar{u}+k_2^2u)}+
{ t^{m'}t^{n'}e^{-i(k_1+k_2,x)_\perp}\over uk_2^2(k_1^2\bar{u}+k_2^2u)}
-{e^{-i(k_1,x)_\perp-i(k_2,y)_\perp}\over \bar{u}uk_1^2k_2^2}t^{m'}t^{n'}
-{e^{-i(k_2,x)_\perp-i(k_1,y)_\perp}\over \bar{u}uk_1^2k_2^2}t^{n'}t^{m'}
\nonumber\\
&&\hspace{52mm}
+~{t^{m'}t^{n'}e^{-i(k_1+k_2,y)_\perp}\over \bar{u}k_1^2(k_1^2\bar{u}+k_2^2u)}
+{t^{n'}t^{m'}e^{-i(k_1+k_2,y)_\perp}\over uk_2^2(k_1^2\bar{u}+k_2^2u)}\Bigg)\Bigg]
U^\dagger_y\Big\}
\nonumber
\end{eqnarray}
This expression agrees with the sum of ``box topology''  diagrams in Ref. \cite{balbel}.

Now we observe that each three-gluon vertex diagram is equal to its own cross diagram (the same cannot be said for 
box diagrams). Thus we may redefine the "effective vertex" (\ref{effective1}) in the following way
\begin{eqnarray}
&&\hspace{-12mm}
{\cal S}^{m'n'}(k_1,k_2;x,y)=\nonumber\\
&&\hspace{-12mm}
t^bf^{bm'n'}
{(e^{-i(k_1+k_2,x)_\perp}-e^{-i(k_1+k_2,y)_\perp})
\over (k_1+k_2)^2(k_1^2\bar{u}+k_2^2u)}
\Big[(k_1^2-k_2^2)\delta_{ij}-{2\over u} k_{1i}(k_1+k_2)_j+
{2\over\bar{u}}(k_1+k_2)_ik_{2j}\Big]
\nonumber\\
&&\hspace{-12mm}
-~i2k_{1i}k_{2j}\Bigg(
{t^{n'}t^{m'}e^{-i(k_1+k_2,x)_\perp}\over \bar{u}k_1^2(k_1^2\bar{u}+k_2^2u)}+
{ t^{m'}t^{n'}e^{-i(k_1+k_2,x)_\perp}\over uk_2^2(k_1^2\bar{u}+k_2^2u)}
-{e^{-i(k_1,x)_\perp-i(k_2,y)_\perp}\over \bar{u}uk_1^2k_2^2}t^{m'}t^{n'}
-{e^{-i(k_2,x)_\perp-i(k_1,y)_\perp}\over \bar{u}uk_1^2k_2^2}t^{n'}t^{m'}
\nonumber\\
&&\hspace{52mm}
+~{t^{m'}t^{n'}e^{-i(k_1+k_2,y)_\perp}\over \bar{u}k_1^2(k_1^2\bar{u}+k_2^2u)}
+{t^{n'}t^{m'}e^{-i(k_1+k_2,y)_\perp}\over uk_2^2(k_1^2\bar{u}+k_2^2u)}\Bigg)
\end{eqnarray}
which corresponds to writing each contribution of the three-gluon vertex diagrams 
as a sum of two equal terms.

A similar expression can be written for the "effective vertex" (\ref{effective2})
and therefore the sum of all diagrams with two gluon-shockwave intersections can be written as 
\begin{eqnarray}
&&\hspace{-12mm}
\langle{\rm Tr}\{\hat{U}_x \hat{U}^\dagger_y\}_{\rm Fig. \ref{2cutdms}}
~=~
{g^2\over 8\pi^2}
\!\int_0^\sigma\!{d\alpha\over\alpha}\!\int_0^1\!du~\bar{u} u\!\int d^2zd^2z'
\!\int\! \dhd^2q_1\dhd^2q_2\int\! \dhd^2k_1\dhd^2k_2
~e^{-i(q_1-k_1,z)_\perp-i(q_2-k_2,z')_\perp}
\nonumber\\
&&\hspace{-12mm}
\times~U_z^{mm'}U_{z'}^{nn'}
{\rm Tr}\Big\{\big[{\cal S}^{\dagger mn}(q_1,q_2;x,y)U_x\big]
\big[ {\cal S}^{ m'n'}(k_1,k_2;x,y)U^\dagger_y\big]\Big\}
\label{twocutsum}
\end{eqnarray}

Separating the contributions of different color structures one obtains
\begin{eqnarray}
&&\hspace{-2mm}
\langle{\rm Tr}\{\hat{U}_x \hat{U}^\dagger_y\}_{\rm Fig. \ref{2cutdms}}
\label{momresult}\\
&&\hspace{-2mm}
=~
{g^2\over 8\pi^2}
\!\int_0^\sigma\!{d\alpha\over\alpha}\!\int_0^1\!{du\over \bar{u} u}~\int d^2zd^2z'
\int\! \dhd^2q_1\dhd^2q_2\int\! \dhd^2k_1\dhd^2k_2
~U_z^{mm'}U_{z'}^{nn'}
\nonumber\\
&&\hspace{-2mm}
\times~
{\rm Tr}\Big\{t^af^{amn}
{(e^{i(q_1,X)+i(q_2,X')}-x\leftrightarrow y)\over (q_1^2\bar{u}+q_2^2{u})}
\Big[{(q_1^2-q_2^2)\bar{u} u\delta_{ij}-2\bar{u} q_{1i}(q_1+q_2)_j+
2u(q_1+q_2)_iq_{2j}\over (q_1+q_2)^2}
-u{q_{1i}q_{2j}\over q_1^2}+\bar{u}{q_{1i}q_{2j}\over q_2^2}\Big]
\nonumber\\
&&\hspace{-2mm}
-~t^af^{amn}{q_{1i}q_{2j}\over q_1^2q_2^2}(e^{i(q_1,X)+i(q_2,Y')}-x\leftrightarrow y)
+i\{t^m,t^n\}{q_{1i}q_{2j}\over q_1^2q_2^2}
(e^{i(q_1,X)}-e^{i(q_1,Y)})(e^{i(q_2,X')}-e^{i(q_2,Y')})\Big\}U_x
\nonumber\\
&&\hspace{-2mm}
\times~
\Big\{t^bf^{bm'n'}{(e^{-i(k_1,X)-i(k_2,X')}-x\leftrightarrow y)
\over (k_1^2\bar{u}+k_2^2u)}
\Big[{(k_1^2-k_2^2)\bar{u} u\delta_{ij}-2\bar{u} k_{1i}(k_1+k_2)_j+
2u(k_1+k_2)_ik_{2j}\over (k_1+k_2)^2}
-u{k_{1i}k_{2j}\over k_1^2}+\bar{u}{k_{1i}k_{2j}\over k_2^2}\Big]    
\nonumber\\
&&\hspace{-2mm}
-~t^bf^{bm'n'}{k_{1i}k_{2j}\over k_1^2k_2^2}(e^{-i(k_1,X)-i(k_2,Y')}
-x\leftrightarrow y)
-i\{t^{m'},t^{n'}\}{k_{1i}k_{2j}\over k_1^2k_2^2}
(e^{-i(k_1,X)}-e^{-i(k_1,Y)})(e^{-i(k_2,X')}-e^{-i(k_2,Y')})\Big\}U_y          
\nonumber
\end{eqnarray}

This result agrees with Ref. \onlinecite{balbel}.

Performing the Fourier transformation

\begin{eqnarray}
&&\hspace{-15mm}
\int\dhd^2 q_1\dhd^2 q_2~e^{i(q_1,x_1)+i(q_2,x_2)}~
{q_{1i}q_{2j}
\over q_1^2(q_1^2\bar{u}+q_2^2u)}
~=~-
{x_{1i}x_{2j}
\over 4\pi^2x_2^2(ux_1^2+\bar{u}x_2^2)}
\nonumber\\
&&\hspace{-15mm}
\int\dhd^2 q_1\dhd^2 q_2~e^{i(q_1,x_1)+i(q_2,x_2)}~
{\delta_{ij}(q_1^2-q_2^2)-{2\over u}q_{1i}(q_1+q_2)_j+{2\over \bar{u}}(q_1+q_2)_iq_{2j}
\over(q_1+q_2)^2(q_1^2\bar{u}+q_2^2u)}
\nonumber\\
&&\hspace{-15mm}
=~
{-(x_1^2-x_2^2)\delta_{ij}+
{2\over u}(x_1-x_2)_ix_{2_j}+{2\over \bar{u}}x_{1i}(x_1-x_2)_j
\over 4\pi^2(x_1-x_2)^2(ux_1^2+\bar{u}x_2^2)}
\label{furie}
\end{eqnarray}
we get
\begin{eqnarray}
&&\hspace{-4mm}
{d\over d\ln\sigma} \langle {\rm Tr}\{\hat{U}_x\hat{U}^{\dagger}_y\}
\rangle_{\rm Fig. \ref{2cutdms}}~=~
{\alpha_s^2\over 8\pi^4}
\int_0^1 du~u\bar{u}\!\int \!d^2 zd^2 z'~U_z^{bb'}U_{z'}^{cc'}
\nonumber\\ 
&&\hspace{-4mm}
\times~{\rm Tr}\Big\{t^af^{abc}\Bigg[
{{X_{ij}\over (z-z')^2}+{X_iX'_j\over uX^2}
-{X_iX'_j\over \bar{u}{X'}^2}\over  uX^2+\bar{u}{X'}^2}
-{X_iY'_j\over \bar{u} u X^2{Y'}^2}-(x\leftrightarrow y)\Bigg]
+i{\{t^b,t^c\}\over\bar{u} u}\Big({X_i\over X^2}-{Y_i\over Y^2}\Big)
\Big({X'_j\over {X'}^2}-{Y'_j\over {Y'}^2}\Big)
\Big\}U_x\nonumber\\ 
&&\hspace{-4mm}\times~
\Big\{t^{a'}f^{a'b'c'}\Bigg[{{X_{ij}\over (z-z')^2}
+ {X_iX'_j\over uX^2}
-{X_iX'_j\over \bar{u}{X'}^2}\over  uX^2+\bar{u}{X'}^2}
-{X_iY'_j\over \bar{u} u X^2{Y'}^2}-(x\leftrightarrow y)\Bigg]
-i{\{t^{b'},t^{c'}\}\over\bar{u} u}\Big({X_i\over X^2}-{Y_i\over Y^2}\Big)
\Big({X'_j\over {X'}^2}-{Y'_j\over {Y'}^2}\Big)
\Big\}U^{\dagger}_y
\label{2cutsum}
\end{eqnarray}
where we introduced the notations
\begin{eqnarray}
X_{ij}&\equiv&(X^2-{X'}^2)\delta_{ij}+{2\over u}(z-z')_iX'_j+{2\over\bar{u}}X_i(z-z')_j
\nonumber\\
Y_{ij}&\equiv&(Y^2-{Y'}^2)\delta_{ij}+{2\over u}(z-z')_iY'_j+
{2\over \bar{u}}Y_i(z-z')_j
\label{XY}
\end{eqnarray}
%

\subsection{Subtraction of the $(LO)^2$ contribution \label{sect:losubtraction}}

It is easy to see that result (\ref{2cutsum}) for the sum of diagrams in Fig. \ref{2cutdms} diverges as $u\rightarrow 0$ and $u\rightarrow 1$. If we put a lower cutoff $\alpha>\sigma'$ on the 
$\alpha$ integrals we would get a contribution $\sim \ln^2{\sigma\over\sigma'}$ coming from the region $\alpha_2\gg\alpha_1>\sigma'$ (or $\alpha_1\gg\alpha_2>\sigma'$ )
which corresponds to the the square of the leading-order BK kernel rather than to the NLO
kernel. To get the NLO kernel we need to subtract this $(LO)^2$ contribution. 
Indeed, the operator form of the evolution equation for the color dipole up to the 
next-to-leading order looks like
\begin{eqnarray}
&&\hspace{-6mm}
{\partial\over\partial\eta}{\rm Tr}\{\hat{U}_x\hat{U}^\dagger_y\}=
K_{\rm LO}{\rm Tr}\{\hat{U}_x\hat{U}^\dagger_y\}+K_{\rm NLO}{\rm Tr}\{\hat{U}_x\hat{U}^\dagger_y\}
\label{nloeveq}
\end{eqnarray}
where $\eta=\ln\sigma$. 
Our goal is to find $K_{\rm NLO}$ by considering the l.h.s. of this equation in the 
external shock-wave background so 
\begin{eqnarray}
&&\hspace{-6mm}
 \langle K_{\rm NLO}{\rm Tr}\{\hat{U}_x\hat{U}^\dagger_y\}\rangle_{\rm shockwave}=
{\partial\over\partial\eta}\langle{\rm Tr}\{\hat{U}_x\hat{U}^\dagger_y\}\rangle_{\rm shockwave}-
\langle K_{\rm LO}{\rm Tr}\{\hat{U}_x\hat{U}^\dagger_y\}\rangle_{\rm shockwave}
\label{subtraction}
\end{eqnarray}
The subtraction (\ref{subtraction}) leads to the $\Big[{1\over u}\Big]_+$ prescription
for the terms divergent as ${1\over u}$ 
(and similarly ${1\over \bar{u}}\rightarrow \Big[{1\over \bar{u}}\Big]_+$ for the contribution divergent as $u\rightarrow 1$). Here we define $\Big[{1\over u}\Big]_+$  in the usual way
\begin{equation}
\int_0^1\! du ~f(u)\Big[{1\over u}\Big]_+~\equiv~\int_0^1\! du ~{f(u)-f(0)\over u},~~~~~
\int_0^1\! du ~f(u)\Big[{1\over \bar{u}}\Big]_+~\equiv~\int_0^1\! du ~{f(u)-f(1)\over \bar{u}}
\label{pluscription}
\end{equation}

To illustrate this prescription, consider  the divergent terms in Eq. (\ref{2cutsum})
proportional to $(X,Y)(Y',z-z')$ or $(X',Y')(Y,z-z')$
\begin{eqnarray}
&&\hspace{-2mm} 
{d\over d\eta}\langle{\rm Tr}\{\hat{U}_x\hat{U}^{\dagger}_y\}\rangle~=~
{\alpha_s^2\over 4\pi^4}
\int_0^1 \!{du\over \bar{u}}\!\int \!d^2 zd^2 z'~U_z^{bb'}U_{z'}^{cc'}
{(X,Y)(Y',z-z')\over (z-z')^2{Y'}^2}\Bigg[{1\over  uX^2+\bar{u}{X'}^2}
\nonumber\\ 
&&\hspace{-2mm}
\times~\Big[f^{abc}{\rm Tr}\{t^aU_x\Big(
 {uf^{a'b'c'}t^{a'}\over  uY^2+\bar{u}{Y'}^2}
- {i\over Y^2}\{t^{b'},t^{c'}\}\Big)U^\dagger_y\}  
+ f^{a'b'c'} {\rm Tr}\{\Big(
t^a {uf^{abc}\over uY^2+\bar{u}{Y'}^2}
+{i\over Y^2}\{t^b,t^c\}\Big)U_xt^{a'}U^\dagger_y\}   \Big]
\nonumber\\
&&\hspace{-2mm} 
+~{1\over X^2(uY^2+\bar{u}{Y'}^2)}{\rm Tr}\{(f^{abc}t^a+i\{t^b,t^c\})U_xf^{a'b'c'}t^{a'}U^\dagger_y  
+t^a f^{abc}U_x( f^{a'b'c'} t^{a'}-i\{t^{b'},t^{c'}\})U^\dagger_y\} \Bigg]
\nonumber\\
&&\hspace{-2mm} 
+~{\alpha_s^2\over 4\pi^4}
\int_0^1 \!{du\over u}\!\int \!d^2 zd^2 z'~U_z^{bb'}U_{z'}^{cc'}
{(X',Y')(Y,z-z')\over (z-z')^2Y^2}\Bigg[{1\over  uX^2+\bar{u}{X'}^2}
\nonumber\\ 
&&\hspace{-2mm}
\times~\Big[f^{abc}{\rm Tr}\{-t^aU_x\Big(
 {\bar{u}f^{a'b'c'}t^{a'}\over (uY^2+\bar{u}{Y'}^2)}
+ {i\over {Y'}^2}\{t^{b'},t^{c'}\}\Big)U^\dagger_y\}  
+ f^{a'b'c'} {\rm Tr}\{\Big(
-t^a {\bar{u}Y_iY'_jf^{abc}\over uY^2+\bar{u}{Y'}^2}
+{i\over {Y'}^2}\{t^b,t^c\}\Big)U_xt^{a'}U^\dagger_y\}   \Big]
\nonumber\\ 
&&\hspace{-2mm}
+~{1\over  {X'}^2( uY^2+\bar{u}{Y'}^2)}{\rm Tr}\{(-f^{abc}t^a+i\{t^b,t^c\})U_xf^{a'b'c'}t^{a'}U^\dagger_y  
-t^a f^{abc}U_x( f^{a'b'c'} t^{a'}+i\{t^{b'},t^{c'}\})U^\dagger_y\}\Bigg]
\label{subtr1}
\end{eqnarray}
Note that the second term is equal to the first one after the replacement $u\leftrightarrow \bar{u}$, $z\leftrightarrow z'$ and $b\leftrightarrow c, b'\leftrightarrow c'$.

It is convenient to return back to the notation $\alpha_1$ and $\alpha_2=\sigma-\alpha_1$
(after ${d\over d\ln\sigma}$ the value of $\alpha$ is set equal to $\sigma$).
\begin{eqnarray}
&&\hspace{-2mm} 
{d\over d\eta}\langle{\rm Tr}\{\hat{U}_x\hat{U}^{\dagger}_y\}\rangle~=~
{\alpha_s^2\over 2\pi^4}
\!\int_0^\sigma\! {d\alpha_2\over \alpha_2}\!\int \!d^2 zd^2 z'~U_z^{bb'}U_{z'}^{cc'}
{(X,Y)(Y',z-z')\over (z-z')^2{Y'}^2}
\label{subtr2}\\
&&\hspace{12mm}
\times~
\Bigg[
{\sigma\over  \alpha_1X^2+\alpha_2{X'}^2}\Big[f^{abc}{\rm Tr}\{t^aU_x\Big(
 {\alpha_1f^{a'b'c'}t^{a'}
\over \alpha_1Y^2+\alpha_2{Y'}^2}
-~ {i\over Y^2}\{t^{b'},t^{c'}\}\Big)U^\dagger_y\}  
+ ~f^{a'b'c'} {\rm Tr}\{\Big(
 {\alpha_1f^{abc}t^a\over \alpha_1Y^2+\alpha_2{Y'}^2}
\nonumber\\ 
&&\hspace{22mm}
+~{i\over Y^2}\{t^b,t^c\}\Big)U_xt^{a'}U^\dagger_y\}   \Big]          
+~{2i\sigma\over X^2(\alpha_1Y^2+\alpha_2{Y'}^2)}{\rm Tr}\{t^ct^bU_xf^{a'b'c'}t^{a'}U^\dagger_y  
-t^a f^{abc}U_xt^{b'}t^{c'}U^\dagger_y\} \Bigg]    
\nonumber
\end{eqnarray}
The corresponding term in $K_{\rm LO} {\rm Tr}\{\hat{U}_x\hat{U}^{\dagger}_y\}$ is (see Eq. (\ref{bk}))
\begin{eqnarray}
&&\hspace{-2mm} 
K_{\rm LO} {\rm Tr}\{\hat{U}_x\hat{U}^{\dagger}_y\}~=~-{4\alpha_s\over\pi^2}\int\! d^2z~{(x-z,y-z)\over (x-z)^2(y-z)^2}
 {\rm Tr}\{t^a\hat{U}_x t^b \hat{U}^{\dagger}_y\} {\rm Tr}\{t^a\hat{U}_zt^b\hat{U}^{\dagger}_z\}
\label{subtr3}
\end{eqnarray}
The relevant term in the ``matrix element'' $\langle K_{\rm LO} {\rm Tr}\{\hat{U}_x\hat{U}^{\dagger}_y\}\rangle$ in the external shock-wave background 
comes from $\hat{U}_x,\hat{U}^\dagger_z$ taken in the leading order in $\alpha_s$ (so that $\hat{U}_x\rightarrow U_x,~\hat{U}^\dagger_z\rightarrow U^\dagger_z$) and $\hat{U}_z\otimes\hat{U}^\dagger_y$
taken in the first order in $\alpha_s$ 
\begin{eqnarray}
&&\hspace{-2mm} 
\langle \hat{U}_z\otimes \hat{U}^{\dagger}_y\rangle~\sim~
-{\alpha_s\over\pi^2}\int_0^\sigma {d\alpha_2\over \alpha_2}\!\int\! d^2z'~{(z-z',y-z')\over (z-z')^2(y-z')^2}
(t^cU_z \otimes t^{c'} U^{\dagger}_y+U_zt^{c'} \otimes U^{\dagger}_yt^c)U_{z'}^{cc'},
 \label{subtr4}
\end{eqnarray}
or vice versa:  $\hat{U}_x\rightarrow U_x,~\hat{U}_z\rightarrow U_z$ and
\begin{eqnarray}
&&\hspace{-2mm} 
\langle \hat{U}^\dagger_z\otimes \hat{U}^{\dagger}_y \rangle~\sim~
{\alpha_s\over\pi^2}\int_0^\sigma {d\alpha_2\over \alpha_2}\!\int\! d^2z'~{(z-z',y-z')\over (z-z')^2(y-z')^2}
(t^{c'} U^\dagger_z \otimes U^{\dagger}_yt^c+ U^\dagger_z t^c\otimes t^{c'} U^\dagger_y)U_{z'}^{cc'}
 \label{subtr5}
\end{eqnarray}
Here we have used the leading-order equations for Wilson lines with arbitrary color indices
\cite{npb96, physlett01}.  Substituting eqs. (\ref{subtr4}) and (\ref{subtr5}) in Eq. (\ref{subtr3}) we obtain
\begin{equation}
\hspace{-0mm} 
\langle K_{\rm LO} {\rm Tr}\{\hat{U}_x\hat{U}^{\dagger}_y\}\rangle~=~
{2\alpha_s^2\over\pi^4}\!\int_0^\sigma\! {d\alpha_2\over \alpha_2}\!\int\! d^2z d^2z'~
{(X,Y)(Y',z-z')\over X^2Y^2{Y'}^2(z-z')^2}
 {\rm Tr}\{(
 if^{a'b'c'}t^ct^bU_xt^{b'}U^\dagger_y
 -if^{abc}t^aU_x t^{b'}t^{c'} U^{\dagger}_y )
 U_z^{bb'}U_{z'}^{cc'}\}
 \label{subtr6}
\end{equation}
From Eq. (\ref{subtraction}) we get 
\begin{eqnarray}
&&\hspace{-4mm}
 \langle K_{\rm NLO}{\rm Tr}\{\hat{U}_x\hat{U}^\dagger_y\}\rangle
 ~=~
{\alpha_s^2\over 2\pi^4}
\!\int_0^\sigma\! {d\alpha_2\over \alpha_2}\!\int \!d^2 zd^2 z'~U_z^{bb'}U_{z'}^{cc'}
{(X,Y)(Y',z-z')\over (z-z')^2{Y'}^2}\Bigg[
{\sigma\over  \alpha_1X^2+\alpha_2{X'}^2}
 \label{subtr7}\\
&&\hspace{-4mm}
\times~\Big[f^{abc}{\rm Tr}\{t^aU_x\Big(
 {\alpha_1f^{a'b'c'}t^{a'}
\over \alpha_1Y^2+\alpha_2{Y'}^2}
- {i\over Y^2}\{t^{b'},t^{c'}\}\Big)U^\dagger_y\} + 
f^{a'b'c'} {\rm Tr}\{\Big(
 {\alpha_1f^{abc}t^a\over \alpha_1Y^2+\alpha_2{Y'}^2}
 +{i\over Y^2}\{t^b,t^c\}\Big)U_xt^{a'}U^\dagger_y\}   \Big]    
\nonumber\\ 
&&\hspace{57mm}      
+~{2i\sigma\over X^2(\alpha_1Y^2+\alpha_2{Y'}^2)}{\rm Tr}\{t^ct^bU_xf^{a'b'c'}t^{a'}U^\dagger_y  
-t^a f^{abc}U_xt^{b'}t^{c'}U^\dagger_y\} \Bigg]    
\nonumber\\ 
&&\hspace{-4mm}
-~{2\alpha_s^2\over\pi^4}\!\int_0^\sigma\! {d\alpha_2\over \alpha_2}\!\int\! d^2z~
{(X,Y)(Y',z-z')\over  X^2Y^2{Y'}^2(z-z')^2}
 {\rm Tr}\{
 if^{a'b'c'}t^ct^bU_xt^{b'}U^\dagger_y
 -if^{abc}t^aU_x t^{b'}t^{c'} U^{\dagger}_y
 \}U_z^{bb'}U_{z'}^{cc'}
\nonumber\\
&&\hspace{-4mm}
=~
{\alpha_s^2\over 2\pi^4}
\!\int_0^1{du\over \bar{u}}\!\int \!d^2 zd^2 z'~U_z^{bb'}U_{z'}^{cc'}
{(X,Y)(Y',z-z')\over (z-z')^2{Y'}^2}
\Bigg[f^{abc}f^{a'b'c'}{\rm Tr}\{t^aU_xt^{a'}U^\dagger_y\} 
 \Big[{2u\over  (uX^2+\bar{u}{X'}^2)(uY^2+\bar{u}{Y'}^2)}-{2\over X^2Y^2}\Big]
\nonumber\\ 
&&\hspace{-4mm}
-~{i\over Y^2}\Big[{1\over  uX^2+\bar{u}{X'}^2}-{1\over X^2}\Big]f^{abc}
{\rm Tr}\{t^aU_x\{t^{b'},t^{c'}\}U^\dagger_y\} 
+{i\over Y^2}\Big[{1\over  uX^2+\bar{u}{X'}^2}-
{1\over X^2}\Big]f^{a'b'c'}{\rm Tr}\{\{t^b,t^c\}U_xt^{a'}U^\dagger_y\}
\nonumber\\ 
&&\hspace{51mm}    
+~{2i\over X^2}\Big[{1\over uY^2+\bar{u}{Y'}^2}-{1\over Y^2}\Big]{\rm Tr}\{t^ct^bU_xf^{a'b'c'}t^{a'}U^\dagger_y  
-t^a f^{abc}U_xt^{b'}t^{c'}U^\dagger_y\}  
\Bigg]
\nonumber
\end{eqnarray}
which corresponds to the $\Big[{1\over\bar{u}}\Big]_+$ 
prescription (\ref{pluscription}) (the same prescription was used in Ref. \cite{balbel}). Note that the ``plus'' prescription (\ref{pluscription}) is a consequence of  the ``rigid'' cutoff
$|\alpha|<\sigma$ (\ref{cutoff}); with the ``smooth'' cutoff (\ref{defy}) we would get different results - see Appendix B.
 
\subsection{Assembling the result for 1$\rightarrow$3 dipoles transition}
There are four color structures in the r.h.s. of Eq. (\ref{2cutsum}). 
Three of them reduce to
\begin{eqnarray}
&&\hspace{-2mm}
f^{abc}f^{a'b'c'}U_z^{bb'}U_{z'}^{cc'}{\rm Tr}\{t^aU_xt^{a'}U^\dagger_y\}
~=~{1\over 4}{\rm Tr}\{U_xU^\dagger_z\}{\rm Tr}\{U_zU^\dagger_{z'}\}{\rm Tr}\{U_{z'}U^\dagger_y\}
-{1\over 4}{\rm Tr}\{U_xU^\dagger_z U_{z'}U^\dagger_yU_zU^\dagger_{z'}\}
+(z\leftrightarrow z')
\nonumber\\ 
&&\hspace{-2mm}
if^{abc}U_z^{bb'}U_{z'}^{cc'}{\rm Tr}(t^aU_x\{t^{b'},t^{c'}\}U^\dagger_y)
~=~-{1\over 4}{\rm Tr}\{U_xU^\dagger_z\}{\rm Tr}\{U_zU^\dagger_{z'}\}{\rm Tr}\{U_{z'}U^\dagger_y\}+{1\over 4}{\rm Tr}\{U_xU^\dagger_z U_{z'}U^\dagger_yU_zU^\dagger_{z'}\}-(z\leftrightarrow z')
\nonumber\\ 
&&\hspace{-2mm}
if^{a'b'c'}U_z^{bb'}U_{z'}^{cc'}{\rm Tr}(\{t^b,t^c\}U_xt^{a'}U^\dagger_y)
=~{1\over 4}{\rm Tr}\{U_xU^\dagger_z\}{\rm Tr}\{U_zU^\dagger_{z'}\}{\rm Tr}\{U_{z'}U^\dagger_y\}+{1\over 4}{\rm Tr}\{U_xU^\dagger_z U_{z'}U^\dagger_yU_zU^\dagger_{z'}\}-(z\leftrightarrow z')
\label{colors}
\end{eqnarray}
We will not need the explicit form of the  fourth color structure 
$U_z^{aa'}U_{z'}^{bb'}{\rm Tr}(\{t^a,t^b\}U_x\{t^{a'},t^{b'}\}U^\dagger_y)$
since it is multiplied by pure LO$^2$ integral 
$\int\!{du\over\bar{u}u}=\int\!{du\over\bar{u}}+\int\!{du\over u}$ 
and does not contribute to the NLO kernel.

Performing integration over $u$ using the prescription (\ref{pluscription})  after some algebra 
we get
\begin{eqnarray}
&&\hspace{-2mm} 
\int_0^1 du~u\bar{u}\Big[{1\over  uX^2+\bar{u}{X'}^2}
\Big({X_{ij}\over (z-z')^2}+{X_iX'_j\over uX^2}
-{X_iX'_j\over \bar{u}{X'}^2}\Big)
-{X_iY'_j\over \bar{u} u X^2{Y'}^2}-(x\leftrightarrow y)
\Big]^2
\nonumber\\ 
&&\hspace{-12mm}
=~ 
{1\over (z-z')^4}
\Big[-4+2{X^2{Y'}^2+{X'}^2Y^2-4\Delta^2(z-z')^2\over X^2{Y'}^2-{X'}^2Y^2}\ln{X^2{Y'}^2\over {X'}^2Y^2}
\Big]
\nonumber\\ 
&&\hspace{-12mm}
+~
\Big({(x-y)^4\over X^2{Y'}^2-{X'}^2Y^2}\Big[
{1\over X^2{Y'}^2}+{1\over Y^2{X'}^2}\Big]
+{(x-y)^2\over (z-z')^2}\Big[{1\over X^2{Y'}^2}-{1\over {X'}^2Y^2}\Big]\Big)
\ln{X^2{Y'}^2\over {X'}^2Y^2}
\label{integralsym}
\end{eqnarray}
and
\begin{eqnarray}
&&\hspace{-2mm} 
\int_0^1 du
\Big[{1\over  uX^2+\bar{u}{X'}^2}
\Big({X_{ij}\over (z-z')^2}+{X_iX'_j\over uX^2}
-{X_iX'_j\over \bar{u}{X'}^2}\Big)
-{X_iY'_j\over \bar{u} u X^2{Y'}^2}-(x\leftrightarrow y)\Big]
\Big({X_i\over X^2}-{Y_i\over Y^2}\Big)
\Big({X'_j\over {X'}^2}-{Y'_j\over {Y'}^2}\Big)
\nonumber\\
&&\hspace{-2mm} 
=~-{(x-y)^4\over 2X^2Y^2{X'}^2{Y'}^2}\ln {X^2{Y'}^2\over {X'}^2Y^2}
+{(x-y)^2\over 2(z-z')^2}\Big[{1\over X^2{Y'}^2}+{1\over{X'}^2Y^2}\Big] \ln {X^2{Y'}^2\over {X'}^2Y^2}
\label{integralantisym}
\end{eqnarray}
so the two-cut contribution (\ref{2cutsum}) reduces to
\begin{eqnarray}
&&\hspace{-2mm} 
{d\over d\eta}\langle {\rm Tr}\{\hat{U}_x\hat{U}^{\dagger}_y\}\rangle_{\rm Fig.\ref{2cutdms}}~=~
{\alpha_s^2\over 16\pi^4}
\int \!d^2 zd^2 z'
\Bigg[
\Big\{-{4\over (z-z')^4}+\Big(2{X^2{Y'}^2+{X'}^2Y^2-4(x-y)^2(z-z')^2
\over  (z-z')^4[X^2{Y'}^2-{X'}^2Y^2]}
\nonumber\\ 
&&\hspace{-2mm}
+~{(x-y)^4\over X^2{Y'}^2-{X'}^2Y^2}\Big[
{1\over X^2{Y'}^2}+{1\over Y^2{X'}^2}\Big]
+{(x-y)^2\over (z-z')^2}\Big[{1\over X^2{Y'}^2}-{1\over {X'}^2Y^2}\Big]\Big)
\ln{X^2{Y'}^2\over {X'}^2Y^2}\Big\}
\nonumber\\ 
&&\hspace{42mm}
\times~[{\rm Tr}\{U_xU^\dagger_z\}{\rm Tr}\{U_zU^\dagger_{z'}\}{\rm Tr}\{U_{z'}U^\dagger_y\}
-{\rm Tr}\{U_xU^\dagger_z U_{z'}U^\dagger_yU_zU^\dagger_{z'}\}]
\nonumber\\ 
&&\hspace{-2mm}
+~\Big\{
-{(x-y)^4\over  X^2{Y'}^2{X'}^2Y^2}
+{(x-y)^2\over (z-z')^2 }\Big({1\over X^2{Y'}^2}+{1\over Y^2{X'}^2}\Big)\Big\}\ln{X^2{Y'}^2\over {X'}^2Y^2}
{\rm Tr}\{U_x U^\dagger_z\}{\rm Tr}\{U_zU^\dagger_{z'}\}{\rm Tr}\{U_{z'}U^\dagger_y\}\Bigg]
\label{2cut}
\end{eqnarray}
This result agrees with the 1$\rightarrow$3 dipoles kernel calculated in Ref. \cite{balbel}.

\subsection{Subtraction of the UV part}
The integral in the r.h.s. of Eq. (\ref{2cut}) diverges as $z\rightarrow z'$. It is convenient to separate the divergent term by  subtracting and adding the contribution at $z=z'$:
\begin{eqnarray}
&&\hspace{-2mm}
{\rm Tr}\{U_xU^\dagger_z\}{\rm Tr}\{U_zU^\dagger_{z'}\}{\rm Tr}\{U_{z'}U^\dagger_y\}
-{\rm Tr}\{U_xU^\dagger_z U_{z'}U^\dagger_yU_zU^\dagger_{z'}\}]
\\ 
&&\hspace{-2mm}
=~[{\rm Tr}\{U_xU^\dagger_z\}{\rm Tr}\{U_zU^\dagger_{z'}\}{\rm Tr}\{U_{z'}U^\dagger_y\}
-{\rm Tr}\{U_xU^\dagger_z U_{z'}U^\dagger_yU_zU^\dagger_{z'}\} -(z'\rightarrow z)]
+[{\rm Tr}\{U_xU^\dagger_z\}{\rm Tr}\{U_zU^\dagger_y\}-{\rm Tr}\{U_xU^\dagger_y\}]
\nonumber
\end{eqnarray}
For the last line in the r.h.s. of Eq. (\ref{2cut}) the subtraction is redundant since 
\begin{eqnarray}
&&\hspace{-6mm}
\int\! d^2z'\Big\{
-{(x-y)^4\over  X^2{Y'}^2{X'}^2Y^2}
+{(x-y)^2\over (z-z')^2 }\Big({1\over X^2{Y'}^2}+{1\over Y^2{X'}^2}\Big)\Big\}\ln{X^2{Y'}^2\over {X'}^2Y^2}
~=~0
\label{pochemyzero1}
\end{eqnarray}
The easiest way to prove this is to set $y=0$ and make an inversion 
$x\rightarrow 1/\tilde{x}$ so the integral (\ref{pochemyzero1}) reduces to 
\begin{eqnarray}
&&\hspace{-6mm}
\int\! d^2\tilde{z}'
{(\tilde{x}-\tilde{z},\tilde{x}-\tilde{z}')\over (\tilde{x}-\tilde{z}')^2(\tilde{z}-\tilde{z}')^2 }
\ln{(\tilde{x}-\tilde{z})^2\over (\tilde{x}-\tilde{z}')^2}~=~0
\label{pochemyzero2}
\end{eqnarray}
Thus, we obtain 
\begin{eqnarray}
&&\hspace{-2mm} 
{d\over d\eta}\langle {\rm Tr}\{\hat{U}_x\hat{U}^{\dagger}_y\}\rangle_{\rm Fig.\ref{2cutdms}}
~=~
{\alpha_s^2\over 16\pi^4}
\int \!d^2 zd^2 z'
\Bigg[
\Big(-{4\over (z-z')^4}+\Big\{2{X^2{Y'}^2+{X'}^2Y^2-4(x-y)^2(z-z')^2\over  (z-z')^4[X^2{Y'}^2-{X'}^2Y^2]}\nonumber\\ 
&&\hspace{-2mm}
+~{(x-y)^4\over X^2{Y'}^2-{X'}^2Y^2}\Big[
{1\over X^2{Y'}^2}+{1\over Y^2{X'}^2}\Big]
+{(x-y)^2\over (z-z')^2}\Big[{1\over X^2{Y'}^2}-{1\over {X'}^2Y^2}\Big]\Big\}
\ln{X^2{Y'}^2\over {X'}^2Y^2}\Big)
\nonumber\\ 
&&\hspace{42mm}
\times~[{\rm Tr}\{U_xU^\dagger_z\}{\rm Tr}\{U_zU^\dagger_{z'}\}{\rm Tr}\{U_{z'}U^\dagger_y\}
-{\rm Tr}\{U_xU^\dagger_z U_{z'}U^\dagger_yU_zU^\dagger_{z'}\}-(z'\rightarrow z)]
\nonumber\\ 
&&\hspace{-2mm}
+~\Big\{
-{(x-y)^4\over  X^2{Y'}^2{X'}^2Y^2}
+{(x-y)^2\over (z-z')^2 }\Big({1\over X^2{Y'}^2}+{1\over Y^2{X'}^2}\Big)\Big\}\ln{X^2{Y'}^2\over {X'}^2Y^2}
{\rm Tr}\{U_x U^\dagger_z\}{\rm Tr}\{U_zU^\dagger_{z'}\}{\rm Tr}\{U_{z'}U^\dagger_y\}\Bigg]
\nonumber\\ 
&&\hspace{-2mm}
+~
{\alpha_s^2\over 16\pi^4}
\int \!d^2 z[{\rm Tr}\{U_xU^\dagger_z\}{\rm Tr}\{U_zU^\dagger_y\}-{\rm Tr}\{U_xU^\dagger_y\}]
\!\int\!d^2 z'
\Big[
-{4\over (z-z')^4}+\Big\{2{X^2{Y'}^2+{X'}^2Y^2-4(x-y)^2(z-z')^2\over  (z-z')^4[X^2{Y'}^2-{X'}^2Y^2]}\nonumber\\ 
&&\hspace{-2mm}
+~{(x-y)^4\over X^2{Y'}^2-{X'}^2Y^2}\Big[
{1\over X^2{Y'}^2}+{1\over Y^2{X'}^2}\Big]
+{(x-y)^2\over (z-z')^2}\Big[{1\over X^2{Y'}^2}-{1\over {X'}^2Y^2}\Big]\Big\}
\ln{X^2{Y'}^2\over {X'}^2Y^2}\Big]
\end{eqnarray}
The first term is now finite while the second term contains the UV divergent contribution which reflects the usual UV divergency of the one-loop diagrams. 
To find the second term we use the dimensional regularization in the transverse space and set $d_\perp=2-\epsilon$. Because the Fourier transforms (\ref{furie}) are more complicated at $d_\perp\neq 2$ it is convenient to return back to Eq. (\ref{twocutsum}) and calculate the subtracted term in the the momentum representation. 
    The calculation is performed in Appendix A and here we only quote the final result
     (\ref{appendixresult})
\begin{eqnarray}
&&\hspace{-2mm} 
{d\over d\eta}\langle{\rm Tr}\{\hat{U}_x\hat{U}^{\dagger}_y\}\rangle_{\rm Fig.\ref{2cutdms}}~=~
{\alpha_s^2\over 16\pi^4}
\int \!d^2 zd^2 z'
\Bigg[
\Big(-{4\over (z-z')^4}+\Big\{2{X^2{Y'}^2+{X'}^2Y^2-4(x-y)^2(z-z')^2\over  (z-z')^4[X^2{Y'}^2-{X'}^2Y^2]}\label{2cutfinal}\\ 
&&\hspace{-2mm}
+~{(x-y)^4\over X^2{Y'}^2-{X'}^2Y^2}\Big[
{1\over X^2{Y'}^2}+{1\over Y^2{X'}^2}\Big]
+{(x-y)^2\over (z-z')^2}\Big[{1\over X^2{Y'}^2}-{1\over {X'}^2Y^2}\Big]\Big\}
\ln{X^2{Y'}^2\over {X'}^2Y^2}\Big)
\nonumber\\ 
&&\hspace{42mm}
\times~[{\rm Tr}\{U_xU^\dagger_z\}{\rm Tr}\{U_zU^\dagger_{z'}\}{\rm Tr}\{U_{z'}U^\dagger_y\}
-{\rm Tr}\{U_xU^\dagger_z U_{z'}U^\dagger_yU_zU^\dagger_{z'}\}-(z'\rightarrow z)]
\nonumber\\ 
&&\hspace{-2mm}
+~\Big\{
-{(x-y)^4\over  X^2{Y'}^2{X'}^2Y^2}
+{(x-y)^2\over (z-z')^2 }\Big({1\over X^2{Y'}^2}+{1\over Y^2{X'}^2}\Big)\Big\}\ln{X^2{Y'}^2\over {X'}^2Y^2}
{\rm Tr}\{U_x U^\dagger_z\}{\rm Tr}\{U_zU^\dagger_{z'}\}{\rm Tr}\{U_{z'}U^\dagger_y\}\Bigg]
\nonumber\\ 
&&\hspace{-2mm}
-~
{\alpha_s^2N_c\over 8\pi^3}\!\int\! d^2z
{(x-y)^2\over X^2Y^2}\Big[{11\over 3}\ln{X^2Y^2\over (x-y)^2}\mu^2
+{67\over 9}-{\pi^2\over 3}\Big]
[{\rm Tr}\{U_xU^\dagger_z\}{\rm Tr}\{U_zU^\dagger_y\}
-{1\over N_c}{\rm Tr}\{U_xU^\dagger_y\}]
\nonumber
\end{eqnarray}
 where $\mu$ is the normalization scale in the $\overline{MS}$ scheme.

\section{Diagrams with one gluon-shockwave intersection}
\subsection{``Running coupling'' diagrams}
\begin{figure}

\vspace{-42mm}
\includegraphics[width=1.13\textwidth]{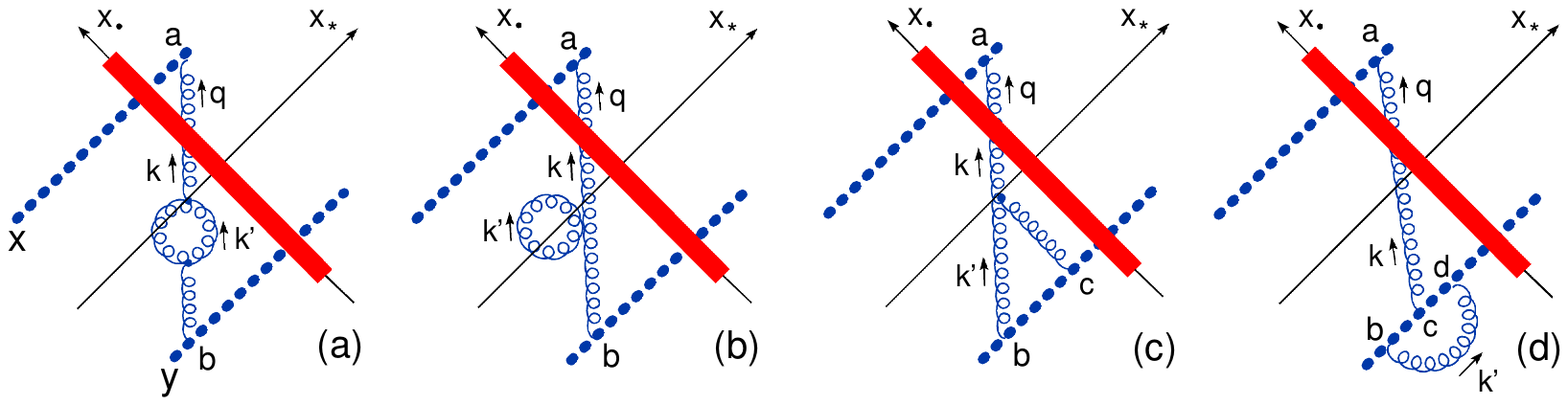}

\vspace{-220mm}
\includegraphics[width=1.13\textwidth]{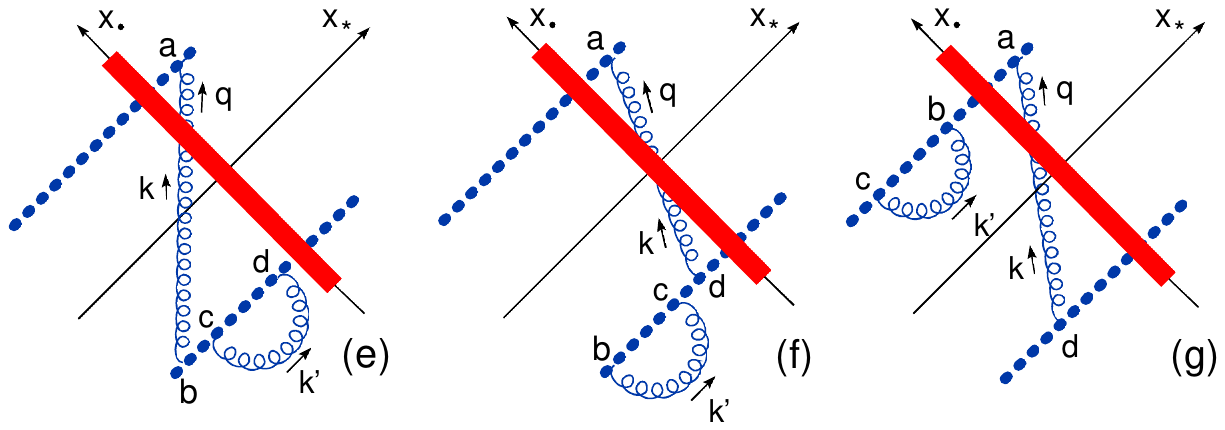}

\vspace{-170mm}
\caption{``Running coupling'' diagrams\label{run}.}
\end{figure}

The relevant diagrams are shown in Fig. \ref{run} (plus permutations).
Let us start from the sum of diagrams  Fig. \ref{run} a and b. It has the form:

\begin{eqnarray}
&&\hspace{-2mm}
\int_0^\infty\!\!
du\int^0_{-\infty}\!\!dv~\langle \hat{A}^a_{\bullet}(up_1+x_{\perp})
\hat{A}^b_{\bullet}(vp_1+y_{\perp})\rangle_{\rm Fig.\ref{run}a+b}~ \label{self1}\\
&&\hspace{-2mm}
=
~g^2N_c{s\over 2}\!\int\! 
\dhd^2k_\perp\dhd ^2k'_\perp{\dhd^2q_\perp\over q_\perp^2}\!\int\! d^2z~U_z^{ab}~
e^{i(q,x-z)_{\perp}-i(k,y-z)_{\perp}}
 \!\int_0^\infty\!\!{\dhd\alpha\over\alpha} \!\int\!\dhd\alpha'\!
\int\!{\dhd\beta\dhd\beta'\over\beta -i\epsilon}~
\nonumber\\
&&\hspace{-2mm}
\times~\Bigg[
{(q+{2(k,q)_\perp\over\alpha s}p_2)_{\lambda}
\over (k^2+i\epsilon)^2}
{d_{\mu\mu'}(k-k')\over (k-k')^2+i\epsilon}\Gamma^{\mu\nu\lambda}
((\alpha-\alpha')p_1+(k-k')_{\perp},\alpha'p_1+k'_{\perp},-\alpha p_1-k_\perp)
{d_{\mu\mu'}(k')\over {k'}^2+i\epsilon}
(k+2\beta p_2)_{\lambda'}
\nonumber\\
&&\hspace{-2mm}
\times~ \Gamma^{\mu'\nu'\lambda'}
((\alpha-\alpha')p_1+(k-k')_\perp,\alpha'p_1+k'_\perp, -\alpha p_1-k_\perp) ~-~2 {(k+2\beta p_2)_{\nu}(q_{\mu}+{2(k,q)_\perp\over\alpha s} p_{2\mu})
\over
(k^2+i\epsilon)^2}
{g^{\mu\nu}d_{\xi}^{\xi}(k')-d^{\mu\nu}(k')\over {k'}^2+i\epsilon}\Bigg]
\nonumber
\end{eqnarray}
where the first term in the square brackets comes from Fig. \ref{run}a and the second from Fig. \ref{run}b. 
We use the principal-value prescription for the ${1/\alpha'}$ terms in $d_{\mu\nu}(k')$ in loop integrals.

 To regularize the UV divergence we change the dimension of the transverse space to 2-$\varepsilon$.  After some algebra one obtains

\begin{eqnarray}
&&\hspace{-8mm}
\int_0^\infty\!\!
du\int^0_{-\infty}\!\!dv~\langle \hat{A}^a_{\bullet}(up_1+x_{\perp})
\hat{A}^b_{\bullet}(vp_1+y_{\perp})\rangle_{\rm Fig.\ref{run}a+b}
 \label{self2}\\
&&\hspace{-8mm}
=
~g^2N_c\mu^{2\varepsilon}\!\int\! \dhd^{2-\varepsilon} k\dhd^{2-\varepsilon}k'\dhd^{2-\varepsilon}q
\!\int\! d^{2-\varepsilon}z~U_z^{ab}
 \!\int_0^\infty\!\dhd\alpha 
\int\dhd\beta\dhd\beta'~
{e^{i(q,X)_{\perp}-i(k,Y)_{\perp}}
\over 
\alpha(\beta -i\epsilon)
(\alpha\beta s-k_{\perp}^2+i\epsilon)^2q^2}
\nonumber\\
&&\hspace{-8mm}
\times~
{s\over 2}\int\!\dhd\alpha'\dhd\beta'
{1\over(\alpha'\beta' s-{k'}_{\perp}^2+i\epsilon)
[(\alpha-\alpha')(\beta-\beta')s-(k-k')_{\perp}^2+i\epsilon]}
\nonumber\\
&&\hspace{-8mm}
\times~\Bigg\{-\varepsilon
[(\alpha-2\alpha')\beta s+{k'}_{\perp}^2-(k-k')_{\perp}^2]
[{\alpha-2\alpha'\over\alpha}(k,q)_{\perp}+(2k'-k,q)_{\perp}]+
2{(\alpha-2\alpha')^2\over\alpha}(k,q)_{\perp}\beta s
\nonumber\\
&&\hspace{-8mm}
+~
{(\alpha-2\alpha')\over\alpha}(k,q)_{\perp}(2k'-k,k)_{\perp}+
(\alpha-2\alpha')\beta s(2k'-k,q)_{\perp}+(q,k)_{\perp}(k-2k',k-k')_{\perp}
+(q,k')_{\perp}(k,2k'-k)_{\perp}
\nonumber\\
&&\hspace{-8mm}
+~{\alpha+\alpha'\over\alpha-\alpha'}
\Bigg[{\alpha-2\alpha'\over\alpha}(q,k)_{\perp}(k,k-k')_{\perp}+
(q,2k'-k)_{\perp}(k,k-k')_{\perp}+
(q,k)_{\perp}({k'}^2_{\perp}-k^2_{\perp})+(q,k-k')_{\perp}(k,2k-k')_{\perp}
\nonumber\\
&&\hspace{-8mm}
+~(\alpha-2\alpha')(q,k-k')_{\perp}\beta s+
(q,k-k')_{\perp}({k'}^2_{\perp}-(k-k')^2_{\perp})-
(q,k')_{\perp}(k',k-k')_{\perp}+(q,k-k')_{\perp}(k-k')^2_{\perp}\Bigg]
\nonumber\\
&&\hspace{-8mm}
+~{\alpha'-2\alpha\over\alpha'}
\Bigg[{\alpha-2\alpha'\over\alpha}(q,k)_{\perp}(k,k')_{\perp}+
(q,2k'-k)_{\perp}(k,k')_{\perp}-(q,k')_{\perp}(k,k+k')_{\perp}
+(q,k)_{\perp}(k',2k-k')_{\perp}
\nonumber\\
&&\hspace{-8mm}
+~(\alpha-2\alpha')(q,k')_{\perp}\beta s+
(q,k')_{\perp}(k,2k'-k)_{\perp}-(q,k')_{\perp}{k'}^2_{\perp}+
(q,k-k')_{\perp}(k',k-k')_{\perp}\Bigg]
\nonumber\\
&&\hspace{-8mm}
+~{\alpha+\alpha'\over\alpha-\alpha'}~{\alpha'-2\alpha\over\alpha'}
\Big[(k,k')_{\perp}(q,k-k')_{\perp}+(q,k')_{\perp}(k,k-k')_{\perp}\Big]
+[(k-k')_{\perp}^2+{k'}^2](q,k)_{\perp}+
\nonumber\\
&&\hspace{-8mm}
+~4\alpha(q,k)_{\perp}\Big[{\alpha-\alpha'\over\alpha'}+
{\alpha\over\alpha-\alpha'}\Big]
\beta s- 
4\alpha(q,k)_{\perp}
\Big[{{k'}_{\perp}^2\over\alpha-\alpha'}+{(k-k')_{\perp}^2\over\alpha'}\Big] 
\Bigg\}
\nonumber
\end{eqnarray}
where we have omitted the contribution 
\begin{equation}
\int\!\dhd\alpha'\dhd\beta'\dhd^2k'
{1\over\alpha'(\alpha'\beta' s-{k'}_{\perp}^2+i\epsilon)}
~=~\int d^4k'{1\over {k'}^2+i\epsilon}~V.p.{1\over (k',p_2)}~=~0 
\end{equation}

Taking residues at $\beta=0$ and $\beta'={{k'}_{\perp}^2\over \alpha's}$ 
and changing to variable $u={\alpha'\over\alpha}$ we obtain

\begin{eqnarray}
&&\hspace{-2mm}
\int_0^\infty\!\!
du\int^0_{-\infty}\!\!dv~\langle \hat{A}^a_{\bullet}(up_1+x_{\perp})
\hat{A}^b_{\bullet}(vp_1+y_{\perp})\rangle_{\rm Fig.\ref{run} a+b}
\label{self3}\\
&&\hspace{-2mm}=~
~-{g^2N_c\over 8\pi^2}\mu^{2\varepsilon}\!\int\! \dhd^{2-\varepsilon} k\dhd^{2-\varepsilon}k'
\dhd^{2-\varepsilon}q\!\int\! d^{2-\varepsilon}z~U_z^{ab}
 \!\int_0^\sigma\! {d\alpha \over\alpha}\!\int_0^1 du
~{e^{i(q,x-z)_{\perp}-i(k,y-z)_{\perp}}
\over 
 k^4q^2
[{k'}^2\bar{u}+(k-k')^2u]}
\nonumber\\
&&\hspace{-2mm}
\times~\Bigg\{
-2\varepsilon(2k'-k,k)(q,k'-ku)+(1-2u)(k,q)(2k'-k,k)+(q,k)(k-2k',k-k')+(q,k')(k,2k'-k)
\nonumber\\
&&\hspace{-2mm}
+~{1+u\over \bar{u}}\Big[(1-2u)(q,k)(k,k-k')+2(q,k){k'}^2-2(k,k')_\perp(q,k')
\Big]
\nonumber\\
&&\hspace{-2mm}
-~{2-u\over u}
\Big[(1-2u)(q,k)(k,k')+2(q,k)(k-k',k')-2(q,k')(k-k',k)\Big]
\nonumber\\
&&\hspace{-2mm}
-~{(1+u)(2-u)\over u\bar{u}}[(k,k')(q,k-k')+(q,k')(k,k-k')]+(q,k)
\Big[(k-k')^2+{k'}^2-4{{k'}^2\over\bar{u}}-4{(k-k')^2\over u}\Big]\Bigg\}
\nonumber
\end{eqnarray}
Using the ``plus'' - prescription (\ref{pluscription}) to subtract the (LO)$^2$ contribution we get
\begin{eqnarray}
&&\hspace{-2mm}
{d\over d\eta}\langle
{\rm Tr}\{\hat{U}_x \hat{U}^\dagger_y\}\rangle_{\rm Fig.~\ref{run} a+b}
\label{vkladotself}\\
&&\hspace{-2mm}
=~-
2\alpha_s^2N_c\mu^{2\varepsilon}\!\int\! d^{2-\varepsilon}z
[{\rm Tr}\{U_x U^\dagger_z\}{\rm Tr}\{U_z U^\dagger_y\}
-{1\over N_c}{\rm Tr}\{U_x U^\dagger_y\}]\!
\int\!\dhd^{2-\varepsilon}k\dhd^{2-\varepsilon}k'\dhd^{2-\varepsilon}q
~{e^{i(q,X)-i(k,Y)}\over k^2q^2}
\nonumber\\
&&\hspace{-2mm}
\times~
\Bigg[
\Big({(q,2k-k')\over {k'}^2}-{(q,k+k')\over(k-k')^2}\Big)\ln{(k-k')^2\over{k'}^2}
+\!\int_0^1\! du~{(2-\varepsilon)(q,ku-k')(k,k-2k')+2(q,k)k^2
\over k^2[{k'}^2\bar{u}+(k-k')^2u]}\Bigg]
\nonumber
\end{eqnarray}

Next we calculate diagram shown in Fig. \ref{run}c.

\begin{eqnarray}
&&\hspace{-2mm}
g^3\!\int_0^\infty\!\!dt\!\int^0_{-\infty}\!\!du\!\int^0_u\!\!dv~ \langle\hat{A}^a_{\bullet}(tp_1+x)\hat{A}^b_{\bullet}(up_1+y)
\hat{A}^c_{\bullet}(vp_1+y)\rangle_{\rm Fig.\ref{run}c}~
\label{vert1}\\
&&\hspace{-2mm}
=~-2\,g^4\,f^{lbc}\!
\int \dhd \alpha\, \dhd\beta\,\dhd \alpha' \,\dhd \beta'\,\dhd \beta''
\,\dhd^{2-\varepsilon}q\,\dhd^{2-\varepsilon}k\,\dhd^{2-\varepsilon}k'\,e^{(iq,x-z)_\perp-i(k,y-z)_\perp}
\frac{\theta(\alpha)U^{al}_z}{q^2_\perp(\alpha\beta s-k^2+i\epsilon)}
\frac{(k'_\perp+2\beta' p_2)_\mu}{\alpha'\beta' s-k'^2+i\epsilon}
\nonumber\\
&&\hspace{-2mm}
\times~
\frac{\big(q_\perp+{2\over \alpha s}(q,k)_\perp p_2\big)_\lambda}{\alpha'(\alpha-\alpha')}
\frac{((k-k')_\perp+2\beta'' p_2)_\nu}{(\alpha-\alpha')\beta'' s-(k-k')^2_\perp+i\epsilon}
\frac{\Gamma^{\mu\nu\lambda}\big(\alpha' p_1+k'_\perp,(\alpha-\alpha')p_1+(k-k')_\perp,-\alpha p_1-k_\perp\big)}
{(\beta'-i\epsilon)(\beta''+\beta'-i\epsilon)(\beta-\beta'-\beta''-i\epsilon)}\nonumber\\
\nonumber
\end{eqnarray}
There are 2 regions of integration over $\alpha$'s: $\alpha>|\alpha'|$ and $\alpha<|\alpha'|$. 
Taking relevant residues, we obtain

\begin{eqnarray}
&&\hspace{-2mm}
-{g^4\over 2\pi^2}f^{abl}\mu^{2\varepsilon}\!\int_0^\infty{d\alpha\over\alpha}
\!\int\dhd^{2-\varepsilon}
k\dhd^{2-\varepsilon}k'\dhd^{2-\varepsilon}q \!\int\!d^2z_\perp ~U_z^{cl}
~{e^{i(q,X)-i(k,Y)}\over {k'}^2q^2k^2}
\int_0^1 du~
\Big\{{1\over {k'}^2\bar{u}+(k-k')^2u}
\Big[
{(q,k')\over \bar{u}}[{k'}^2+k^2]
\nonumber\\
&&\hspace{-2mm}
+~{{k'}^2\over u}(q,2k-k')-2(q,k)(k',k-k')\Big]
+{1\over (k-k')^2\bar{u}+
k^2u}\Big[{1\over u}[(q,k){k'}^2+(q,k')k^2]-(k',2k-k')(q,k)
\Big]\Big\}
\label{vert2}
\end{eqnarray}
where we have introduced the variable $u=|\alpha'|/\alpha$ as usual. After integration
over $u$ with help of Eq. (\ref{pluscription}) this reduces to

\begin{eqnarray}
&&\hspace{-2mm}
-{g^4\over 2\pi^2}f^{abl}\mu^{2\varepsilon}\!\int_0^\infty{d\alpha\over\alpha}
\int\dhd^{2-\varepsilon}
k\dhd^{2-\varepsilon}k'\dhd^{2-\varepsilon}q \!\int\!d^{2-\varepsilon}z_\perp ~U_z^{cl}
\Big\{
\Big[{(q,k')\over (k-k')^2}[{k'}^2+k^2]+(q,k'-2k)\Big]\ln{(k-k')^2\over{k'}^2}
\nonumber\\
&&\hspace{-2mm}
+~\Big[{1\over (k-k')^2}[(q,k){k'}^2+(q,k')k^2]+(q,k)\Big]\ln{(k-k')^2\over k^2}
-\int_0^1\! du~{2(q,k)(k',k-k')\over {k'}^2\bar{u}+(k-k')^2u}
\Big\}~{e^{i(q,X)-i(k,Y)}\over {k'}^2q^2k^2}
\label{vert3}
\end{eqnarray}
and therefore

\begin{eqnarray}
&&\hspace{-22mm}
{d\over d\eta}\langle{\rm Tr}\{\hat{U}_x\hat{U}^{\dagger}_y\}
\rangle_{\rm Fig.~\ref{run}c}
=~-{g^4N_c\over 8\pi^2}\mu^{2\varepsilon}\!\int\! d^{2-\varepsilon}z[{\rm Tr}\{U_xU^{\dagger}_z\}{\rm Tr}\{U_zU^{\dagger}_y\}
-{1\over N_c}{\rm Tr}\{U_xU^{\dagger}_y\}]\!    
\nonumber\\
&&\hspace{-22mm}
\times\int\!\dhd^{2-\varepsilon}k\dhd^{2-\varepsilon}k'\dhd^{2-\varepsilon}q  
~{e^{i(q,X)-i(k,Y)}\over k^2q^2}
~\Big\{\Big[{(q,k')\over (k-k')^2}+{(q,k')k^2\over  {k'}^2(k-k')^2}+
{(q,k'-2k)\over  {k'}^2}\Big]\ln{(k-k')^2\over {k'}^2}\nonumber\\
&&\hspace{-22mm}-~
\Big[{(q,k)\over {k'}^2}+{(q,k)\over (k-k')^2}+{(q,k-k')k^2\over  {k'}^2(k-k')^2}
\Big]\ln{k^2\over {k'}^2}
-2\!\int_0^1\!du~{(q,k)(k',k-k')\over {k'}^2[{k'}^2\bar{u}+(k-k')^2u]}\Big\}
\label{vert4}
\end{eqnarray}

Next we calculate the sum of diagrams in Fig. \ref{run} d,e, and f. 
The contribution of the diagram shown in Fig. \ref{run}d is
\begin{eqnarray}
&&\hspace{-1mm}
\langle{\rm Tr}\{\hat{U}_x\hat{U}^\dagger_y\}\rangle_{\rm Fig.~\ref{run} d}
\label{vert5}\\
&&\hspace{-1mm}
=
~\!\int\! \dhd\alpha\dhd\alpha'\dhd\beta\dhd\beta'\!\int\! \dhd^{2-\varepsilon}k 
\dhd^{2-\varepsilon} k'\dhd^{2-\varepsilon}q
\!\int\! d^{2-\varepsilon}z~U_z^{ab}~
{4g^4\mu^{2\varepsilon}\theta(\alpha)(q,k)_\perp e^{i(q,x-z)_\perp-i(k,y-z)_\perp}
{\rm Tr}\{t^aU_x t^ct^bt^cU^\dagger_y\}  
\over \alpha\alpha'(\beta+\beta'-i\epsilon)(\beta-i\epsilon)
(\alpha'\beta's-{k'}_{\perp}^2+i\epsilon)
(\alpha\beta s-k_{\perp}^2+i\epsilon)
q_{\perp}^2}            
\nonumber\\
&&\hspace{-1mm}
=                                      
~-{g^4\over\pi^2}\mu^{2\varepsilon}\!\int_0^\sigma\!{d\alpha\over\alpha}
\!\int_0^1\! du\!\int\!\dhd^{2-\varepsilon}q
\dhd^{2-\varepsilon}k \dhd^{2-\varepsilon} k'\int\! d^{2-\varepsilon}z~U^{ab}_z
~{(q,k)_\perp e^{i(q,x-z)_{\perp}-i(k,y-z)_{\perp}}
\over uk^2(uk^2+\bar{u}{k'}^2) q^2}                     
{\rm Tr}\{t^aU_x t^ct^bt^cU^\dagger_y\}
\nonumber
\end{eqnarray}
where we took residues at $\beta={k_\perp^2\over\alpha s}$,  $\beta'={{k'}_\perp^2\over\alpha' s}$ 
and introduced the variable $u={\alpha\over\alpha+\alpha'}$.
It should be noted that the cutoff $\alpha<\sigma$ in the r.h.s. of this equation translates into
$\int_0^\infty\!d\alpha d\alpha'~\theta(\sigma-\alpha-\alpha')$ while our cutoff (\ref{cutoff}) corresponds
to $\int_0^\infty\!d\alpha d\alpha'~\theta(\sigma-\alpha)\theta(\sigma-\alpha')$. Fortunately, the difference 
\begin{equation}
\int_0^\infty\!{d\alpha d\alpha'\over\alpha'(\alpha{k'}^2+\alpha' k^2)}
~[\theta(\sigma-\alpha)\theta(\sigma-\alpha')-\theta(\sigma-\alpha-\alpha')]
~=~{1\over {k'}^2}\!\int_0^1\! {dv\over v}\ln{{k'}^2+k^2v\over {k'}^2\bar{v}+k^2v}
\end{equation}
does not contain $\ln\sigma$ and hence does not
contribute to the NLO kernel. Similarly, one can impose the cutoff $\alpha_1+\alpha_2<\sigma$ instead of the cutoff $\alpha_1,\alpha_2<\sigma$ in other diagrams whenever convenient.

Before calculating the diagrams in Fig. \ref{run}e and Fig. \ref{run}f
it is convenient to make the replacement
\begin{equation}
\int^0_{-\infty}\! \! du\!\int_u^0\! \! dv\int_v^0\! \! dt~ \hat{A}^a(u)\hat{A}^b(v)\hat{A}^c(t)
~\rightarrow~{1\over 2}\int^0_{-\infty}\! \! du\!\int_u^0\!\!  dv dt ~\hat{A}^a(u)\hat{A}^b(v)\hat{A}^c(t)
\end{equation}
which can be performed since the color indices $b$ and $c$ in $...t^bt^c...$ are contracted. 
For the diagram in Fig. \ref{run}e we get
\begin{eqnarray}
&&\hspace{-1mm}
\langle{\rm Tr}\{\hat{U}_x\hat{U}^\dagger_y\}\rangle_{\rm Fig. \ref{run}e}~=
~2g^4\mu^{2\varepsilon}c_F{\rm Tr}\{t^aU_x t^bU^\dagger_y\}
\label{vert6}\\
&&\hspace{-1mm}
\times~
\!\int\! \dhd\alpha\dhd\alpha'\dhd\beta\dhd\beta'\!\int\! \dhd^{2-\varepsilon}k 
\dhd^{2-\varepsilon} k'
\dhd^{2-\varepsilon}q\!\int\! d^{2-\varepsilon}z~ U_z^{ab}
{e^{i(q,X)_{\perp}-i(k,Y)_{\perp}}
\over\alpha'(\beta-i\epsilon)^2 
({k'}^2+i\epsilon)}
\Big({\beta'\over \beta+\beta'-i\epsilon}+{\beta'\over \beta-\beta'-i\epsilon}\Big)
 {\theta(\alpha)(k,q)_{\perp}\over \alpha
(k^2+i\epsilon)q_{\perp}^2}
\nonumber\\
&&\hspace{-1mm}
=~
4g^4\mu^{2\varepsilon}c_F{\rm Tr}\{t^aU_x t^bU^\dagger_y\}
\!\int\! \dhd\alpha\dhd\alpha'\dhd\beta\dhd\beta'\!\int\! \dhd^{2-\varepsilon}k 
\dhd^{2-\varepsilon} k'
\dhd^{2-\varepsilon}q\!\int\! d^{2-\varepsilon}z~ U_z^{ab}
e^{i(q,X)_{\perp}-i(k,Y)_{\perp}}
\nonumber\\
&&\hspace{15mm} 
\times~
{\beta'\over  \alpha\alpha'(\beta-i\epsilon) 
(\alpha'\beta's-{k'}_{\perp}^2+i\epsilon)( \beta+\beta'-i\epsilon)(\beta-\beta'-i\epsilon)}
 {\theta(\alpha)(k,q)_{\perp}\over
(\alpha\beta s-k_{\perp}^2+i\epsilon)
q_{\perp}^2}
\nonumber
\end{eqnarray}
where $c_F={N_c^2-1\over 2N_c}$. 
Taking residues at $\beta'=-\beta$, $\beta={k_\perp^2\over\alpha s}$ at $\alpha'>0$ and 
$\beta'=\beta$, $\beta={k_\perp^2\over\alpha s}$ at $\alpha'<0$ we obtain
\begin{eqnarray}
&&\hspace{-0mm}
\langle{\rm Tr}\{U_xU^\dagger_y\}\rangle_{\rm Fig.~\ref{run}e}
\nonumber\\
&&\hspace{0mm} 
=~{g^4\over\pi^2}c_F\mu^{2\varepsilon}\!\int_0^\sigma \!{d\alpha\over\alpha}\!\int_0^1\! du\!\int\!
\dhd^{2-\varepsilon}k \dhd^{2-\varepsilon} k'
~{e^{i(q,X)_{\perp}-i(k,Y)_{\perp}}\over uk^2(uk^2+\bar{u}{k'}^2)}
\int \dhd^{2-\varepsilon}q\!\int\! d^{2-\varepsilon}z
{(k,q)_{\perp}\over q_{\perp}^2}
U_z^{db}{\rm Tr}\{t^dU_x t^bU^\dagger_y\}
\label{vert7}
\end{eqnarray}
The diagram in Fig. \ref{run}f yields
\begin{eqnarray}
&&\hspace{-15mm}
\langle{\rm Tr}\{\hat{U}_x\hat{U}^\dagger_y\}\rangle_{\rm Fig.~\ref{run}f}
\label{vert8}\\
&&\hspace{-15mm}
=
~g^4\mu^{2\varepsilon}\!\int\! \dhd\alpha\dhd\beta\dhd^{2-\varepsilon}k
\int\dhd\alpha'\dhd\beta'\dhd^{2-\varepsilon}k'
{1\over\alpha'(\beta-i\epsilon)(\beta'-i\epsilon)
(\alpha'\beta' s-{k'}_{\perp}^2+i\epsilon)}
\nonumber\\
&&\hspace{-15mm}\times~
 2\theta(\alpha)\!\int \!\dhd^{2-\varepsilon}q\!\int \!d^{2-\varepsilon}z
e^{i(q,X)_{\perp}-i(k,Y)_\perp}{(k,q)_{\perp}\over \alpha
(\alpha\beta s-k_{\perp}^2+i\epsilon)
q_{\perp}^2}U_z^{ab}{\rm Tr}\{t^aU_xt^ct^ct^bU^\dagger_y\}
\nonumber\\
&&\hspace{-15mm}
=-{g^4\over 2\pi^2}\mu^{2\varepsilon}\!\int_0^\sigma\!{d\alpha\over\alpha}\!\int_0^1\!{du\over\bar{u}u}
\!\int\!\dhd^{2-\varepsilon}\dhd^{2-\varepsilon}k' 
\! \int \dhd^{2-\varepsilon}q\!\int \!d^{2-\varepsilon}z~
e^{i(q,x-z)-i(k,y-z)}{(k,q)_{\perp}\over k^2{k'}^2q_{\perp}^2}
U_z^{ab}{\rm Tr}\{t^aU_xt^ct^ct^bU^\dagger_y\}
\nonumber
\end{eqnarray}
Adding Eqs. (\ref{vert5}),  (\ref{vert7}), (\ref{vert8}) and integrating over $u$ using Eq. (\ref{pluscription}) we get
\begin{eqnarray}
&&\hspace{-1mm}
{d\over d\eta}\langle{\rm Tr}\{\hat{U}_x\hat{U}^\dagger_y\}
\rangle_{\rm Fig.~\ref{run}d+e+f}
\label{vert9}\\
&&\hspace{-1mm}
=                                      
~-{g^4N_c\over 4\pi^2}\mu^{2\varepsilon}\!\int\!\dhd^{2-\varepsilon}q
\dhd^{2-\varepsilon}k \dhd^{2-\varepsilon} k'\int\! d^{2-\varepsilon}z
~{(q,k)\over  k^2{k'}^2q^2}\ln{k^2\over{k'}^2}~e^{i(q,X)-i(k,Y)}
[{\rm Tr}\{U_xU^\dagger_z\}{\rm Tr}\{U_zU^\dagger_y\}-{1\over N_c}{\rm Tr}\{U_xU^\dagger_y\}]
\nonumber
\end{eqnarray}
Note that the diagram in Fig.~\ref{run}f does not contribute to the NLO kernel.

The contribution of the last ``running coupling'' diagram shown in Fig. \ref{run}g has the form
\begin{eqnarray}
&&\hspace{-1mm}
\langle{\rm Tr}\{\hat{U}_x\hat{U}^\dagger_y\}\rangle_{\rm Fig.~\ref{run}g}
\label{XXVIIa}\\
&&\hspace{-1mm}
=~-{g^4\over 2}{\rm Tr}\{t^aU_xt^bt^ct^dU^\dagger_y\}\!\int_0^\infty\!\! du\!\int^0_{-\infty}\!\! dv
\langle \hat{A}^a_\bullet (up_1+x_\perp)\hat{A}^d_\bullet (vp_1+y_\perp)\rangle \int_{-\infty}^0\!\! dt dw
\langle\hat{A}^b(tp_1+x_\perp)\hat{A}^c(wp_1+x_\perp)\rangle
\nonumber
\end{eqnarray}
where we have again replaced $\int_{-\infty}^0\! dt \int_{-\infty}^t\! dw~
\hat{A}^b(t)\hat{A}^c(w)$ by ${1\over 2}\!\int_{-\infty}^0\! dtdw~ \hat{A}^b(t)\hat{A}^c(w)$.  
Using the Eq. (\ref{bk1}) we get
\begin{eqnarray}
&&\hspace{-1mm}
\langle{\rm Tr}\{\hat{U}_x\hat{U}^\dagger_y\}\rangle_{\rm Fig.~\ref{run}g}
\label{XXVIIb}\\
&&\hspace{-1mm}
=~{ig^4\over \pi}{\rm Tr}\{t^aU_xt^bt^ct^dU^\dagger_y\}(x|{p_i\over p_\perp^2}U^{ad}{p_i\over p_\perp^2}|y)
\int\!\dhd\alpha'\dhd\beta'{\beta'\over\alpha'({\beta'}^2+\epsilon^2)}
(x|{1\over\alpha'\beta' s-p_\perp^2+i\epsilon}|x)
 \nonumber\\
&&\hspace{-3mm}  
=~{g^4\over 2\pi^2}
{\rm Tr}\{t^aU_xt^bt^ct^dU^\dagger_y\}
(x|{p_i\over p_\perp^2}U^{ad}{p_i\over p_\perp^2}|y)(x|{1\over p_\perp^2}|x)
\!\int_0^\sigma\!{d\alpha\over\alpha}{d\alpha'\over\alpha'}
 \nonumber
\end{eqnarray}
which is obviously a (LO)$^2$ term which does not contribute to the NLO kernel.

\begin{figure}

\vspace{-30mm}
\includegraphics[width=1.1\textwidth]{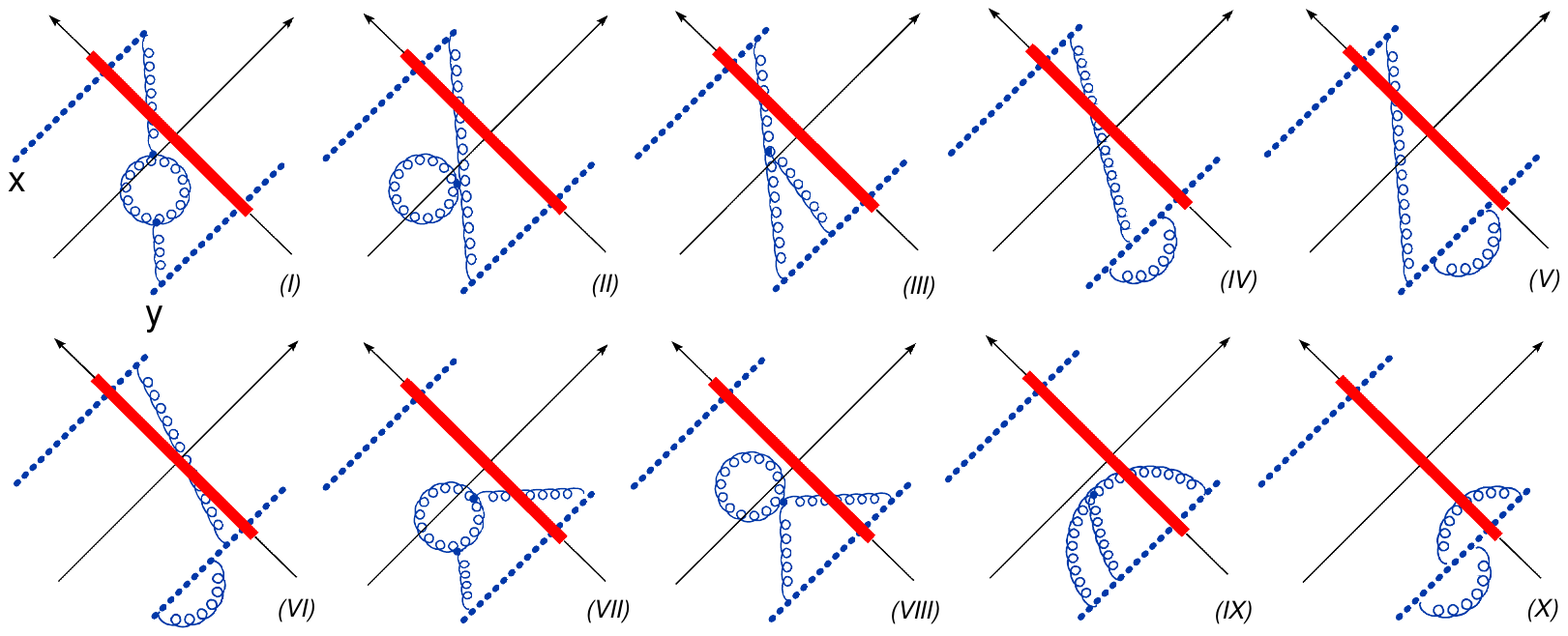}

\vspace{-193mm}
\includegraphics[width=1.1\textwidth]{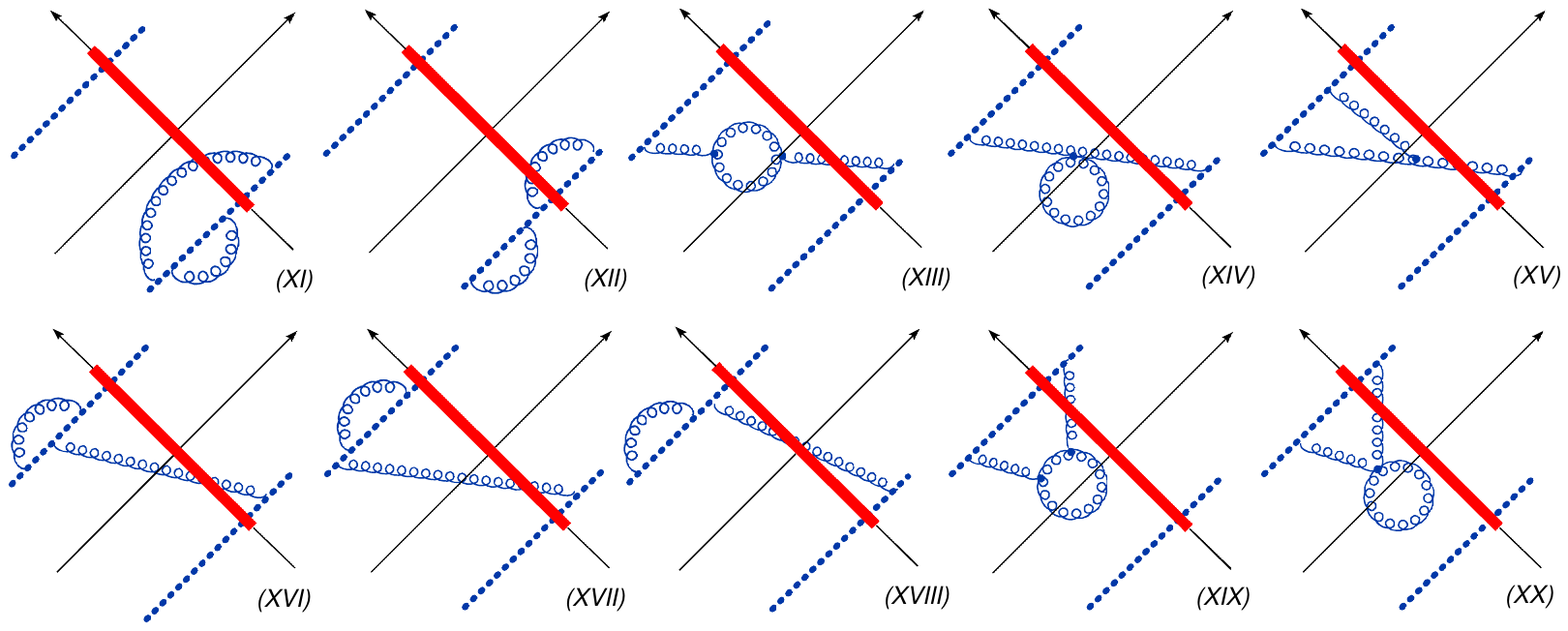}

\vspace{-192mm}
\includegraphics[width=1.1\textwidth]{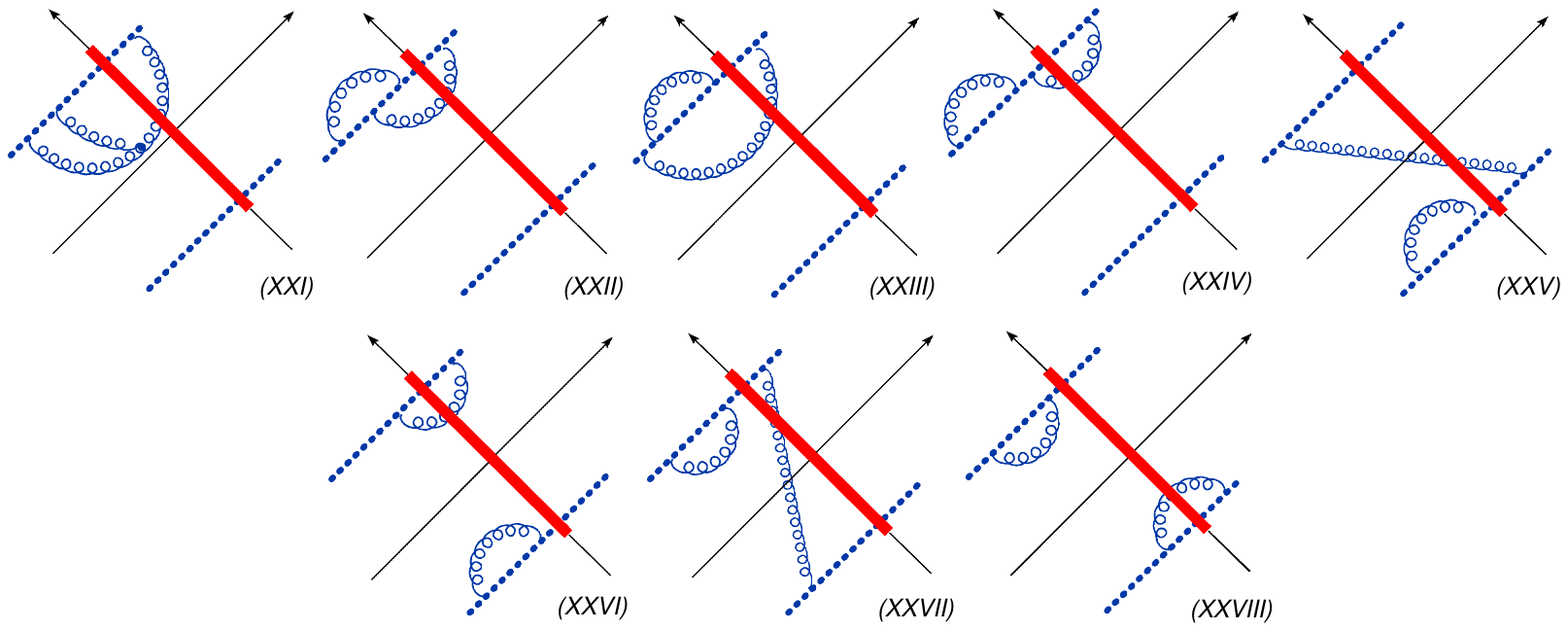}

\vspace{-160mm}
\caption{The full set of ``running coupling'' diagrams. \label{1cutdms}}
\end{figure}

Combining expressions (\ref{vkladotself}), (\ref{vert4}), and (\ref{vert9}) we get
\begin{eqnarray}
&&\hspace{-2mm}
{d\over d\eta}\langle{\rm Tr}\{\hat{U}_x \hat{U}^\dagger_y\}
\rangle_{\rm Fig.~\ref{run}}
\label{vklad1}\\
&&\hspace{-2mm}
=~-
2\alpha_s^2N_c\mu^{2\epsilon}\!\int\! d^{2-\varepsilon}z
[{\rm Tr}\{U_x U^\dagger_z\}{\rm Tr}\{U_z U^\dagger_y\}
-{1\over N_c}{\rm Tr}\{U_x U^\dagger_y\}]\!
\int\!\dhd^{2-\varepsilon}k\dhd^{2-\varepsilon}k'\dhd^{2-\varepsilon}q
~{e^{i(q,X)-i(k,Y)}\over k^2q^2}
\nonumber\\
&&\hspace{-2mm}
\times~
\Bigg\{
\Big[{k^2\over {k'}^2}(q,2k-k')-{k^2\over(k-k')^2}(q,k+k')\Big]\ln{(k-k')^2\over{k'}^2}
+
\!\int_0^1\! du~{(2-\varepsilon)(q,ku-k')(k,k-2k')+2(q,k)k^2
\over {k'}^2\bar{u}+(k-k')^2u}
\nonumber\\
&&\hspace{-2mm}
+~\Big[{(q,k')\over (k-k')^2}+{(q,k')k^2\over  {k'}^2(k-k')^2}+
{(q,k'-2k)\over  {k'}^2}\Big]\ln{(k-k')^2\over {k'}^2}\nonumber\\
&&\hspace{-2mm}-~
\Big[{(q,k)\over {k'}^2}+{(q,k)\over (k-k')^2}+{(q,k-k')k^2\over  {k'}^2(k-k')^2}
\Big]\ln{k^2\over {k'}^2}
-2\!\int_0^1\!du~{(q,k)(k',k-k')\over {k'}^2[{k'}^2\bar{u}+(k-k')^2u]}
+{2(q,k)\over  {k'}^2}\ln{k^2\over{k'}^2}\Bigg\}
 \nonumber\\
&&\hspace{-2mm}
=~-2\alpha_s^2N_c\mu^{2\varepsilon}\!\int\! d^{2-\varepsilon}z
[{\rm Tr}\{U_x U^\dagger_z\}{\rm Tr}\{U_z U^\dagger_y\}
-{1\over N_c}{\rm Tr}\{U_x U^\dagger_y\}]\!
\int\!\dhd^{2-\varepsilon}k\dhd^{2-\varepsilon}k'\dhd^{2-\varepsilon}q
{e^{i(q,X)-i(k,Y)}\over k^2q^2}
\Bigg\{\ln{k^2\over {k'}^2}
\nonumber\\
&&\hspace{-2mm}
\times~{3(q,k')k^2-4(q,k)(k,k')\over  {k'}^2(k-k')^2}
+\!\int_0^1\! du~{(2-\varepsilon)(q,ku-k')(k,k-2k')+2(q,k)k^2\over k^2[{k'}^2\bar{u}+(k-k')^2u]}~
-\int_0^1\!du~{2(q,k)(k',k-k')\over {k'}^2[{k'}^2\bar{u}+(k-k')^2u]}
\Bigg\} 
\nonumber
\end{eqnarray}
Using the  integral over $k'$ 

\begin{eqnarray}
&&\hspace{-12mm}
\int\!\dhd^{2-\varepsilon} k'_\perp
~\Bigg\{{3(q,k')k^2-4(q,k)(k,k')\over  {k'}^2(k-k')^2}\ln{k^2\over {k'}^2}
+\!\int_0^1\! du~{(2-\varepsilon)(q,ku-k')(k,k-2k')+2(q,k)k^2\over k^2[{k'}^2\bar{u}+(k-k')^2u]}~
\nonumber\\
&&\hspace{-12mm}
-\int_0^1\!du~{2(q,k)(k',k-k')\over {k'}^2[{k'}^2\bar{u}+(k-k')^2u]}\Bigg\}~=~{(q,k)\over 4\pi}\Bigg\{{\Gamma(\varepsilon/ 2)\over (k^2)^{\varepsilon/ 2}}
{\Gamma^2(1-{\varepsilon\over 2})\over\Gamma(2-\varepsilon)}
\Big[{11\over 3}-\varepsilon{\pi^2\over 6}\Big]+{1\over 9}\Bigg\}
 \nonumber
\end{eqnarray}
one reduces the r.h.s. of Eq. (\ref{vklad1}) to
\begin{eqnarray}
&&\hspace{-2mm}
{d\over d\eta}\langle{\rm Tr}\{\hat{U}_x \hat{U}^\dagger_y\}
\rangle_{\rm Fig.~\ref{run}}
~=~-
{\alpha_s^2N_c\over 2\pi}\mu^{2\varepsilon}\!\int\! d^{2-\varepsilon}z
[{\rm Tr}\{U_x U^\dagger_z\}{\rm Tr}\{U_z U^\dagger_y\}
-{1\over N_c}{\rm Tr}\{U_x U^\dagger_y\}]\!
 \nonumber\\
&&\hspace{-2mm}
\times\int\!\dhd^{2-\varepsilon}k\dhd^{2-\varepsilon}q
~e^{i(q,X)-i(k,Y)}~{(q,k)\over k^2q^2}
\Big\{{\Gamma(\varepsilon/ 2)\over (k^2)^{\varepsilon/ 2}}
{\Gamma^2(1-{\varepsilon\over 2})\over\Gamma(2-\varepsilon)}
\Big[{11\over 3}-\varepsilon{\pi^2\over 6}\Big]+{1\over 9}\Big\}
\label{vklad1otve}
\end{eqnarray}
Next we subtract the counterterm 
\begin{eqnarray}
&&\hspace{-12mm}
-{22\over 3}{\alpha^2_s N_c\over\pi \varepsilon}
\mu^\varepsilon\!\int\! d^{2-\varepsilon}z
[{\rm Tr}\{U_x U^\dagger_z\}{\rm Tr}\{U_z U^\dagger_y\}
-{1\over N_c}{\rm Tr}\{U_x U^\dagger_y\}]\!
\!\int\! \dhd^{2-\varepsilon}q~\dhd^{2-\varepsilon}k~e^{i(q,X)-i(k,Y)}
{(q,k)\over q^2k^2}
\label{ct}
\end{eqnarray}
corresponding to the poles $1/\varepsilon$ in the loop diagrams 
in Fig. \ref{run} (we use the $\overline{MS}$ scheme).  We obtain
\begin{eqnarray}
&&\hspace{-2mm}
{d\over d\eta}\langle{\rm Tr}\{\hat{U}_x \hat{U}^\dagger_y\}
\rangle_{\rm Fig.~\ref{run}}
~=~-
{\alpha_s^2N_c\over 2\pi}\!\int\! d^{2-\varepsilon}z
[{\rm Tr}\{U_x U^\dagger_z\}{\rm Tr}\{U_z U^\dagger_y\}
-{1\over N_c}{\rm Tr}\{U_x U^\dagger_y\}]\!
 \nonumber\\
&&\hspace{-2mm}
\times\int\!\dhd^{2-\varepsilon}k\dhd^{2-\varepsilon}k'\dhd^{2-\varepsilon}q
~e^{i(q,X)-i(k,Y)}~{(q,k)\over k^2q^2}
\Big\{{11\over 3}\ln{\mu^2\over k^2}+{67\over 9}-{\pi^2\over 3}\Big\}     
\label{vklad1otvet}
\end{eqnarray}
The complete set of running-coupling diagrams is presented in Fig. \ref{1cutdms}.

%
The contribution of diagrams in Fig. \ref{1cutdms} VII-XII differs from Eq. (\ref{vklad1otvet}) by the replacement 
$e^{i(q,X)}\rightarrow e^{i(q,Y)}$ and sign change. 
There is also a symmetric set of diagrams XIII-XXIV obtained by the reflection of  diagrams I-XII with respect to $x_\ast$ axis. The result is obtained by 
the  substitution $e^{-i(k,X)}\leftrightarrow e^{-i(k,Y)}$ and therefore the contribution 
of all diagrams in Fig. (\ref{1cutdms}) takes the form
\begin{eqnarray}
&&\hspace{-2mm}
{d\over d\eta}\langle{\rm Tr}\{\hat{U}_x \hat{U}^\dagger_y\}
\rangle_{\rm Fig.~\ref{1cutdms}~I+...+XXIV}
~=~
{\alpha_s^2N_c\over 2\pi}\!\int\! d^2z
[{\rm Tr}\{U_x U^\dagger_z\}{\rm Tr}\{U_z U^\dagger_y\}
-{1\over N_c}{\rm Tr}\{U_x U^\dagger_y\}]\!
 \nonumber\\
&&\hspace{-2mm}                            
\times\int\!\dhd^2k\dhd^2k'\dhd^2q
~[e^{i(q,X)}-e^{i(q,Y)}][e^{-i(k,X)}-e^{-i(k,Y)}]{(q,k)\over k^2q^2}
\Big\{{11\over 3}\ln{\mu^2\over k^2}+{67\over 9}-{\pi^2\over 3}\Big\}               
\label{vklad1fina}
\end{eqnarray}
The  remaining diagrams XXV-XXVIII contribute only to the (LO)$^2$. We have shown this for diagram XXVII (Fig. \ref{run}g), see Eq. (\ref{XXVIIb}). 
The diagram in Fig. \ref{1cutdms} XXV is obtained 
from the equation  (\ref{XXVIIb}) by the replacement $x\leftrightarrow y$, and the diagrams XXVI and XXVIII by the replacements 
$(x|{p_i\over p_\perp^2}U{p_i\over p_\perp^2}|y)(x|{1\over p_\perp^2}|x)$ by $(x|{p_i\over p_\perp^2}U{p_i\over p_\perp^2}|x)(y|{1\over p_\perp^2}|y)$ and
 $(y|{p_i\over p_\perp^2}U{p_i\over p_\perp^2}|y)(x|{1\over p_\perp^2}|x)$, respectively.
 Thus, the diagrams XXV-XXVII do not contribute to the NLO kernel.

There is another set of diagrams obtained by the reflection of diagrams shown in Fig. (\ref{1cutdms}) 
with respect to the shock-wave line. Their contribution is obtained from Eq. (\ref{vklad1fina}) by the replacement
 $q\leftrightarrow k$ in the logarithm so the final result for the sum of all ``running coupling'' 
 diagrams of Fig.~\ref{1cutdms} type has the form
\begin{eqnarray}
&&\hspace{-4mm}
{d\over d\eta}\langle {\rm Tr}\{\hat{U}_x\hat{U}^\dagger_y\}\rangle_{\rm Fig.~\ref{1cutdms}~total}
~=~
{\alpha_s^2N_c\over 2\pi}\!\int\! d^2z
[{\rm Tr}\{U_x U^\dagger_z\}{\rm Tr}\{U_z U^\dagger_y\}
-{1\over N_c}{\rm Tr}\{U_x U^\dagger_y\}]\!
\label{vklad1final}\\
&&\hspace{44mm}                            
\times\int\!\dhd^2k\dhd^2k'\dhd^2q
~[e^{i(q,X)}-e^{i(q,Y)}][e^{-i(k,X)}-e^{i(k,Y)}]{(q,k)\over k^2q^2}
\Big\{{11\over 3}\ln{\mu^4\over q^2k^2}+{134\over 9}+{2\pi^2\over 3}\Big\}    
 \nonumber\\
&&\hspace{-4mm}         
 =~
{\alpha_s^2N_c\over 8\pi^3}\!\int\! d^2z
[{\rm Tr}\{U_x U^\dagger_z\}{\rm Tr}\{U_z U^\dagger_y\}
-{1\over N_c}{\rm Tr}\{U_x U^\dagger_y\}]\!
~\Big\{{(x-y)^2\over X^2Y^2} \Big[{11\over 3}\ln {X^2Y^2\over \mu^{-4}}+{134\over 9}-{2\pi^2\over 3}\Big]
+{11\over 3}\Big[{1\over X^2}-{1\over Y^2}\Big]\ln{X^2\over Y^2}\Big\}      
 \nonumber               
\end{eqnarray}
%

\subsection{Diagrams for 1$\rightarrow$2 dipoles transition}

\begin{figure}

\vspace{-25mm}
\includegraphics[width=1.15\textwidth]{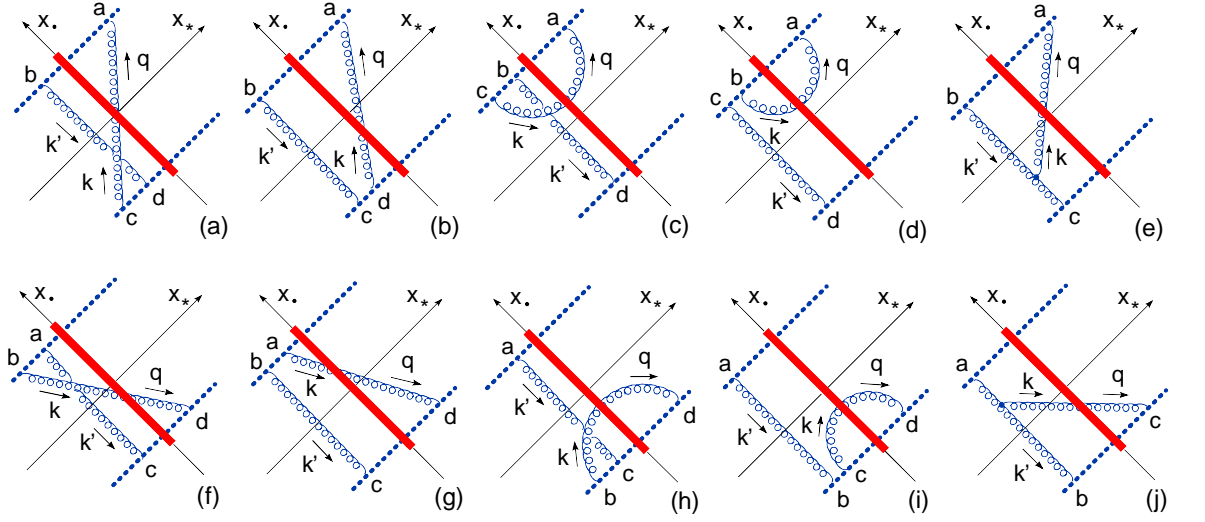}

\vspace{-161mm}
\caption{1$\rightarrow$2 dipoles transition diagrams. \label{1cutadd}}
\end{figure}

There is one more class of diagrams with one gluon-shockwave intersection shown in Fig. \ref{1cutadd}.
These diagrams are UV-convergent so we do not need to change  the dimension of the transverse space to $2-\varepsilon$. First we calculate diagrams shown in Fig. \ref{1cutadd}a,b. 

\begin{eqnarray}
&&\hspace{-1mm}
\langle {\rm Tr}\{\hat{U}_x\hat{U}^\dagger_y\}\rangle_{\rm Fig.~\ref{1cutadd}a+b}
\label{add1}\\
&&\hspace{-1mm}
=~4{\rm Tr}\{t^aU_xt^bt^ct^dU^\dagger_y\}\!\int\!d^{2-\varepsilon}z
\!\int\!\dhd\alpha_1\dhd\beta_1 \dhd\alpha_2\dhd\beta_2\!\int\!\dhd^{2-\varepsilon}k_1
\dhd^{2-\varepsilon} k_2~ e^{-i(k_1,y)_\perp-i(k_2,y)_\perp}
\Big[{\delta^{bd}U^{ac}_z\over (\beta_1-i\epsilon)(\beta_1+\beta_2-i\epsilon)}
\nonumber\\
&&\hspace{-1mm}
+~{\delta^{bc}U^{ad}_z\over (\beta_2-i\epsilon)(\beta_1+\beta_2-i\epsilon)}\Big]
\theta(\alpha_1)\int \dhd^2q
e^{i(q,x-z)_\perp+i(k_1,z)_\perp}{(q_1,k_1)_\perp\over 
\alpha_1(\alpha_1\beta_1 s-k_{1\perp}^2+i\epsilon)
q_\perp^2}
{e^{i(k_2,x)_\perp}\over\alpha_2(\alpha_2\beta_2 s-k_{2\perp}^2+i\epsilon)}
\nonumber\\
&&\hspace{-1mm}
=~-{g^4\over \pi^2}{\rm Tr}\{t^aU_xt^bt^ct^dU^\dagger_y\}\!\int\!d^{2-\varepsilon}z
\!\int_0^\sigma\!{d\alpha\over\alpha}\!\int_0^1\! du\!\int\!\dhd^{2-\varepsilon}k
\dhd^{2-\varepsilon} k'~ e^{-i(k,y)_\perp-i(k',y)_\perp}
\nonumber\\
&&\hspace{41mm}
\times~\Big[{\delta^{bd}U^{ac}_z\over k^2\bar{u}(k^2\bar{u}+{k'}^2u)}+{\delta^{ab}U^{dc}_z\over {k'}^2u(k^2\bar{u}+{k'}^2u) }\Big]
\!\int \dhd^2q
e^{i(q,x-z)_\perp+i(k,z)_\perp+i(k',x)_\perp}{(q,k)_\perp\over q_\perp^2}
\nonumber\\
&&\hspace{-1mm}
=~-{g^4\over \pi^2}{\rm Tr}\{t^aU_xt^bt^ct^dU^\dagger_y\}\!\int\!d^2z
\!\int_0^\infty\!\! {d\alpha\over\alpha}\!\int\!\dhd^2k\dhd^2 k' 
\dhd^2q
~{(q,k)\over q^2}~ e^{i(q,X)-i(k,Y)_\perp+i(k',x-y)_\perp}
{\delta^{bc}U^{ad}_z-\delta^{bd}U^{ac}_z\over k^2{k'}^2}\ln{k^2\over{k'}^2}
\nonumber
\end{eqnarray}
and therefore
\begin{eqnarray}
&&\hspace{-3mm}
{d\over d\eta}\langle {\rm Tr}\{\hat{U}_x\hat{U}^\dagger_y\}\rangle_{\rm Fig.~\ref{1cutadd}a+b}
\label{add2}\\
&&\hspace{-3mm}
=~-{g^4N_c\over 4\pi^2}\!\int\!d^2z
[{\rm Tr}\{U_xU^\dagger_z\}{\rm Tr}\{U_zU^\dagger_y\}-{1\over N_c}{\rm Tr}\{U_xU^\dagger_y\}]
\!\int\!\dhd^2k\dhd^2 k' \dhd^2q
~{(q,k)\over q^2k^2{k'}^2}~ e^{i(q,x-z)-i(k,y-z)_\perp+i(k',x-y)_\perp}
\ln{k^2\over{k'}^2}
\nonumber
\end{eqnarray}
The contribution of diagrams shown in Fig. \ref{1cutadd}c,d is obtained from Eq. (\ref{add2}) 
by the replacement $x\leftrightarrow y$ in the left part of the graph and the sign change so 
that $e^{-ik(y-z)+i(k',x-y)}\rightarrow -e^{-ik(x-z)-i(k',x-y)}$. The sum of the diagrams in
Fig. \ref{1cutadd}a-d takes the form
\begin{eqnarray}
&&\hspace{-3mm}
{d\over d\eta}\langle {\rm Tr}\{\hat{U}_x\hat{U}^\dagger_y\}\rangle_{\rm Fig.~\ref{1cutadd}a-d}
~=~-{g^4N_c\over 4\pi^2}\!\int\!d^2z
[{\rm Tr}\{U_xU^\dagger_z\}{\rm Tr}\{U_zU^\dagger_y\}
-{1\over N_c}{\rm Tr}\{U_xU^\dagger_y\}]
\label{sumad}\\
&&\hspace{33mm}
\times~
\!\int\!\dhd^2k\dhd^2k' \dhd^2q
~{(q,k)\over q^2k^2{k'}^2}~ e^{i(q,x-z)}\Big(e^{-i(k,y-z)_\perp-i(k',x-y)_\perp}-x\leftrightarrow y\Big)
\ln{k^2\over{k'}^2}
\nonumber
\end{eqnarray}
%
Next relevant diagram is shown in Fig. \ref{1cutadd}e
\begin{eqnarray}
&&\hspace{-1mm}g^3
\!\int_0^{\infty}\!du\!\int^0_{-\infty}\!dv\!\int^0_{-\infty}\!dt~
\langle \hat{A}^a_{\bullet}(up_1+x)\hat{A}^b_{\bullet}(vp_1+x)\hat{A}^c_{\bullet}(tp_1+y)\rangle_{\rm Fig. \ref{1cutadd}e}
\label{flast1}\\
&&\hspace{-1mm}=~
2g^4\!\int\! \dhd\alpha_1 \dhd\beta_1 \dhd\alpha_2 \dhd\beta_2
\dhd^2 k^\perp_1\dhd^2 k^\perp_2~
 {\theta(\alpha_1+\alpha_2)f^{bcd}
(k^{\perp}_{1\mu}+2\beta_1 p_{2\mu})(k^{\perp}_{2\nu}+2\beta_2 p_{2\nu}) 
\over\alpha_1\alpha_2
(\beta_1-i\epsilon)(\alpha_1\beta_1 s-k_{1\perp}^2+i\epsilon)
(\beta_2-i\epsilon)(\alpha_2\beta_2 s-k_{2\perp}^2+i\epsilon)}
\nonumber\\
&&\hspace{-1mm}
\times~
\int\!d^2z~U_z^{ad}\!\int \dhd^2 q~
{e^{i(q-k_1,x-z)_\perp-ik_2(y-z)_\perp}
\Big[q+{2(k_1+k_2,q)_{\perp}\over(\alpha_1+\alpha_2) s}
p_2\Big]_\lambda
\over  
q^2[(\alpha_1+\alpha_2)(\beta_1+\beta_2)s-(k_1+k_2)^2_{\perp}+i\epsilon]} 
\Gamma^{\mu\nu\lambda}(k_1,k_2,-k_1-k_2)~
\nonumber\\
\nonumber
\end{eqnarray}
There 
are three regions of integration over $\alpha$'s: $\alpha_1,\alpha_2>0$, $\alpha_1>-\alpha_2>0$ and
$\alpha_2>-\alpha_1>0$. Going to the variables 
$\alpha=\alpha_1+\alpha_2,~u=\alpha_2/\alpha$ in the first region, 
$\alpha=\alpha_1,~u=-\alpha_2/\alpha$ in the second  and 
$\alpha=\alpha_2,~u=-\alpha_1/\alpha$ in the third, we obtain

\begin{eqnarray}
&&\hspace{-1mm}
g^3\!\int_0^{\infty}\!du\!\int^0_{-\infty}\!dv\!\int^0_{-\infty}\!dt~
\langle \hat{A}^a_{\bullet}(up_1+x)\hat{A}^b_{\bullet}(vp_1+x)\hat{A}^c_{\bullet}(tp_1+y)\rangle_{\rm Fig. \ref{1cutadd}e}
\label{flast2}\\
&&\hspace{-1mm}
=~
{1\over 2\pi^2}\!\int_0^\infty\! {d\alpha\over\alpha}
\!\int_0^1\!du\!\int\! \dhd^2 k^\perp_1\dhd^2 k^\perp_2~f^{bcd}
U_z^{ad}\int\! {\dhd^2q\over q^2}
~e^{i(q-k_1,x-z)_\perp-i(k_2,y-z)_\perp}
\nonumber\\
&&\hspace{-1mm}
\times~ \Bigg[k^{\perp}_{1\mu}k^{\perp}_{2\nu}\Big(q_{\lambda}+{2(k_1+k_2,q)_{\perp}\over\alpha s}
p_{2\lambda}\Big)
{\Gamma^{\mu\nu\lambda}(\alpha \bar{u} p_1+k^\perp_1,\alpha u p_1+k^\perp_2,
-\alpha p_1-k^\perp_1-k^\perp_2)\over  
\bar{u} u k_{1\perp}^2k_{2\perp}^2(k_1+k_2)^2_\perp} 
\nonumber\\
&&\hspace{-1mm}-~
\bar{u} k^\perp_{1\mu}\Big(k^\perp_{2\nu}+2{(k_1+k_2)_\perp^2\over \alpha\bar{u} s}
p_{2\nu}\Big)
\Big(q_{\lambda}+{2(k_1+k_2,q)_{\perp}\over \alpha\bar{u} s}
p_{2\lambda}\Big){\Gamma^{\mu\nu\lambda}(\alpha p_1+k^\perp_1,-u\alpha p_1+k^\perp_2,
-\alpha\bar{u}-k^\perp_1-k^\perp_2)
\over uk_{1\perp}^2(k_1+k_2)_\perp^2[u(k_1+k_2)_\perp^2+\bar{u} k_{2\perp}^2]}
\nonumber\\
&&\hspace{-1mm}-~
\bar{u} \Big(k^\perp_1+2{(k_1+k_2)_\perp^2\over \alpha\bar{u} s}p_2\Big)_\mu 
k^\perp_{2\nu}\Big(q+{2(k_1+k_2,q)_{\perp}\over \alpha\bar{u} s}p_2\Big)_{\lambda}
{\Gamma^{\mu\nu\lambda}(-u\alpha p_1+k^\perp_1,\alpha p_1+k^\perp_2,
-\alpha\bar{u}-k^\perp_1-k^\perp_2)
\over uk_{2\perp}^2(k_1+k_2)_\perp^2[u(k_1+k_2)_\perp^2+\bar{u} k_{1\perp}^2]}
\Bigg]
\nonumber
\end{eqnarray}
Using the formula
\begin{eqnarray}
&&\hspace{-1mm}
\Big(k_1^{\perp}+{2A\over s}p_2\Big)_{\mu}
\Big(k_2^{\perp}+{2B\over s}p_2\Big)_{\nu}
 \Big(q+{2C\over s}p_2\Big)_{\lambda}\Gamma^{\mu\nu\lambda}
(\alpha_1p_1+k_{1\perp},\alpha_2p_1+k_{2\perp},-(\alpha_1+\alpha_2)p_1
-(k_1+k_2)_{\perp})~=
\nonumber\\
&&\hspace{-1mm}
-C(\alpha_1-\alpha_2)(k_1,k_2)_{\perp}-
A(\alpha_1+2\alpha_2)(q,k_2)_{\perp}+B(2\alpha_1+\alpha_2)(q,k_1)_{\perp}
-[(q,k_1)_{\perp}(k_2,k_1+k_2)_{\perp}-(k_1\leftrightarrow k_2)]
\nonumber
\end{eqnarray}
we get
\begin{eqnarray}
&&\hspace{-1mm}
\langle{\rm Tr}\{\hat{U}_x\hat{U}^\dagger_y\}\rangle_{\rm Fig.~\ref{1cutadd}c}
~=~{g^4N_c\over 8\pi^2}
\!\int\! \dhd^2 k_1\dhd^2k_2{\dhd^2 q\over q^2}\!\int\! d^2z
\int_0^\sigma\!{d\alpha\over\alpha}\int_0^1 du~
e^{i(q-k_1,x-z)-i(k_2,y-z)}
\\
&&\hspace{-1mm}\times~[{\rm Tr}\{U_xU^\dagger_z\}{\rm Tr}\{U_zU^\dagger_y\}
-{1\over N_c}{\rm Tr}\{U_xU^\dagger_y\}]
\Bigg\{{(q,k_1)(k_2,k_1+k_2)-(q,k_2)(k_1,k_1+k_2)\over \bar{u}u k_1^2k_2^2(k_1+k_2)^2}
 \nonumber\\
 &&\hspace{-1mm}
+~
{(1+\bar{u})(k_1+k_2)^2(q,k_1)-(1+u)(q,k_1+k_2)(k_1,k_2)-
\bar{u}[(q,k_1)(k_2,k_1+k_2)-(q,k_2)(k_1,k_1+k_2)]
\over uk_1^2(k_1+k_2)^2(k_2^2\bar{u}+(k_1+k_2)^2u)}
\nonumber\\
&&\hspace{-1mm}+~
{-(1+\bar{u})(k_1+k_2)^2(q,k_2)+(1+u)(q,k_1+k_2)(k_1,k_2)-
\bar{u}[(q,k_1)(k_2,k_1+k_2)-(q,k_2)(k_1,k_1+k_2)]
\over uk_2^2(k_1+k_2)^2(k_1^2\bar{u}+(k_1+k_2)^2u)}\Bigg\}
\nonumber
\end{eqnarray}
Performing the integration over u (with prescription (\ref{pluscription})) we obtain 
\begin{eqnarray}
&&\hspace{-1mm}
{d\over d\eta}\langle{\rm Tr}\{\hat{U}_x\hat{U}^\dagger_y\}\rangle_{\rm Fig.~\ref{1cutadd}c}
\label{fig1cutaddc}\\
&&\hspace{-1mm}
=~{g^4N_c\over 8\pi^2}\!\int\! d^2z~
[{\rm Tr}\{U_xU^\dagger_z\}{\rm Tr}\{U_zU^\dagger_y\}
-{1\over N_c}{\rm Tr}\{U_xU^\dagger_y\}]
\!\int\! {\dhd^2k_1\dhd^2k_2\over k_1^2k_2^2(k_1+k_2)^2}
{\dhd^2 q\over q^2}
~e^{i(q-k_1,X)-i(k_2,Y)}
U_z^{cl}
\nonumber\\
&&\hspace{-1mm}\times~
\Big\{(k_1+k_2)^2
\Big[(q,k_2)\ln{(k_1+k_2)^2\over k_1^2}-(q,k_1)\ln{(k_1+k_2)^2\over k_2^2}
\Big]-(q,k_1+k_2)(k^2_1+k_2^2)\ln{k_1^2\over k_2^2}\Big\}
\nonumber\\
&&\hspace{-1mm}
=~{g^4N_c\over 8\pi^2}\!\int\! d^2z~
[{\rm Tr}\{U_xU^\dagger_z\}{\rm Tr}\{U_zU^\dagger_y\}
-{\rm Tr}\{U_xU^\dagger_y\}]
\!\int\! \dhd^2k\dhd^2k'
{\dhd^2 q\over q^2}
~e^{i(q-k',x-z)-i(k-k',y-z)}
U_z^{cl}
\nonumber\\
&&\hspace{-1mm}\times~
\Big\{ {(q,k-k')\over {k'}^2(k-k')^2}\ln{k^2\over {k'}^2}-{(q,k')\over {k'}^2(k-k')^2}\ln{k^2\over (k-k')^2}
+{(q,k)\over k^2{k'}^2}\ln{(k-k')^2\over {k'}^2}+{(q,k)\over k^2(k-k')^2}\ln{(k-k')^2\over {k'}^2}\Big\}
\nonumber
\end{eqnarray}
where we made the change of variables $k_1\rightarrow k'$ and $k_2\rightarrow k-k'$.

The sum of diagrams shown in Fig. \ref{1cutadd}a-e can be represented as
\begin{eqnarray}
&&\hspace{-3mm}
{d\over d\eta}\langle{\rm Tr}\{\hat{U}_x\hat{U}^\dagger_y\}\rangle_{\rm Fig.~\ref{1cutadd}a-e}
=~2\alpha_s^2N_c\!\int\!d^2z
[{\rm Tr}\{U_xU^\dagger_z\}{\rm Tr}\{U_zU^\dagger_y\}
-{1\over N_c}{\rm Tr}\{U_xU^\dagger_y\}]
\!\int\!\dhd^2k\dhd^2k' \dhd^2q
\nonumber\\
&&\hspace{-1mm}
\times~e^{i(q,x-z)} \Big(e^{-i(k,y-z)_\perp-i(k',x-y)_\perp}-x\leftrightarrow y\Big)
\Big[{(q,k-k')\over {k'}^2(k-k')^2}+{(q,k)\over k^2(k-k')^2}-{(q,k)\over k^2{k'}^2}\Big]\ln{k^2\over{k'}^2}
\label{sum1cutadd}
\end{eqnarray}
Note that the expressions (\ref{sumad}) and (\ref{fig1cutaddc}) are IR divergent  as  ${k'}\rightarrow 0$ but their sum (\ref{sum1cutadd}) is IR stable.
Once again, the contribution of the diagrams in Fig. \ref{1cutadd}f-k are obtained by 
replacement $e^{iq(x-z)}\rightarrow -e^{iq(y-z)}$ so the contribution 
of diagrams of Fig. \ref{1cutadd} a-k has the form
\begin{eqnarray}
&&\hspace{-3mm}
{d\over d\eta}\langle{\rm Tr}\{\hat{U}_x\hat{U}^\dagger_y\}\rangle_{\rm Fig.~\ref{1cutadd}a-k}
=~2\alpha_s^2N_c\!\int\!d^2z
[{\rm Tr}\{U_xU^\dagger_z\}{\rm Tr}\{U_zU^\dagger_y\}
-{1\over N_c}{\rm Tr}\{U_xU^\dagger_y\}]
\!\int\!\dhd^2k\dhd^2k' \dhd^2q
\nonumber\\
&&\hspace{-1mm}
\times~(e^{i(q,x-z)} -e^{i(q,y-z)})\Big(e^{-i(k,y-z)_\perp-i(k',x-y)_\perp}-x\leftrightarrow y\Big)
\Big[{(q,k-k')\over {k'}^2(k-k')^2}+{(q,k)\over k^2(k-k')^2}-{(q,k)\over k^2{k'}^2}\Big]\ln{k^2\over{k'}^2}
\label{sum1cutadda}
\end{eqnarray}

Performing the Fourier transformation with the help of the formula

\begin{eqnarray}
&&\hspace{-12mm}
\int\! \dhd k\dhd k'~{e^{-i(k,y)-i(k',x-y)}\over {k'}^2}
\Big({(k-k')_i\over (k-k')^2}-{k_i\over k^2}
+{k_i{k'}^2\over k^2(k-k')^2}\Big)\ln{k^2\over{k'}^2}
\label{fypie}\\
&&\hspace{-12mm}=~
{i\over 16\pi^2}\Big({x_i\over x^2}-{y_i\over y^2}\Big)
\ln{(x-y)^2\over x^2}\ln{(x-y)^2\over y^2}
\nonumber\\
&&\hspace{-12mm}
+~{i\over 8\pi^2}\Big({(x,y)\over y^2}y_i-x_i\Big){1\over i\kappa}
\Bigg\{\int_0^1\! du~
\Bigg[{\ln u\over u-{(x,y)-i\kappa\over x^2}}-{\ln u\over u-{(x,y)+i\kappa\over x^2}}\Bigg]
+{1\over 2}\ln{x^2\over y^2}\ln{(x-y,y)+i\kappa\over  (x-y,y)-i\kappa}\Bigg\}
\nonumber\\
&&\hspace{-12mm}
+~{i\over 8\pi^2}\Big({(x,y)\over x^2}x_i-y_i\Big){1\over i\kappa}
\Bigg\{\int_0^1\! du~
\Bigg[{\ln u\over u-{(x,x-y)-i\kappa\over (x-y)^2}}
-{\ln u\over u-{(x,x-y)+i\kappa\over (x-y)^2}}\Bigg]
-{1\over 2}\ln{(x-y)^2\over x^2}\ln{(x,y)+i\kappa\over  (x,y)-i\kappa}\Bigg\}
\nonumber
\end{eqnarray}
(here $\kappa=\sqrt{x^2y^2-(x,y)^2}$) one obtains
\begin{equation}
\hspace{-0mm}
{d\over d\eta}\langle{\rm Tr}\{\hat{U}_x\hat{U}^\dagger_y\}\rangle_{\rm Fig.~\ref{1cutadd}a-k}
~=~-{\alpha_s^2N_c\over 8\pi^3}\!\int\!d^2z
[{\rm Tr}\{U_xU^\dagger_z\}{\rm Tr}\{U_zU^\dagger_y\}
-{1\over N_c}{\rm Tr}\{U_xU^\dagger_y\}]
{(x-y)^2\over X^2Y^2}\ln{X^2\over (x-y)^2}\ln{Y^2\over (x-y)^2}
\label{1cutaddcoord}
\end{equation}
Note that the two last terms in the r.h.s. of Eq. (\ref{fypie}) do not contribute. 

The contribution of the diagram obtained by reflection of Fig. \ref{1cutadd} with respect to the shock wave differs from Eq. (\ref{sum1cutadd}) by replacement $q\leftrightarrow k$ which doubles result 
(\ref{1cutaddcoord}). The final expression for the contribution of all ``dipole recombination diagrams''  of Fig. \ref{1cutadd} type has the form
\begin{equation}
\hspace{-0mm}
{d\over d\eta}\langle {\rm Tr}\{\hat{U}_x\hat{U}^\dagger_y\}\rangle_{\rm Fig.~\ref{1cutadd}~total}
=~-{\alpha_s^2N_c\over 4\pi^3}\!\int\!d^2z
[{\rm Tr}\{U_xU^\dagger_z\}{\rm Tr}\{U_zU^\dagger_y\}
-{1\over N_c}{\rm Tr}\{U_xU^\dagger_y\}]
{(x-y)^2\over X^2Y^2}\ln{X^2\over (x-y)^2}\ln{Y^2\over (x-y)^2}
\label{1cutaddfinal}
\end{equation}
%

\section{Assembling the NLO kernel}
Adding  results (\ref{2cut}), (\ref{vklad1final}) and (\ref{1cutaddfinal}) one obtains the 
contribution of the diagrams with one and two gluon intersections with the shock wave in the form:
\begin{eqnarray}
&&\hspace{-2mm}
{d\over d\eta}\langle {\rm Tr}\{\hat{U}_x \hat{U}^\dagger_y\}
\rangle_{\rm Fig. \ref{2cutdms}+Fig. \ref{1cutdms}+Fig. \ref{1cutadd}}    
~ =~
{\alpha_s^2N_c\over 8\pi^3}\!\int\! d^2z
~\Big\{{(x-y)^2\over X^2Y^2} \Big[{11\over 3}\ln (x-y)^2\mu^2+{67\over 9}-{\pi^2\over 3}\Big]
\label{12cuts}\\ 
&&\hspace{-2mm}
+~{11\over 3}\Big[{1\over X^2}-{1\over Y^2}\Big]\ln{X^2\over Y^2}   
 -2{(x-y)^2\over X^2Y^2}\ln{X^2\over (x-y)^2}\ln{Y^2\over (x-y)^2}\Big\}   
 [{\rm Tr}\{U_x U^\dagger_z\}{\rm Tr}\{U_z U^\dagger_y\}
-{1\over N_c}{\rm Tr}\{U_x U^\dagger_y\}]
\nonumber\\
&&\hspace{-2mm} 
+~{\alpha_s^2\over 16\pi^4}
\int \!d^2 zd^2 z'
\Bigg[
\Big(-{4\over (z-z')^4}+\Big\{2{X^2{Y'}^2+{X'}^2Y^2-4(x-y)^2(z-z')^2\over  (z-z')^4[X^2{Y'}^2-{X'}^2Y^2]}\nonumber\\ 
&&\hspace{-2mm}
+~\Big({(x-y)^4\over X^2{Y'}^2-{X'}^2Y^2}\Big[
{1\over X^2{Y'}^2}+{1\over Y^2{X'}^2}\Big]
+{(x-y)^2\over (z-z')^2}\Big[{1\over X^2{Y'}^2}-{1\over {X'}^2Y^2}\Big]\Big\}
\ln{X^2{Y'}^2\over {X'}^2Y^2}\Big)
\nonumber\\ 
&&\hspace{42mm}
\times~[{\rm Tr}\{U_xU^\dagger_z\}{\rm Tr}\{U_zU^\dagger_{z'}\}{\rm Tr}\{U_{z'}U^\dagger_y\}
-{\rm Tr}\{U_xU^\dagger_z U_{z'}U^\dagger_yU_zU^\dagger_{z'}\}-(z'\rightarrow z)]
\nonumber\\ 
&&\hspace{-2mm}
+~\Big\{
-{(x-y)^4\over  X^2{Y'}^2{X'}^2Y^2}
+{(x-y)^2\over (z-z')^2 }\Big({1\over X^2{Y'}^2}+{1\over Y^2{X'}^2}\Big)\Big\}\ln{X^2{Y'}^2\over {X'}^2Y^2}
{\rm Tr}\{U_x U^\dagger_z\}{\rm Tr}\{U_zU^\dagger_{z'}\}{\rm Tr}\{U_{z'}U^\dagger_y\}\Bigg]
\nonumber
\end{eqnarray}
%
\begin{figure}

\vspace{-27mm}
\includegraphics[width=1.13\textwidth]{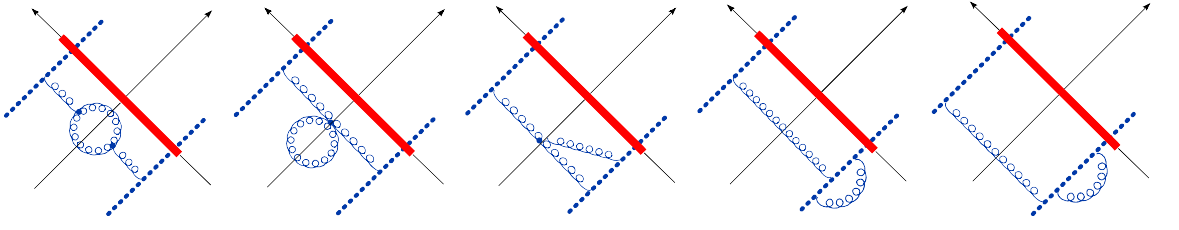}

\vspace{-198mm}
\caption{Typical diagrams without the gluon-shockwave intersection\label{0cuts}.}
\end{figure}
There are also diagrams without gluon-shockwave intersection like the graph shown in Fig. \ref{0cuts}.  They are proportional to the parent dipole ${\rm Tr}\{U_xU^\dagger_y\}$ and their contribution can be found from Eq. (\ref{12cuts}) using the requirement that the r.h.s. of the evolution equation must vanish at $x=y$ (since $U_xU^\dagger_x$=1).  It is easy to see that the replacement 
${\rm Tr}\{U_x U^\dagger_z\}{\rm Tr}\{U_z U^\dagger_y\}
-{1\over N_c}{\rm Tr}\{U_x U^\dagger_y\}$ by  ${\rm Tr}\{U_x U^\dagger_z\}{\rm Tr}\{U_z U^\dagger_y\}
-N_c{\rm Tr}\{U_x U^\dagger_y\}$ fulfills the above requirement so one obtains the final gluon contribution to the NLO kernel in the form
\begin{eqnarray}
&&\hspace{-2mm}
{d\over d\eta}\langle{\rm Tr}\{\hat{U}_x \hat{U}^\dagger_y\} \rangle  
~ =~
{\alpha_s^2N_c\over 8\pi^3}\!\int\! d^2z
~\Big\{{(x-y)^2\over X^2Y^2} \Big[{11\over 3}\ln (x-y)^2\mu^2+{67\over 9}-{\pi^2\over 3}\Big]
\label{nlogluon}\\ 
&&\hspace{-2mm}
+~{11\over 3}\Big[{1\over X^2}-{1\over Y^2}\Big]\ln{X^2\over Y^2}   
 -2{(x-y)^2\over X^2Y^2}\ln{X^2\over (x-y)^2}\ln{Y^2\over (x-y)^2}\Big\}   
 [{\rm Tr}\{U_x U^\dagger_z\}{\rm Tr}\{U_z U^\dagger_y\}
-N_c{\rm Tr}\{U_x U^\dagger_y\}]
\nonumber\\
&&\hspace{-2mm} 
+~{\alpha_s^2\over 16\pi^4}
\int \!d^2 zd^2 z'
\Bigg[
\Big(-{4\over (z-z')^4}+\Big\{2{X^2{Y'}^2+{X'}^2Y^2-4(x-y)^2(z-z')^2\over  (z-z')^4[X^2{Y'}^2-{X'}^2Y^2]}\nonumber\\ 
&&\hspace{-2mm}
+~\Big({(x-y)^4\over X^2{Y'}^2-{X'}^2Y^2}\Big[
{1\over X^2{Y'}^2}+{1\over Y^2{X'}^2}\Big]
+{(x-y)^2\over (z-z')^2}\Big[{1\over X^2{Y'}^2}-{1\over {X'}^2Y^2}\Big]\Big\}
\ln{X^2{Y'}^2\over {X'}^2Y^2}\Big)
\nonumber\\ 
&&\hspace{42mm}
\times~[{\rm Tr}\{U_xU^\dagger_z\}{\rm Tr}\{U_zU^\dagger_{z'}\}{\rm Tr}\{U_{z'}U^\dagger_y\}
-{\rm Tr}\{U_xU^\dagger_z U_{z'}U^\dagger_yU_zU^\dagger_{z'}\}-(z'\rightarrow z)]
\nonumber\\ 
&&\hspace{-2mm}
+~\Big\{
-{(x-y)^4\over  X^2{Y'}^2{X'}^2Y^2}
+{(x-y)^2\over (z-z')^2 }\Big({1\over X^2{Y'}^2}+{1\over Y^2{X'}^2}\Big)\Big\}\ln{X^2{Y'}^2\over {X'}^2Y^2}
{\rm Tr}\{U_x U^\dagger_z\}{\rm Tr}\{U_zU^\dagger_{z'}\}{\rm Tr}\{U_{z'}U^\dagger_y\}\Bigg]
\nonumber
\end{eqnarray}
Promoting Wilson lines in the r.h.s of this equation to operators and adding the quark contribution from Ref. \cite{prd75}
\begin{eqnarray}
&&\hspace{-6mm}
{d\over d\eta}\langle{\rm Tr}\{\hat{U}_x \hat{U}^{\dagger}_y\}\rangle_{\rm quark}~=~{\alpha_s\over 2\pi^2}
\!\int\!d^2z~[{\rm Tr}\{U_x U^{\dagger}_z\}{\rm Tr}\{U_z U^{\dagger}_y\}
-N_c{\rm Tr}\{U_x U^{\dagger}_y\}]
\nonumber\\
&&\hspace{6mm}
\times~
\Big[
-{\alpha_s n_f\over 6\pi}{(x-y)^2\over X^2 Y^2}\Big(\ln (x-y)^2\mu^2+{5\over 3}\Big)+
{\alpha_s n_f\over 6\pi}{X^2-Y^2\over X^2Y^2}\ln{X^2\over Y^2}\Big]
\nonumber\\
&&\hspace{-6mm}
~+{\alpha^2_s\over \pi^4}n_f
{\rm Tr}\{t^aU_xt^bU^{\dagger}_y\}
\!\int\!d^2z d^2z'~
{\rm Tr}\{t^a
U_zt^b U^\dagger_{z'}-t^aU_zt^bU^\dagger_z\}~{1\over (z-z')^4}
\nonumber\\
&&\hspace{6mm}
\times~
\Big\{1
-{{X'}^2Y^2+{Y'}^2X^2-(x-y)^2(z-z')^2\over 2({X'}^2Y^2-{Y'}^2X^2)}
\ln{{X'}^2Y^2\over {Y'}^2X^2}\Big\}
\label{nlobkw}
\end{eqnarray}
 we obtain the full NLO kernel cited in Eq. (\ref{nlobk}).

\section{Comparison to NLO BFKL \label{sect:compare}}
\subsection{Linearized forward kernel}
In this section we compare our kernel to the forward NLO BFKL results\cite{nlobfkl}.
The linearized equation (\ref{nlobk}) has the form
\begin{eqnarray}
&&\hspace{-2mm}
{d\over d\eta}\hat{\cal U}(x,y)~
~
=~{\alpha_sN_c\over 2\pi^2}
\!\int\!d^2z~
{(x-y)^2\over X^2 Y^2}\Big\{1+{\alpha_s\over 4\pi}\Big[b\ln(x-y)^2\mu^2
-b{X^2-Y^2\over (x-y)^2}\ln{X^2\over Y^2}+
({67\over 9}-{\pi^2\over 3})N_c-{10\over 9}n_f
\nonumber\\
&&\hspace{62mm} 
-~
2N_c\ln{X^2\over(x-y)^2}\ln{Y^2\over(x-y)^2}\Big]\Big\}
[\hat{\cal U}(x,z)+\hat{\cal U}(z,y)-\hat{\cal U}(x,y)]   
\nonumber\\
&&\hspace{-2mm} 
+~{\alpha_s^2N_c^2\over 16\pi^4}
\int \!d^2 zd^2 z'
\Bigg[
-{4\over (z-z')^4}+\Big\{2{X^2{Y'}^2+{X'}^2Y^2-4(x-y)^2(z-z')^2\over  (z-z')^4[X^2{Y'}^2-{X'}^2Y^2]}\nonumber\\ 
&&\hspace{-2mm}
+~{(x-y)^4\over X^2{Y'}^2-{X'}^2Y^2}\Big[
{1\over X^2{Y'}^2}+{1\over Y^2{X'}^2}\Big]
+{(x-y)^2\over (z-z')^2}\Big[{1\over X^2{Y'}^2}-{1\over {X'}^2Y^2}\Big]\Big\}
\ln{X^2{Y'}^2\over {X'}^2Y^2}
\nonumber\\ 
&&\hspace{-2mm}
-~{n_f\over N_c^3}
\Big\{{4\over(z-z')^4}
-2{{X'}^2Y^2+{Y'}^2X^2-(x-y)^2(z-z')^2\over (z-z')^4({X'}^2Y^2-{Y'}^2X^2)}
\ln{{X'}^2Y^2\over {Y'}^2X^2}\Big\}
\Bigg]\hat{\cal U}(z,z')
\label{nlolin}
\end{eqnarray}
For the case of forward scattering $\langle\hat{\cal U}(x,y)\rangle=\langle\hat{\cal U}(x-y)\rangle$ and the linearized 
equation (\ref{nlolin}) reduces to

\begin{eqnarray}
&&\hspace{-2mm}
{d\over d\eta}\langle\hat{\cal U}(x)\rangle~
~
=~{\alpha_sN_c\over 2\pi^2}
\!\int\!d^2z~
{x^2\over (x-z)^2 z^2}\Big\{1+{\alpha_s\over 4\pi}\Big[b\ln x^2\mu^2
-b{(x-z)^2-z^2\over x^2}\ln{(x-z)^2\over z^2}+
({67\over 9}-{\pi^2\over 3})N_c-{10\over 9}n_f
\nonumber\\
&&\hspace{62mm} 
-~
2N_c\ln{(x-z)^2\over x^2}\ln{z^2\over x^2}\Big]\Big\}
[\langle\hat{\cal U}(x-z)\rangle+\langle\hat{\cal U}(z)\rangle-\langle\hat{\cal U}(x)\rangle]   
\nonumber\\
&&\hspace{-2mm} 
+~{\alpha_s^2N_c^2\over 16\pi^4}
\int \!d^2 zd^2 z'
\Bigg[-{4\over z^4}+\Big\{2{(x-z-z')^2{z'}^2+(x-z')^2(z+z')^2-4x^2z^2\over 
 z^4[(x-z-z')^2{z'}^2-(x-z')^2(z+z')^2]}+{x^4\over (x-z-z')^2{z'}^2-(x-z')^2(z+z')^2}
 \nonumber\\ 
&&\hspace{-2mm}
\times~\Big[
{1\over (x-z-z')^2{z'}^2}+{1\over (x-z')^2(z+z')^2}\Big]
+{x^2\over z^2}\Big[{1\over (x-z-z')^2{z'}^2}-{1\over (x-z')^2(z+z')^2}\Big]\Big\}
\ln{(x-z-z')^2{z'}^2\over (x-z')^2(z+z')^2}
\nonumber\\ 
&&\hspace{-2mm}
-~{n_f\over N_c^3}
\Big\{{4\over z^4}
-2{(x-z-z')^2{z'}^2+(x-z')^2(z+z')^2-x^2z^2\over z^4[(x-z-z')^2{z'}^2-(x-z')^2(z+z')^2]}
\ln{(x-z-z')^2{z'}^2\over (x-z')^2(z+z')^2}\Big\}
\Bigg]\hat{\cal U}(z)\rangle
\label{nloforward}
\end{eqnarray}

Using the integral $J_{13}$ from hep-ph/9704267 \cite{integral}
 we get
\begin{eqnarray}
&&\hspace{-1mm}
{1\over \pi }\!\!\int\! d^2z' ~
\Big[{x^4\over (x-z-z')^2{z'}^2-(x-z')^2(z+z')^2}+{x^2\over z^2}\Big]
{1\over {z'}^2(x-z-z')^2}\ln{(x-z-z')^2{z'}^2\over (x-z')^2(z+z')^2}~
\nonumber\\
&&\hspace{-1mm}
=~
~{2x^2\over z^2}\Bigg\{ {(x^2-z^2)\over (x-z)^2(x+z)^2}
\Big[\ln{x^2\over z^2}\ln{x^2z^2(x-z)^4\over (x^2+z^2)^4}
+2Li_2\Big(-{z^2\over x^2}\Big)-2Li_2\Big(-{x^2\over z^2}\Big)\Big]
\nonumber\\
&&\hspace{-1mm}
-~\Big(1-{(x^2-z^2)^2\over (x-z)^2(x+z)^2}\Big)\Big[\!\int_0^1-\int_1^\infty\Big]
{du\over (x-zu)^2}\ln{u^2z^2\over x^2}\Bigg\}
\label{integral1}
\end{eqnarray}
\begin{eqnarray}
&&\hspace{-2mm}
{d\over d\eta}\langle\hat{\cal U}(x)\rangle~
~
=~{\alpha_sN_c\over 2\pi^2}
\!\int\!d^2z~
{x^2\over (x-z)^2 z^2}\Big\{1+{\alpha_s\over 4\pi}\Big[b\ln x^2\mu^2
-b{(x-z)^2-z^2\over x^2}\ln{(x-z)^2\over z^2}+
({67\over 9}-{\pi^2\over 3})N_c-{10\over 9}n_f
\nonumber\\
&&\hspace{62mm} 
-~
2N_c\ln{(x-z)^2\over x^2}\ln{z^2\over x^2}\Big]\Big\}
[\langle\hat{\cal U}(x-z)\rangle+\langle\hat{\cal U}(z)\rangle-\langle\hat{\cal U}(x)\rangle]
\nonumber\\
&&\hspace{-2mm} 
+~{\alpha_s^2N_c^2\over 4\pi^3}\!\int\! d^2z~ {x^2\over z^2}
\Bigg\{\Big(1+{n_f\over N_c^3}\Big){3(x,z)^2-2x^2z^2\over16x^2z^2}
\Big({2\over x^2}+{2\over z^2}+{x^2-z^2\over x^2z^2}\Big)\ln{x^2\over z^2}
\nonumber\\
&&\hspace{-1mm}
-~\Big[3+\Big(1+{n_f\over N_c^3}\Big)\Big(1-{(x^2+z^2)^2\over 8x^2z^2}
+{3x^4+3z^4-2x^2z^2\over 16x^4z^4}(x,z)^2\Big)\Big]\!\int_0^\infty\!dt{1\over x^2+t^2z^2}\ln{1+t\over |1-t|} 
\nonumber\\
&&\hspace{-1mm}
+~
{(x^2-z^2)\over (x-z)^2(x+z)^2}
\Big[\ln{x^2\over z^2}\ln{x^2z^2(x-z)^4\over (x^2+z^2)^4}
+2Li_2\Big(-{z^2\over x^2}\Big)-2Li_2\Big(-{x^2\over z^2}\Big)\Big]
\nonumber\\
&&\hspace{-1mm}
-~\Big(1-{(x^2-z^2)^2\over (x-z)^2(x+z)^2}\Big)\Big[\!\int_0^1-\int_1^\infty\Big]
{du\over (x-zu)^2}\ln{u^2z^2\over x^2}\Bigg\}~{\cal U}(z)
\label{nlobfkernel}
\end{eqnarray}
%

\subsection{Comparison of eigenvalues}

To compare the eigenvalues of the Eq. (\ref{nlobfkernel}) with NLO BFKL we expand
${\cal U}(x,0)$ in eigenfunctions
\begin{equation}
\langle\hat{\cal U}(x_\perp,0)\rangle=\sum_{n=-\infty}^\infty \!\int_{-{1\over 2}-i\infty}^{-{1\over 2}+i\infty}
\! {d\gamma\over 2\pi i} ~e^{in\phi}(x_\perp^2\mu^2)^\gamma ~\langle\hat{\cal U}(n,\gamma)\rangle~,
\label{eigenfunctions}
\end{equation}
compute the evolution of $\langle \hat{\cal U}(n,\gamma)\rangle$ from Eq. (\ref{nlobfkernel}) and compare it to the calculation based on the NLO BFKL results from \cite{nlobfkl, lipkot00}. (For the quark part of the NLO BK  kernel the agreement with NLO BFKL was proved in Ref. \cite{kw2}).

The relevant integrals have the form
\begin{eqnarray}
&&\hspace{-1mm}
{1\over 2\pi }\!\int\! d^2z~[2 (z^2/x^2)^\gamma e^{in\phi}-1] {x^2\over(x-z)^2z^2}
~=~\chi(n,\gamma)
\nonumber\\
&&\hspace{-1mm}
{1\over \pi }\!\int\! d^2z~
[2 (z^2/x^2)^\gamma e^{in\phi}-1] \Big({1\over(x-z)^2}- {1\over z^2}\Big)\ln{(x-z)^2\over z^2}
~=~\chi^2(n,\gamma)-\chi'(n,\gamma)-
{4\gamma\chi(\gamma)\over\gamma^2-{n^2\over 4}}
\nonumber\\
&&\hspace{-1mm}
{1\over \pi }\!\int\! d^2z ~ (z^2/x^2)^\gamma
{x^2\over(x-z)^2z^2}
e^{in\phi}\ln{(x-z)^2\over x^2}\ln{z^2\over x^2}
~=~{1\over 2}\chi''(n,\gamma)+\chi'(n,\gamma)\chi(n,\gamma)
\label{relintegral1}
\end{eqnarray}
where 
$\chi(n,\gamma)=2\phi(1)-\psi(\gamma+{n\over 2}) -\psi(1-\gamma+{n\over 2})$, 
and 
\begin{eqnarray}
&&\hspace{-1mm}
{1\over \pi}\int\! d^2z~(z^2/x^2)^{\gamma-1}e^{in\phi}
\Bigg\{\Big(1+{n_f\over N_c^3}\Big){3(x,z)^2-2x^2z^2\over16x^2z^2}
\Big({2\over x^2}+{2\over z^2}+{x^2-z^2\over x^2z^2}\Big)\ln{x^2\over z^2}
\label{relintegral2}\\
&&\hspace{-1mm}
-~\Big[3+\Big(1+{n_f\over N_c^3}\Big)\Big(1-{(x^2+z^2)^2\over 8x^2z^2}
+{3x^4+3z^4-2x^2z^2\over 16x^4z^4}(x,z)^2\Big)\Big]\!\int_0^\infty\!dt{1\over x^2+t^2z^2}\ln{1+t\over |1-t|} \Bigg\}
\nonumber\\
&&\hspace{-1mm}
=~\Big\{-\Big[3+\Big(1+{n_f\over N_c^3}\Big)
{2+3\gamma\bar{\gamma}\over (3-2\gamma)(1+2\gamma)}\Big]\delta_{0n}
+\Big(1+{n_f\over N_c^3}\Big)
{\gamma\bar{\gamma}\over 2(3-2\gamma)(1+2\gamma)}\delta_{2n}\Big\}
{\pi^2\cos\pi\gamma\over(1-2\gamma)\sin^2\pi\gamma}
~\equiv~F(n,\gamma)
\nonumber
\end{eqnarray}
\begin{eqnarray}
&&\hspace{-1mm}
{1\over 2\pi}\int\! d^2z~(z^2/x^2)^{\gamma-1}e^{in\phi}
\Bigg\{ {(x^2-z^2)\over (x-z)^2(x+z)^2}
\Big[\ln{x^2\over z^2}\ln{x^2z^2(x-z)^4\over (x^2+z^2)^4}
+2Li_2\Big(-{z^2\over x^2}\Big)-2Li_2\Big(-{x^2\over z^2}\Big)\Big]
\nonumber\\
&&\hspace{-1mm}
-~\Big(1-{(x^2-z^2)^2\over (x-z)^2(x+z)^2}\Big)\Big[\!\int_0^1-\int_1^\infty\Big]
{du\over (x-zu)^2}\ln{u^2z^2\over x^2}\Bigg\}~=~-\Phi(n,\gamma)-\Phi(n,1-\gamma)
\label{relintegral3}
\end{eqnarray}
where \cite{lipkot00}
\begin{eqnarray}
&&\hspace{-1mm}
\Phi(n,\gamma)~=~\int_0^1\!{dt\over 1+t}~t^{\gamma-1+{n\over 2}}
\Big\{{\pi^2\over 12}-{1\over 2}\psi'\Big({n+1\over 2}\Big)-{\rm Li}_2(t)-{\rm Li}_2(-t)
\nonumber\\
&&\hspace{-1mm}
-~\Big(\psi(n+1)-\psi(1)+\ln(1+t)+\sum_{k=1}^\infty{(-t)^k\over k+n}\Big)\ln t
-\sum_{k=1}^\infty{t^k\over (k+n)^2}[1-(-1)^k]\Big\}
\label{fi}
\end{eqnarray}

The convenient way to  calculate the integrals over angle $\phi$  is to represent 
$\cos n\phi$ as $T_n(\cos\phi)$ and use formulas for the integration of Chebyshev polynomials from Ref. \cite{lipkot00}.

Using integrals (\ref{relintegral1}) - (\ref{relintegral3}) one easily obtains the  evolution equation for ${\cal U}(n,\gamma)$ in the form
\begin{eqnarray}
&&\hspace{-1mm}
{d\over d\eta}\langle\hat{\cal U}(n,\gamma)\rangle~=~{\alpha_s N_c\over \pi}
\Big\{\Big[1-{b\alpha_s\over 4\pi}{d\over d\gamma}
+\Big({67\over 9}-{\pi^2\over 3}\Big)N_c-{10\over 9}{n_f\over N_c^2}\Big]\chi(n,\gamma)
+{\alpha_s b\over 4\pi}
\Big[{1\over 2}\chi^2(n,\gamma)-{1\over 2}\chi'(n,\gamma)
-{2\gamma\chi(n,\gamma)\over\gamma^2-{n^2\over 4} }
\Big]
\nonumber\\
&&\hspace{-1mm}
+~{\alpha_s N_c\over 4\pi}\Big[-\chi"(n,\gamma)-2\chi(n,\gamma)\chi'(n,\gamma)+4\zeta(3)+F(n,\gamma)
-2\Phi(n,\gamma)-2\Phi(n,1-\gamma)\Big]\Big\}\langle\hat{\cal U}(n,\gamma)\rangle
\label{eigenvalue}
\end{eqnarray}
where $\chi'(n,\gamma)\equiv{d\over d\gamma}\chi(n,\gamma)$ etc.

Next we calculate the same thing using NLO BFKL results \cite{nlobfkl,lipkot00}.
The  impact factor $\Phi_A(q)$ for the color dipole ${\cal U}(x,y)$ is proportional to
$\alpha_s(q)(e^{iqx}-e^{iqy})(e^{-iqx}-e^{-iqy})$ so one obtains the cross section of the scattering of color dipole in the form
\begin{eqnarray}
&&\hspace{-1mm}
\langle\hat{\cal U}(x,0)\rangle~=~{1\over 4\pi^2}\!\int\!{d^2 q\over q^2}
{d^2 q'\over {q'}^2}\alpha_s(q)
(e^{iqx}-1)(e^{-iqx}-1)\Phi_B(q')
\!\int_{a-i\infty}^{a+i\infty}\!{d\omega\over 2\pi i}\Big({s\over qq'}\Big)^\omega
G_\omega(q,q')
\label{lip1}
\end{eqnarray}
where $G_\omega(q,q')$ is the partial wave of the forward reggeized gluon scattering amplitude satisfying the equation
\begin{equation}
\omega G_\omega(q,q')=\delta^{(2)}(q-q')+\int\! d^2p K(q,p)G_\omega(p,q') 
\label{wgw}
\end{equation}
and $\Phi_B(q')$ is the target impact factor. The kernel $K(q,p)$ is symmetric with respect to $q\leftrightarrow p$  and the eigenvalues are
\begin{eqnarray}
&&\hspace{-1mm}
\int\! d^2p \Big({p^2\over q^2}\Big)^{\gamma-1}e^{in\phi}K(q,p)~=~
{\alpha_s(q)\over \pi}N_c\Big[\chi(n,\gamma)+{\alpha_sN_c\over 4\pi}\delta(n,\gamma)
\Big],
\label{lip3}\\
&&\hspace{-1mm}
\delta(n,\gamma)~=~-{b\over 2}[\chi'(n,\gamma)+\chi^2(n,\gamma)]
+\Big({67\over 9}-{\pi^2\over 3}-{10\over 9}{n_f\over N_c^3}\Big)\chi(n,\gamma)
+6\zeta(3)
\nonumber\\
&&\hspace{35mm}-\chi"(n,\gamma)+F(n,\gamma)
-2\Phi(n,\gamma)-2\Phi(n,1-\gamma)\Big\}
\nonumber
\end{eqnarray}

The corresponding expression for 
$\langle\hat{\cal U}(n,\gamma)\rangle$ takes the form
\begin{equation}
\hspace{-0mm}
\langle \hat{\cal U}(n,\gamma)\rangle~=~-{1\over 2\pi^2}
\cos{\pi n\over 2}{\Gamma(-\gamma+{n\over 2})\over \Gamma(1+\gamma+{n\over 2})}
\!\int\! {d^2 q\over q^2}
{d^2 q'\over {q'}^2}e^{-in\theta}\alpha_s(q)\Big({q^2\over 4\mu^2}\Big)^\gamma
\Phi_B(q')\!\int_{a-i\infty}^{a+i\infty}\!{d\omega\over 2\pi i}\Big({s\over qq'}\Big)^\omega
G_\omega(q,q')
\label{lip4}
\end{equation}
where $\theta$ is the angle between $\vec{q}$ and $x$ axis.
Using Eq. (\ref{wgw}) we obtain
\begin{eqnarray}
&&\hspace{-1mm}
s{d\over ds}\langle \hat{\cal U}(n,\gamma)\rangle~
\label{lip5}\\
&&\hspace{-1mm}=~-{1\over 2\pi^2}
\cos{\pi n\over 2}{\Gamma(-\gamma+{n\over 2})\over \Gamma(1+\gamma+{n\over 2})}
\!\int\!{d^2 q\over q^2}
{d^2 q'\over {q'}^2}e^{-in\theta}\alpha_s(q)\Big({q^2\over 4\mu^2}\Big)^\gamma
\Phi_B(q')\!\int_{a-i\infty}^{a+i\infty}\!{d\omega\over 2\pi i}\Big({s\over qq'}\Big)^\omega
\int\! d^2p K(q,p)G_\omega(p,q')
\nonumber
\end{eqnarray}
The integration over q can be performed using 
\begin{eqnarray}
&&\hspace{-1mm}
\int\! d^2q ~\alpha_s(q)\Big({q^2\over p^2}\Big)^{\gamma-1}e^{in\phi}K(q,p)~=~
{\alpha_s^2(p)\over \pi}N_c\Big[\chi(n,\gamma)-{b\alpha_s\over 4\pi}\chi'(n,\gamma)+{\alpha_sN_c\over 4\pi}\delta(n,\gamma)
\Big]
\label{lipxz}\\
&&\hspace{-1mm}
\nonumber
\end{eqnarray}
 (recall that $K(q,p)=K(p,q)$ and $\alpha_s(p)=\alpha_s-{b\alpha_s^2\over 4\pi}\ln{p^2\over \mu^2}$
 with our accuracy). The result is
\begin{eqnarray}
&&\hspace{-1mm}
s{d\over ds}\langle \hat{\cal U}(n,\gamma)\rangle~
=~-{\alpha_s\over 2\pi^2}
\cos{\pi n\over 2}{\Gamma(-\gamma+{n\over 2})\over \Gamma(1+\gamma+{n\over 2})}
\!\int\!{d^2 p\over p^2}
{d^2 q'\over {q'}^2}e^{-in\varphi}\Big({p^2\over 4\mu^2}\Big)^\gamma
\Phi_B(q')\label{lip6}\\
&&\hspace{-1mm}
\times~
\!\int_{a-i\infty}^{a+i\infty}\!{d\omega\over 2\pi i}\Big({s\over pq'}\Big)^\omega
G_\omega(p,q'){\alpha_s(p)\over \pi}N_c\Big[\chi(n,\gamma-{\omega\over 2})
-{b\alpha_s\over 4\pi}\chi'(n,\gamma-{\omega\over 2})
+{\alpha_sN_c\over 4\pi}\delta(n,\gamma-{\omega\over 2}))
\Big]
\nonumber
\end{eqnarray}
where the angle $\varphi$ corresponds to $\vec{p}$.
Since $\omega\sim\alpha_s$ we can neglect terms $\sim \omega$ in the argument of $\delta$ and expand $\chi(n,\gamma-{\omega\over 2})\simeq\chi(n,\gamma)
-{\omega\over 2}\chi'(n,\gamma)$. Using again Eq. (\ref{wgw}) in the leading order we
can replace extra $\omega$ by ${\alpha_s\over \pi}N_c\chi(n,\gamma)$ and obtain
\begin{eqnarray}
&&\hspace{-1mm}
s{d\over ds}\langle \hat{\cal U}(n,\gamma)\rangle~=~-{\alpha_s\over 2\pi^2}
\cos{\pi n\over 2}{\Gamma(-\gamma+{n\over 2})\over \Gamma(1+\gamma+{n\over 2})}
\!\int\!{d^2 p\over p^2}
{d^2 q'\over {q'}^2}e^{-in\varphi}\Big({p^2\over 4\mu^2}\Big)^\gamma
\nonumber\\
&&\hspace{-1mm}
\times~
\Phi_B(q')\!\int_{a-i\infty}^{a+i\infty}\!{d\omega\over 2\pi i}\Big({s\over pq'}\Big)^\omega
G_\omega(p,q'){\alpha^2_s(p)\over \pi}N_c\Big[\chi(n,\gamma)
-{b\alpha_s\over 4\pi}\chi'(n,\gamma)+{\alpha_sN_c\over 4\pi}[\delta(n,\gamma)-2\chi(n,\gamma)\chi'(n,\gamma)]
\Big]
\label{lip7}
\end{eqnarray}
Finally, expanding 
 $\alpha_s^2(p)\simeq \alpha_s(p)(\alpha_s-{b\alpha_s^2\over 4\pi}\ln{p^2\over \mu^2})\alpha_s(\mu)$ 
 we obtain

\begin{eqnarray}
&&\hspace{-1mm}
s{d\over ds}\langle \hat{\cal U}(n,\gamma)\rangle~=~-{\alpha_sN_c\over 2\pi^3}
\cos{\pi n\over 2}{\Gamma(-\gamma+{n\over 2})\over \Gamma(1+\gamma+{n\over 2})}
\Big\{\chi(n,\gamma)\Big(1-{b\alpha_s\over 4\pi}{d\over d\gamma}\Big)
-{b\alpha_s\over 4\pi}\chi'(n,\gamma)
\nonumber\\
&&\hspace{-1mm}
+~{\alpha_sN_c\over 4\pi}[\delta(n,\gamma)-2\chi(n,\gamma)\chi'(n,\gamma)]\Big\}
\!\int\!{d^2 p\over p^2}
{d^2 q'\over {q'}^2}e^{-in\varphi}\alpha_s(p)\Big({p^2\over 4\mu^2}\Big)^\gamma
\Phi_B(q')\!\int_{a-i\infty}^{a+i\infty}\!{d\omega\over 2\pi i}\Big({s\over pq'}\Big)^\omega
G_\omega(p,q')
\label{lip8}
\end{eqnarray}
which can be rewritten as an evolution equation
\begin{eqnarray}
&&\hspace{-1mm}
s{d\over ds}\langle \hat{\cal U}(n,\gamma)\rangle~=~
{\alpha_sN_c\over \pi}\Big\{\Big(1
+{b\alpha_s\over 4\pi}\Big[\chi(n,\gamma)-{2\gamma\over \gamma^2-{n^2\over 4}}
-{d\over d\gamma}\Big]\Big)\chi(n,\gamma)
+{\alpha_sN_c\over 4\pi}[\delta(n,\gamma)-2\chi(n,\gamma)\chi'(n,\gamma)]
\Big\}
\langle {\cal U}(n,\gamma)\rangle
\nonumber\\
&&\hspace{-1mm}
~=~{\alpha_s N_c\over \pi}
\Big\{\Big[1-{b\alpha_s\over 4\pi}{d\over d\gamma}
+\Big({67\over 9}-{\pi^2\over 3}\Big)N_c-{10\over 9}{n_f\over N_c^2}\Big]\chi(n,\gamma)
+{\alpha_s b\over 4\pi}
\Big[{1\over 2}\chi^2(n,\gamma)-{1\over 2}\chi'(n,\gamma)-{2\gamma\over\gamma^2-{n^2\over 4} }\chi(\gamma)
\Big]
\nonumber\\
&&\hspace{-1mm}
+~{\alpha_s N_c\over 4\pi}\Big[-\chi"(n,\gamma)-2\chi(n,\gamma)\chi'(n,\gamma)+6\zeta(3)+F(n,\gamma)
-2\Phi(n,\gamma)-2\Phi(n,1-\gamma)\Big]\Big\}\langle \hat{\cal U}(n,\gamma)\rangle
\label{leigenvalue}
\end{eqnarray}
This eigenvalue coincides with Eq. (\ref{eigenvalue}) up to the extra term $2\zeta(3)$. It would correspond to the additional contribution to the r.h.s.  of eq. (\ref{nlobk}) in the form of 
${\alpha_s^2N_c^2\over 4\pi^2}\zeta(3){\rm Tr}U_xU^\dagger_y$ which contradicts the requirement ${d\over d\eta}U_xU^\dagger_y=0$
at $x=y$. A possible reason for the disagreement is the connection between 
the matrix element of the color dipole with a rigid cutoff $\alpha<\sigma$ and the 
cutoff by energy $s$ in Eq. (\ref{lip1}). It is worth noting that the coefficient $6\zeta(3)$ in Eq. (\ref{leigenvalue}) agrees with the $j\rightarrow 1$ asymptotics of the three-loop anomalous dimensions of leading-twist gluon operators \cite{3loops}.

\begin{figure}

\vspace{-12mm}
\includegraphics[width=0.8\textwidth]{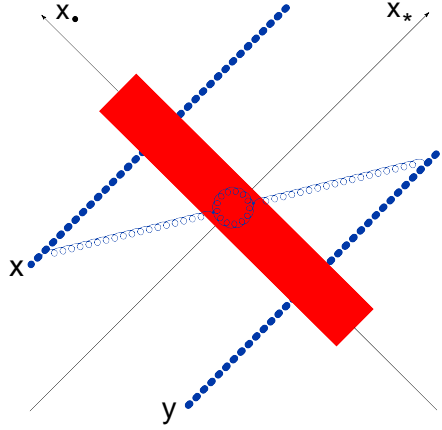}

\vspace{-107mm}
\caption{Gluon loop inside the shock wave \label{loopinside}.}
\end{figure}
It should be emphasized that the coincidence of terms with the nontrivial $\gamma$ dependence proves that there is no additional 
$O(\alpha_s)$  correction
to the vertex of the gluon - shock wave interaction coming from the small loop inside the shock wave, see Fig. \ref{loopinside} 
(In other words, all the effects coming from the small loop in the shock wave are absorbed in the renormalization of coupling constant in the definition of the U operator (\ref{defy})).
In the case of
quark loop, we proved that by the comparison of our results for  ${\rm Tr}\{U_xU^\dagger_y\}$  in the shock-wave background  with explicit light-cone calculation of the behavior of ${\rm Tr}\{U_xU^\dagger_y\}$ as $x\rightarrow y$ 
\cite{prd75}. For the gluon loop, we can use the NLO BFKL results as an independent calculation.
Let us repeat the arguments of Ref. \cite{prd75} for this case. 
The characteristic transverse scale inside the shock wave is small (see the discussion in Ref. \cite{prd75} ) 
 and therefore the contribution of the diagram in Fig. \ref{loopinside} reduces
to the contribution of some operator {\it local} in the transverse space. This would bring the additional terms with the nontrivial $z$ dependence to the kernel which translates into the nontrivial additional $\gamma$-dependent term in the eigenvalues. 
Such terms do not exist and therefore the gluon interaction with the shock wave does not get an extra $O(\alpha_s)$ correction.

\section{Argument of the coupling constant in the BK equation}

In this section we briefly summarize the results of the renormalon-based analysis of the argument of the coupling constant 
carried in Refs. \cite{prd75,kw1}

\begin{figure}
\vspace{-15mm}
\includegraphics[width=\textwidth]{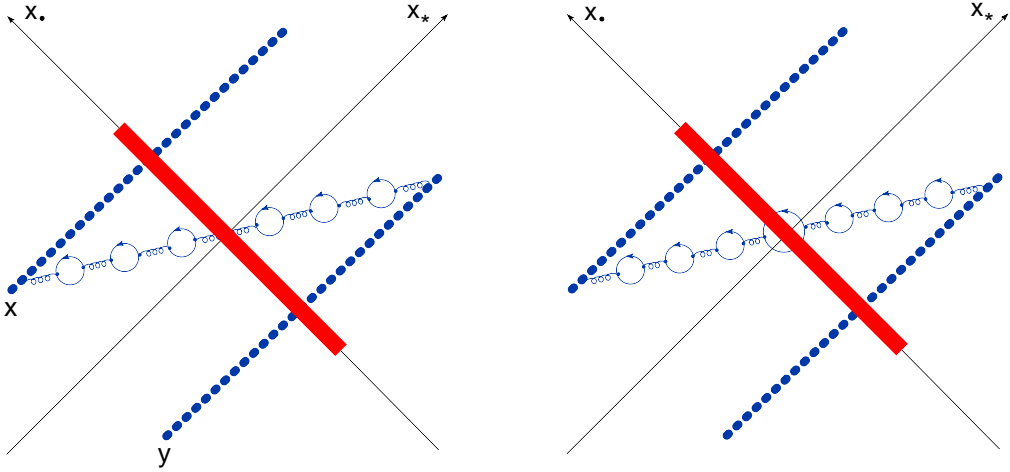}

\vspace{-146mm}
\caption{Renormalon bubble chain of quark loops\label{quarkloops}.}
\end{figure}

To get an argument of coupling constant we can trace the quark part of the $\beta$-function (proportional to $n_f$). In the leading log approximation $\alpha_s\ln {p^2\over\mu^2}\sim 1, ~\alpha_s\ll 1$ the quark part 
 of the $\beta$-function comes from the bubble chain of quark loops in the shock-wave background. 
 We can either have no intersection of quark loop with the shock wave (see  Fig. \ref{quarkloops}a) or we may have one of the loops in the shock-wave background
 (see Fig. \ref{quarkloops}b).

The sum of these diagrams yields
\begin{eqnarray}
&&\hspace{-16mm}
{d\over d\eta}\langle{\rm Tr}\{\hat{U}_x\hat{U}^\dagger_y\}\rangle~=~2\alpha_s{\rm Tr}\{t^aU_xt^bU^\dagger_y\}
\int\! \dhd^2p\dhd^2l~[e^{i(p,x)_\perp}-e^{i(p,y)_\perp}]
[e^{-i(p-l,x)_\perp}-e^{-i(p-l,y)_\perp}]
\nonumber\\
&&\hspace{-16mm}
\times~
{1\over p^2(1+{\alpha_s\over 6\pi}\ln{\mu^2\over p^2})}
\Big(1-{\alpha_sn_f\over 6\pi}\ln{l^2\over \mu^2}\Big)\partial_\perp^2U^{ab}(l)
{1\over (p-l)^2(1+{\alpha_s\over 6\pi}\ln{\mu^2\over (p-l)^2})}
\label{bubblechain1}
\end{eqnarray}
where we have left only the $\beta$-function part of the quark loop.
Replacing the quark part of the $\beta$-function 
$-{\alpha_s\over 6\pi}n_f\ln{p^2\over\mu^2}$ by
the total contribution ${\alpha_s\over 4\pi}b\ln{p^2\over\mu^2}$  we get
\begin{eqnarray}
&&\hspace{-16mm}
{d\over d\eta}\langle{\rm Tr}\{\hat{U}_x\hat{U}^\dagger_y\}\rangle~=~2{\rm Tr}\{t^aU_xt^bU^\dagger_y\}
\nonumber\\
&&\hspace{-16mm}
\times\int\! \dhd^2p\dhd^2q~[e^{i(p,x)_\perp}-e^{i(p,y)_\perp}]
[e^{-i(p-l,x)_\perp}-e^{-i(p-l,y)_\perp}]
{\alpha_s(p^2)\over p^2}
\alpha_s^{-1}(l^2)\partial_\perp^2U^{ab}(q)
{\alpha_s((p-l)^2)\over (p-l)^2}
\label{bubblechain2}
\end{eqnarray}
In principle, one should also include the ``renormalon dressing''  of the double-log and conformal terms 
in Eq. (\ref{nlobk}).   We think, however, that they form a separate contribution 
which has nothing to do with the argument of the BK equation.

To go to the coordinate space, we expand the coupling constants in Eq. (\ref{bubblechain2}) 
in powers of $\alpha_s=\alpha_s(\mu^2)$, i.e. return back to Eq. (\ref{bubblechain1}) 
with ${\alpha_s\over 6\pi}n_f\rightarrow -b{\alpha_s\over 4\pi}$.
Unfortunately, the Fourier transformation to the coordinate space can be performed explicitly only for a couple of  first terms of the expansion 
  $\alpha_s(p^2)\simeq \alpha_s-{b\alpha_s\over 4\pi}\ln p^2/\mu^2+({b\alpha_s\over 4\pi}\ln p^2/\mu^2)^2$.
 In the first order we get the running-coupling part of the NLO BK equation (\ref{nlobk})
\begin{eqnarray}
&&\hspace{-6mm}
{d\over d\eta}\langle{\rm Tr}\{\hat{U}_x \hat{U}^{\dagger}_y\}\rangle~
\label{arg1}\\
&&\hspace{-6mm}
=~{\alpha_s\over 2\pi^2}
\!\int\!d^2z~[{\rm Tr}\{U_x U^{\dagger}_z\}{\rm Tr}\{U_z U^{\dagger}_y\}
-N_c{\rm Tr}\{U_x U^{\dagger}_y\}]
\Big[{(x-y)^2\over X^2 Y^2}
\Big(1+b{\alpha_s \over 4\pi}\ln (x-y)^2\mu^2\Big)-b
{\alpha_s \over 4\pi}{X^2-Y^2\over X^2Y^2}\ln{X^2\over Y^2}\Big]
\nonumber
\end{eqnarray}

The result of the  Fourier transformation up to the second order  has the form \cite{prd75, kw1}
\begin{eqnarray}
&&\hspace{-6mm}
{d\over d\eta}\langle{\rm Tr}\{\hat{U}_x\hat{U}^\dagger_y\}\rangle~
=~{\alpha_s\over 2\pi^2}
\!\int\!d^2z~[{\rm Tr}\{U_x U^{\dagger}_z\}{\rm Tr}\{U_z U^{\dagger}_y\}
-N_c{\rm Tr}\{U_x U^{\dagger}_y\}]\Big\{{(x-y)^2\over X^2 Y^2}
\Big[1+{b\alpha_s\over 4\pi}\Big(\ln (x-y)^2\mu^2+{5\over 3}\Big)
\label{chtoposchitano}\\
&&\hspace{-6mm}
+~\Big({b\alpha_s\over 4\pi}\Big)^2\ln^2 (x-y)^2\mu^2\Big]
+{b\alpha_s\over 4\pi}{1\over X^2}\ln{X^2\over Y^2}
\Big[1+{b\alpha_s\over 4\pi}\ln (x-y)^2\mu^2+
{b\alpha_s\over 4\pi}\ln X^2\mu^2\Big]\Big]
\nonumber\\
&&\hspace{-6mm}
-~{b\alpha_s\over 4\pi}{1\over Y^2}\ln{X^2\over Y^2}\Big[1+{b\alpha_s\over 4\pi}\ln (x-y)^2\mu^2+
{b\alpha_s\over 4\pi}\ln Y^2\mu^2\Big]\Big\}+...
\nonumber
\end{eqnarray}
We extrapolate the $\ln+\ln^2$ terms in the above equation as follows:
\begin{eqnarray}
&&\hspace{-6mm}
{d\over d\eta}{\rm Tr}\{\hat{U}_x \hat{U}^{\dagger}_y\}~
=~{\alpha_s((x-y)^2)\over 2\pi^2}
\!\int\!d^2z~[{\rm Tr}\{\hat{U}_x \hat{U}^{\dagger}_z\}{\rm Tr}\{\hat{U}_z \hat{U}^{\dagger}_y\}
-N_c{\rm Tr}\{\hat{U}_x \hat{U}^{\dagger}_y\}]
\label{runningcoupling}\\
&&\hspace{-6mm}
\times~\Big[{(x-y)^2\over X^2 Y^2}
+{1\over X^2}\Big({\alpha_s(X^2)\over\alpha_s(Y^2)}-1\Big)
+{1\over Y^2}\Big({\alpha_s(Y^2)\over\alpha_s(X^2)}-1\Big)\Big]+...
\end{eqnarray}
where dots stand for the remaining conformal terms and  $\ln^2$ term. (Here we promoted Wilson lines in the r.h.s. to operators). 

When the sizes of the dipoles are very different the kernel of the above equation reduces to 
\begin{eqnarray}
&\hspace{-6mm}
{\alpha_s((x-y)^2)\over 2\pi^2}{(x-y)^2\over X^2 Y^2}
  &|x-y|\ll |x-z|,|y-z|
\nonumber\\
&\hspace{-6mm}
{\alpha_s(X)^2)\over 2\pi^2 X^2 }
  &|x-z|\ll |x-y|,|y-z|
\nonumber\\
&\hspace{-6mm}
{\alpha_s(Y)^2)\over 2\pi^2 Y^2 }
  &|y-z|\ll |x-y|,|x-z|
\label{limits}
\end{eqnarray}
In the earlier paper\cite{prd75} the Eq. (\ref{runningcoupling}) was interpreted 
as an indication that the argument of the coupling constant is the size of the parent dipole $x-y$. We are grateful to G. Salam for pointing out that the proper interpretation is the size of the smallest dipole as follows from Eq. (\ref{limits}).

It is instructive to compare our result to the paper \cite{kw1} where the NLO BK equation
 is rewritten in terms of three effective coupling constants. 
 The authors of Ref. \cite{kw1} 
extrapolate Eq. (\ref{chtoposchitano}) in a different way  
\begin{eqnarray}
&&\hspace{-6mm}
{d\over d\eta}{\rm Tr}\{\hat{U}_x \hat{U}^{\dagger}_y\}~
=~{1\over 2\pi^2}
\!\int\!d^2z~[{\rm Tr}\{\hat{U}_x \hat{U}^{\dagger}_z\}{\rm Tr}\{\hat{U}_z \hat{U}^{\dagger}_y\}
-N_c{\rm Tr}\{\hat{U}_x \hat{U}^{\dagger}_y\}]
\label{ixinterpretation}\\
&&\hspace{-6mm}
\times~\Big[ {1\over X^2}\alpha_s(X^2)
+{1\over Y^2} \alpha_s(Y^2)
- {2(x-z,y-z)\over X^2 Y^2}  {\alpha_s(X^2)  \alpha_s(Y^2)\over \alpha_s(R^2)}
\Big]
\nonumber
\end{eqnarray}
where $R^2$ is some  scale interpolating between $X^2$ and $Y^2$ (the explicit form can be found in Ref. \cite{kw1}). Theoretically,  until 
the Fourier transformations in all orders in $\ln p^2/\mu^2$ are performed, both of these interpretations are models of the high-order behavior of running coupling constant. The convenience of these models can be checked by the numerical estimates of the size of the neglected term(s) in comparison to  terms taken into account by the model, see the discussion in Refs. \cite{runcon}

\section{Conclusions and Outlook}

We have calculated the NLO kernel for the evolution of the color dipole. It consists
of three parts: the running-coupling part proportional to $\beta$-function (see diagrams shown in Fig. \ref{1cutdms}), the conformal
part describing 1 $\rightarrow$ 3 dipoles transition (diagrams in Fig. \ref{2cutdms}) 
and the non-conformal term coming from the 
diagrams in Fig. (\ref{1cutadd}).  The result agrees with the forward NLO BFKL 
kernel \cite{nlobfkl} up to a term proportional $\alpha_s^2\zeta(3)$ times the original dipole. 
We think that the difference could be due to different definitions of the cutoff 
in the longitudinal momenta (see the discussion in previous Section). It would be instructive to get the $j\rightarrow 1$ asymptotics of the anomalous dimensions 
of gluon operators directly from Eq. (\ref{nlobk}), without a Fourier transformation of our result to the momentum space and comparing to
NLO BFKL as it is done in Sect. \ref{sect:compare}. The study is in progress.

There is a recent paper \cite{fadin07} where the dipole form of the non-forward NLO BFKL kernel is calculated using the non-forward NLO BFKL kernel\cite{nfnlobfkl}.  The kernel obtained in \cite{fadin07} is different
from our result (and not conformally invariant). We think that at least part of the difference is coming from the fact that the evolution kernel (\ref{nlobk}) should be compared to the non-symmetric ``evolution'' NLO BFKL kernel $K^{\rm evol}(q,p)$ 
rather that to the symmetric kernel $K(q,p)$ defined by Eq. (\ref{lip1}). 
The kernel $K^{\rm evol}$ corresponds to  the Green function
$\tilde{G}_\omega$ defined by Eq. (\ref{lip1}) with different lower cutoff for the longitudinal integration
\begin{eqnarray}
&&\hspace{-1mm}
\langle \hat{\cal U}(x,0)\rangle~=~{1\over 4\pi^2}\!\int\!{d^2 q\over q^2}
{d^2 q'\over {q'}^2}\alpha_s(q)
(e^{iqx}-1)(e^{-iqx}-1)\Phi_B(q')
\!\int_{a-i\infty}^{a+i\infty}\!{d\omega\over 2\pi i}\Big({s\over {q'}^2}\Big)^\omega
\tilde{G}_\omega(q,q')
\label{getilde}
\end{eqnarray}
The $\tilde{G}_\omega(q,q')$ satisfies the equation (\ref{wgw}) with the kernel
$ K^{\rm evol}$
\begin{equation}
\omega \tilde{G}_\omega(q,q')=\delta^{(2)}(q-q')+
\int\! d^2p K^{\rm evol}(q,p)\tilde{G}_\omega(p,q') 
\label{eqngetilde}
\end{equation}
and the relation between $ K^{\rm evol}(q,p)$ and $K(q,p)$ has the form
(cf. Ref. \cite{nlobfkl})
\begin{equation}
K^{\rm evol}(q,p)=K(q,p)- {1\over 2}\!\int\! d^2q'K(q,q')\ln{q^2\over {q'}^2}K(q',p)
\label{ktilde}
\end{equation}
It is easy to see that the structure (\ref{getilde}) repeats itself after differentiation with respect to $s$ so it can be rewritten as an evolution equation for ${\cal U}(x)$
(whereas the derivative of the original formula (\ref{lip1}) does not have the structure of the evolution equation due to an extra ${1\over |q|^\omega}$). In terms of eigenvalues, the modified kernel (\ref{ktilde}) lead to the shifts of the type
$\chi(n,\gamma)\rightarrow \chi(n,\gamma-{\omega\over 2})$ which we saw in 
Sect. \ref{sect:compare}B.

It should be emphasized that the conformally invariant 
NLO kernel describes  the evolution of the light-like Wilson lines 
with the ``rigid'' cutoff in the longitudinal momenta (\ref{cutoff}). On the contrary,  for  dipoles with the non-light-like slope the sum of the diagrams in Fig. \ref{2cutdms} is not conformally invariant (see Appendix). The reason is that a general Wilson line is a non-local operator which is not conformally invariant to begin with - for example,  the non-light-like Wilson line turns into a circle under the inversion $x^\mu\rightarrow x^\mu/x^2$.
 With the light-like Wilson lines, the situation is different. Formally, a Wilson line
\begin{equation}
[\infty p_1+x_\perp,-\infty p_1+x_\perp]~=~{\rm Pexp}~\Big\{ig\!\int_{-\infty}^\infty\!dx^+~A_+(x^+,x_\perp)\Big\}
\end{equation}
is invariant under the inversion $x^\mu\rightarrow x^\mu/x^2$ (with respect to the point with zero (-) component). Indeed, $(x^+,x_\perp)^2=-x_\perp^2$ so after the inversion $x_\perp\rightarrow x_\perp/x_\perp^2$ and $x^+\rightarrow x^+/x_\perp^2$ and
therefore
\begin{equation}
[\infty p_1+x_\perp,-\infty p_1+x_\perp]~\rightarrow~{\rm Pexp}~\Big\{ig\!\int_{-\infty}^\infty\!d{x^+\over x_\perp^2}~A_+({x^+\over x_\perp^2},x_\perp)\Big\}
~=~[\infty p_1+x_\perp,-\infty p_1+x_\perp]
\end{equation}
Thus, it is not surprising that the bulk of our NLO kernel for the light-like dipoles 
is conformally invariant in the transverse space. The part proportional to the $\beta$-function is not conformally invariant and should not be, but there is another term 
$\sim\ln{(x-y)^2\over(x-z)^2}\ln{(x-y)^2\over(y-z)^2}$ which is not invariant. The reason for that is probably the cutoff $|\alpha|<\sigma$ which can be expressed as a cutoff
in longitudinal coordinate $x^+$, and therefore under the inversion 
$x^+\rightarrow x^+/x_\perp^2$ the cutoff can pick up some logs of transverse separations. 
It is worth noting that conformal and non-conformal terms come from graphs with different topology: the conformal terms come from 1$\rightarrow$3 dipoles diagrams in Fig. (\ref{2cutdms}) which describe the dipole creation while the non-conformal double-log term comes from the1$\rightarrow$2 dipole transitions (see Fig. \ref{1cutadd}) which can be regarded as a combination of dipole creation and dipole recombination. It is possible that in the effective action language, symmetric with respect to the projectile and the target \cite{effaction}, the evolution kernel  is conformally invariant.
 We hope to study this problem in a separate publication.

\section*{Acknowledgments}
The authors are indebted to Yu.V. Kovchegov, L.N. Lipatov, G. Salam  and H. Weigert  for valuable discussions.
I.B.  would like to thank  E. Iancu and other members of theory group at CEA Saclay for for valuable discussions and kind hospitality. 
This work was supported by contract
 DE-AC05-06OR23177 under which the Jefferson Science Associates, LLC operate the Thomas Jefferson National Accelerator Facility.

\section{Appendix A: UV part of the one-to-three dipoles kernel}
As we mentioned above, it is convenient to separate the UV-divrgent and UV-finite parts of 
the Eq. (\ref{2cutsum}) by writing down $U_z^{mm'}U_{z'}^{nn'}=
(U_z^{mm'}U_{z'}^{mm'}-U_z^{mm'}U_z^{mm'})+U_z^{mm'}U_z^{nm'}$.
The contribution of the first part leads to Eq. (\ref{2cut}) while the second UV-divergent term  have the same color structure as the leading-order BK equation.
After replacing $U_z^{mm'}U_{z'}^{nn'}$ by $U_z^{mm'}U_z^{nn'}$, integrating over $u$ with the prescription (\ref{pluscription}) and changing variables to
$k_2=q_2=k'$, $p=q_1+q_2$, $l=q_1-k_1$ (so that
$q_1=p-k'$, $k_1=p-l-k'$ and $k_1+k_2=p-l$)  the Eq. (\ref{2cutsum}) turns into  
\begin{eqnarray}
&&\hspace{-2mm}
\langle{\rm Tr}\{\hat{U}_x 
\hat{U}^\dagger_y\}\rangle_{\rm Fig. \ref{2cutdms}~z'\rightarrow z}
=~
{g^4\over 8\pi^2}\!\int_0^\sigma\!{d\alpha\over\alpha}
\!\int d^2z~
({N_c\over 2}{\rm Tr}\{U_xU^\dagger_z\}{\rm Tr}\{U_zU^\dagger_y\}
-{1\over 2}{\rm Tr}\{U_xU^\dagger_y\} )
\nonumber\\
&&\hspace{-2mm}
\times~\Bigg[\Big\{
\!\int\! \dhd^{2-\varepsilon} p\dhd^{2-\varepsilon}~F_1(p,l)+
\int\! \dhd^2 p\dhd^2l~F_2(p,l)\Big\}
~(e^{i(p,X)}-e^{i(p,Y)})(e^{-i(p-l,X)}-e^{-i(p-l,Y)})
\nonumber\\
&&\hspace{-2mm}
+~(e^{-i(p-l,X)}-e^{-i(p-l,Y)})(e^{i(p-k',X)+i(k',Y)}-e^{i(p-k',Y)+i(k',X)})
\nonumber\\
&&\hspace{32mm}
\times~{(k',p-k')(p-k')^2-2(p-k',p-l-k')(k',p-l-k')\over
(p-l)^2(p-k')^2{k'}^2(p-l-k')^2}\ln{(p-l-k')^2\over {k'}^2}\Bigg]
\nonumber\\
&&\hspace{-2mm}
+~\!\int\! \dhd^2 p\dhd^2l\dhd^2k'\Big\{(e^{i(p,X)}-e^{i(p,Y)})
(e^{-i(p-l-k',X)-i(k',Y)}-e^{-i(p-l-k',Y)-i(k',X)})
\nonumber\\
&&\hspace{32mm}
\times~
{(k',p-l-k')(p-k')^2-2(p-k',p-l-k')(k',p-k')\over
p^2(p-k')^2{k'}^2(p-l-k')^2}\ln{(p-k')^2\over {k'}^2}
\label{a1}
\end{eqnarray}
where

\begin{eqnarray}
&&\hspace{-2mm}
F_1(p,l)~=~
\int\!\dhd^{2-\varepsilon}k'~\Bigg(
{1-{\varepsilon\over 2}\over p^2(p-l)^2}\Bigg\{ -2-{(p-k')^2+(p-k'-l)^2\over (p-k')^2-(p-k'-l)^2}
\ln{(p-k')^2\over(p-k'-l)^2}
+{{k'}^2+(p-k')^2\over (p-k')^2-{k'}^2}
\ln{ (p-k')^2\over {k'}^2}
\nonumber\\
&&\hspace{-2mm}+~{(p-k'-l)^2+{k'}^2\over (p-k'-l)^2-{k'}^2}
\ln{(p-k'-l)^2\over {k'}^2}\Bigg\}
+~{2(p,p-l)\over p^2(p-l)^2}\Bigg\{
\Big({(p-k',p-k'-l)\over (p-k')^2 (p-k'-l)^2}
-{1\over {k'}^2}\Big)\ln{(p-k')^2(p-k'-l)^2\over {k'}^4}
\nonumber\\
&&\hspace{-2mm}-\Big({(p-k',p-k'-l)\over (p-k')^2}+{(p-k',p-k'-l)\over (p-k'-l)^2}+2\Big)
{\ln (p-k')^2/(p-k'-l)^2\over(p-k')^2-(p-k'-l)^2}
\Bigg\}          
+{2\over p^2}{(p,p-l-k')\over (p-l-k')^2{k'}^2} \ln {p^2\over {k'}^2}
\nonumber\\
&&\hspace{-2mm}
+~{2\over (p-l)^2}{(p-l,p-k')\over (p-k')^2{k'}^2}\ln{(p-l)^2\over {k'}^2}
\Bigg)
\label{f1}
\end{eqnarray}
and

\begin{eqnarray}
&&\hspace{-12mm}
F_2(p,l)~=~
\mu^{2\epsilon}\!\int\!\dhd^2k'~\Bigg(2l_i\Bigg[\Big(-{p_i\over p^2}+{(p-k')_i\over (p-k')^2}\Big)
{\ln { (p-k')^2 /(p-l-k')^2}\over {k'}^2[(p-k')^2-(p-k'-l)^2]}
\label{f2}\\
&&\hspace{-12mm}
+~\Big({(p-l)_i\over (p-l)^2}-{(p-l-k')_i\over (p-k'-l)^2}\Big)
{1/{k'}^2\over (p-k')^2-(p-k'-l)^2}
\ln { (p-k')^2\over (p-l-k')^2}
\Bigg]
\nonumber\\
&&\hspace{-12mm}
+~{2\over p^2}\Bigg[
{2(l,p-k'-l)(p,k')/{k'}^2\over (p-k'-l)^2[(p-k')^2-(p-k'-l)^2]}
\ln { (p-k')^2\over (p-k'-l)^2}
-{(p-k',p-l-k')(p,k')\over {k'}^2(p-k')^2(p-k'-l)^2}
\ln { (p-k')^2\over {k'}^2}
\nonumber\\
&&\hspace{-12mm}
+~{(l,k')/{k'}^2\over (p-k')^2-(p-k'-l)^2}
\ln { (p-k')^2\over (p-l-k')^2}
-{(p-k'-l,k')\over 2{k'}^2(p-l-k')^2}\ln { (p-k')^2\over {k'}^2}
+{(p-l,k')/{k'}^2\over (p-l-k')^2-{k'}^2}\ln { (p-k'-l)^2\over {k'}^2}
\nonumber\\
&&\hspace{-12mm}
+~{2\over (p-k')^2-(p-k'-l)^2}
\ln { (p-k')^2\over (p-l-k')^2}-
{2\ln (p-l-k')^2/ {k'}^2\over (p-l-k')^2-{k'}^2}
+
{(p,p-l-k')\over (p-l-k')^2{k'}^2}\ln {(p-k')^2\over p^2}  \Bigg]
\nonumber\\
&&\hspace{-12mm}
+~{2\over (p-l)^2}\Bigg[ 
{-2(l,p-k')(p-l,k')/{k'}^2\over (p-k')^2[(p-k')^2-(p-k'-l)^2]}
\ln { (p-k')^2\over (p-k'-l)^2}
-{(p-k',p-l-k')(p-l,k')\over {k'}^2(p-k')^2(p-k'-l)^2}
\ln { (p-k'-l)^2\over {k'}^2}
\nonumber\\
&&\hspace{-12mm}
-~{(l,k')/{k'}^2\over (p-k')^2-(p-k'-l)^2}
\ln { (p-k')^2\over (p-l-k')^2}
+{(p,k')/{k'}^2\over (p-k')^2-{k'}^2}\ln { (p-k')^2\over {k'}^2}
-{(p-k',k')\over 2{k'}^2(p-k')^2}\ln { (p-l-k')^2\over {k'}^2}
\nonumber\\
&&\hspace{-12mm}
+~{2\over (p-k')^2-(p-k'-l)^2}
\ln { (p-k')^2\over (p-l-k')^2}-
{2\ln (p-k')^2/ {k'}^2\over (p-k')^2-{k'}^2}
+{(p-l,p-k')\over (p-k')^2{k'}^2}\ln {(p-l-k')^2\over (p-l)^2}  
\Bigg]\Bigg)
\end{eqnarray}
We need to perform the integration over $k'$. Let us start with the UV-divergent term $\sim F_1$. Using the integrals
\begin{eqnarray}
&&\hspace{-2mm}
4\pi\!\int\!\dhd^dk'{(l-k')_i\over {k'}^2(l-k')^2}\ln{p^2\over {k'}^2}~=~
l_i{\Gamma({d\over 2})\Gamma({d\over 2}-1)\over\Gamma(d-1)}
{\Gamma(2-{d\over 2})\over l^{4-d}}\Big[\ln{p^2\over l^2}+{2\over d-2}
+\psi(2-{d\over 2})+\psi(d-1)-\psi({d\over 2})-\psi(1)\Big]
\nonumber\\
&&\hspace{-2mm}
4\pi\!\int\!\dhd^dk'{(k',k'-l)\over {k'}^2(l-k')^2}\ln{(p-k')^2\over {k'}^2}~=~{1\over 2}\ln{p^2\over l^2}\ln{(p-l)^2\over l^2}
\nonumber\\
&&\hspace{-2mm}
4\pi\!\int\!\dhd^dk'{(p-k',p-k'-l)\over (p-k')^2 (p-k'-l)^2}
\ln{(p-k')^2(p-k'-l)^2\over  {k'}^4}~
=~-\ln{p^2\over l^2}\ln{(p-l)^2\over l^2}
\nonumber\\
&&\hspace{-2mm}
4\pi\!\int\!\dhd^dk'{(p,p-l-k')\over (p-l-k')^2{k'}^2}\ln {p^2\over {k'}^2}  
\nonumber\\
&&\hspace{-2mm}=~ 
(p,p-l){\Gamma({d\over 2})\Gamma({d\over 2}-1)\over\Gamma(d-1)}
{\Gamma(2-{d\over 2})\over |p-l|^{4-d}}\Big[\ln{p^2\over (p-l)^2}+{2\over d-2}
+\psi(2-{d\over 2})+\psi(d-1)-\psi({d\over 2})-\psi(1)\Big]
\label{kintegrals1}
\end{eqnarray}
one obtains
\begin{eqnarray}
&&\hspace{-2mm}
F_1(p,l)~=~{1\over 4\pi}\Bigg\{ {2(p,p-l)\over p^2(p-l)^2}
{\Gamma^2(1-{\varepsilon\over 2})\over \Gamma(2-\varepsilon)}\Gamma(\varepsilon/2)
\Big(-4+{1-{\varepsilon\over 2}\over 3-\varepsilon}\Big)+
 {2(p,p-l)\over p^2(p-l)^2}\Big({11\over 3}\ln {l^2\over\mu^2}
 \nonumber\\
&&\hspace{-12mm}
-~\ln{p^2\over l^2}\ln{(p-l)^2\over l^2}-\ln^2{(p-l)^2\over p^2}+{\pi^2\over 3}\Big)
-{\ln p^2/l^2\over 3(p-l)^2}-{\ln(p-l)^2/l^2\over 3p^2}+O(\varepsilon)\Bigg\}
\label{f1a}
\end{eqnarray}
Let us at first consider the UV-divergent contribution
\begin{eqnarray}
&&\hspace{-12mm}
{d\over d\sigma}\langle{\rm Tr}\{\hat{U}_x 
\hat{U}^\dagger_y\}\rangle_{\rm UV}
~=~
{\alpha_s^2\over \pi}\mu^{2\varepsilon}
\!\int d^{2-\varepsilon}z
({N_c\over 2}{\rm Tr}\{U_xU^\dagger_z\}{\rm Tr}\{U_zU^\dagger_y\}
-{1\over 2}{\rm Tr}\{U_xU^\dagger_y\} )
\int\! \dhd^{2-\varepsilon} p~\dhd^{2-\varepsilon}l
\nonumber\\
&&\hspace{-12mm}
\times
~(e^{i(p,X)}-e^{i(p,Y)})(e^{-i(p-l,X)}-e^{-i(p-l,Y)}){(p,p-l)\over p^2(p-l)^2}
\Big[{\Gamma^2(1-{\varepsilon\over 2})\over \Gamma(2-\varepsilon)}\Gamma(\varepsilon/2)
\Big(-4+{1-{\varepsilon\over 2}\over 3-\varepsilon}\Big)+{11\over 3}\ln{l^2\over\mu^2}
+O(\varepsilon)\Big]
\nonumber\\
&&\hspace{-12mm}
\label{uvpart}
\end{eqnarray}
To this contribution we should add the counterterm corresponding 
to quark and gluon loops lying inside the shock wave.
The rigorous calculation of the counterterm was performed in Ref. \cite{prd75}
and the result is 
\begin{eqnarray}
&&\hspace{-12mm}
{d\over d\sigma}\langle{\rm Tr}\{\hat{U}_x 
\hat{U}^\dagger_y\}\rangle_{\rm CT}~
~
=~-b{\alpha_s^2\over \pi}{2\over\varepsilon}\!\int\! d^{2-\varepsilon}z_\perp 
({N_c\over 2}{\rm Tr}\{U_xU^\dagger_z\}{\rm Tr}\{U_zU^\dagger_y\}
-{1\over 2}{\rm Tr}\{U_xU^\dagger_y\} )
\nonumber\\
&&\hspace{-12mm}
\times~\int\! \dhd^dp~\dhd^dl~
(e^{i(p,X)}-e^{i(p,Y)})(e^{-i(p-l,X)}-e^{-i(p-l,Y)})
{(p,p-l)\over p^2(p-l)^2}
\label{ct1}
\end{eqnarray}
where we need the gluon part of $b$ ($={11\over 3}N_c$). 
After subtraction of the counterterm (\ref{ct1})  the UV-divergent contribution (\ref{uvpart}) reduces to
\begin{eqnarray}
&&\hspace{-12mm}
{d\over d\sigma}\langle{\rm Tr}\{\hat{U}_x 
\hat{U}^\dagger_y\}\rangle_{\rm UV-CT}
~=~{\alpha_s^2\over \pi}
\!\int d^2z
({N_c\over 2}{\rm Tr}\{U_xU^\dagger_z\}{\rm Tr}\{U_zU^\dagger_y\}
-{1\over 2}{\rm Tr}\{U_xU^\dagger_y\} )
\nonumber\\
&&\hspace{-12mm}
\times
\int\! \dhd^2 p~\dhd^2 l
~(e^{i(p,X)}-e^{i(p,Y)})(e^{-i(p-l,X)}-e^{-i(p-l,Y)}){(p,p-l)\over p^2(p-l)^2}
\Big[{11\over 3}\ln{l^2\over\mu^2}-{67\over 9}\Big]
\nonumber\\
&&\hspace{-12mm}
\label{uvpart1}
\end{eqnarray}
so one obtains the regularized $F_1$ in the form
\begin{equation}
\hspace{-2mm}
F_1^{\rm reg}(p,l)~=~{1\over 4\pi}\Bigg[{2(p,p-l)\over p^2(p-l)^2}
\Big({11\over 3}\ln {l^2\over\mu^2}-{67\over 9}
-~\ln{p^2\over l^2}\ln{(p-l)^2\over l^2}-\ln^2{(p-l)^2\over p^2}+{\pi^2\over 3}\Big)
-{\ln p^2/l^2\over 3(p-l)^2}-{\ln(p-l)^2/l^2\over 3p^2}\Bigg]
\label{f1reg}
\end{equation}
It is convenient to calculate first the Fourier transform with $e^{i(p,X)-i(p-l,Y)}$.
Using the integrals
\begin{eqnarray}
&&\hspace{-12mm}
\!\int\! \dhd^2 p~\dhd^2l
~e^{i(p,\Delta)+i(l,Y)}~{(p,p-l)\over p^2(p-l)^2}\ln{l^2\over\mu^2}~=~
-{1\over 4\pi^2}{(X,Y)\over X^2Y^2}\ln{X^2Y^2\over\Delta^2}\mu^2
\nonumber\\
&&\hspace{-12mm}
\int\!\dhd^2 p~ \dhd^2 l~e^{i(p,\Delta)+i(l,Y)}{(p,p-l)\over p^2(p-l)^2} 
\ln{p^2\over l^2}\ln{(p-l)^2\over l^2}
~=~ {1\over 4\pi^2}{(X,Y)\over X^2Y^2}\ln{X^2\over \Delta^2}\ln{Y^2\over \Delta^2}
\nonumber\\
&&\hspace{-12mm}
\int\!\dhd^2 p~ \dhd^2 l~e^{i(p,\Delta)+i(l,Y)}{(p,p-l)\over p^2(p-l)^2} 
\ln^2{(p-l)^2\over p^2}
~=~{1\over 4\pi^2}
{(X,Y)\over X^2Y^2}\ln^2{X^2\over Y^2}
\end{eqnarray}
we get
\begin{eqnarray}
&&\hspace{-2mm}
\!\int\! \dhd^2 p~\dhd^2l~e^{i(p,\Delta)+i(l,Y)}F_1^{\rm reg}(p,l)
~=
~-{11\over 24\pi^3}{(X,Y)\over X^2Y^2}\Big(\ln{X^2Y^2\over \Delta^2}\mu^2+{67\over 33}\Big)
\label{f1contribution}\\
&&\hspace{22mm}
-~{1\over 16\pi^3}{(X,Y)\over X^2Y^2}\Big[\ln^2{X^2\over \Delta^2}+\ln^2{Y^2\over \Delta^2}
+\ln^2{X^2\over Y^2}-{2\pi^2\over 3}\Big]
-{1\over 48\pi^3}\Big[{1\over X^2}\ln{Y^2\over \Delta^2}
+{1\over Y^2}\ln{X^2\over \Delta^2}\Big]
\nonumber
\end{eqnarray}
Hereafter we use the notation $\Delta\equiv X-Y=x-y$.

Next we calculate the $F_2$ contribution. We need the following Fourier integrals:
\begin{eqnarray}
&&\hspace{-12mm}
\int\! \dhd^2 p\dhd^2l
~e^{i(p,\Delta)+i(l,Y)}
\int\!\dhd^2k'~2l_i\Bigg[\Big(-{p_i\over p^2}+{(p-k')_i\over (p-k')^2}\Big)
{\ln { (p-k')^2 /(p-l-k')^2}\over {k'}^2[(p-k')^2-(p-k'-l)^2]}
\nonumber\\
&&\hspace{-12mm}
+~\Big({(p-l)_i\over (p-l)^2}-{(p-l-k')_i\over (p-k'-l)^2}\Big)
{1/{k'}^2\over (p-k')^2-(p-k'-l)^2}
\ln { (p-k')^2\over (p-l-k')^2}
\Bigg]
~
=~{(X,Y)\over 16\pi^3X^2Y^2}
 \Big[\ln^2{X^2\over Y^2}+{2\pi^2\over 3}\Big]
 \label{term2}
\end{eqnarray}
\begin{eqnarray}
&&\hspace{-2mm}
\int\!\dhd^2 p~\dhd^2 l ~e^{i(p,\Delta)+i(l,Y)}~
\Bigg[{4\over p^2}\int\!\dhd k'
{(l,p-l-k')(p,k')\over {k'}^2(p-k'-l)^2((p-k')^2-(p-k'-l)^2)}\ln {(p-k')^2\over (p-k'-l)^2}
\nonumber\\
&&\hspace{-2mm}
-~{4\over (p-l)^2}\int\!\dhd k'
{(l,p-k')(p-l,k')\over {k'}^2(p-k')^2((p-k')^2-(p-k'-l)^2)}\ln {(p-k')^2\over (p-k'-l)^2}
\Bigg]
\nonumber\\
&&\hspace{-2mm}=~
{1\over 16\pi^3}
\Bigg\{-{\pi^2\over 3}{(X+Y)^2\over X^2Y^2}+~{2\over Y^2}\!\int_0^1\!du{\ln u\over u-{X^2\over X^2-Y^2}}
+~{2\over X^2}\!\int_0^1\!du{\ln u\over u+{Y^2\over X^2-Y^2}}
\nonumber\\
&&\hspace{-2mm}
+~{i\kappa\over X^2Y^2}
\Bigg(2\!\int_0^1\!du\Big[{\ln u\over u-{(X,\Delta)-i\kappa\over \Delta^2}}
+{\ln u\over u+{(Y,\Delta)+i\kappa\over \Delta^2}}-c.c.\Big]
+\ln{X^2Y^2\over\Delta^4}\ln{(X,Y)+i\kappa
\over (X,Y)-i\kappa}\Bigg)\Bigg\}
\label{term3}     
\end{eqnarray}
where $\kappa=\sqrt{X^2Y^2-(X,Y)^2}$, and
\begin{eqnarray}
&&\hspace{-2mm}
\int\! \dhd^2 p\dhd^2l
~e^{i(p,\Delta)+i(l,Y)}\int\!\dhd^2k'
~\Bigg[
-{2(p-k',p-l-k')(p,k')\over p^2{k'}^2(p-k')^2(p-k'-l)^2}
\ln { (p-k')^2\over {k'}^2} 
\label{term4}\\
&&\hspace{82mm}- ~{2(p-k',p-l-k')(p-l,k')\over (p-l)^2{k'}^2(p-k')^2(p-k'-l)^2}
\ln { (p-k'-l)^2\over {k'}^2}
\Bigg]
\nonumber\\
&&\hspace{-2mm}
=~{1\over 32\pi^3X^2Y^2}\Big[X^2\ln^2{X^2\over \Delta^2}+Y^2\ln^2{Y^2\over \Delta^2}
+2(X,Y)\ln{X^2\over\Delta^2}\ln{Y^2\over\Delta^2}\Big]
\nonumber\\
&&\hspace{22mm}
+~{i\kappa\over 16\pi^3X^2Y^2}
\!\int_0^1\!du\Big[{\ln u\over u-{(\Delta,X)+i\kappa\over \Delta^2}}
+{\ln u\over u+{(\Delta,Y)-i\kappa\over \Delta^2}}-c.c.
-{1\over  2}\ln{X^2Y^2\over\Delta^4}\ln{(X,Y)+i\kappa\over (X,Y)-i\kappa}\Big]
\nonumber
\end{eqnarray}
\begin{eqnarray}
&&\hspace{-6mm}
\int\! \dhd^2 p\dhd^2l
~e^{i(p,\Delta)+i(l,Y)}
\!\int\!\dhd^2k'\Big({2(l,k')p^{-2}\over {k'}^2[(p-k')^2-(p-k'-l)^2]}
\ln { (p-k')^2\over (p-l-k')^2}
+{2(p-l,k')p^{-2}\over {k'}^2[(p-k'-l)^2-{k'}^2]}
\nonumber\\
&&\hspace{-6mm}
\times~\ln { (p-l-k')^2\over {k'}^2}-{2(l,k')(p-l)^{-2}\over {k'}^2[(p-k')^2-(p-k'-l)^2]}
\ln { (p-k')^2\over (p-l-k')^2}
+{2(p,k')(p-l)^{-2}\over {k'}^2[(p-k')^2-{k'}^2]}\ln { (p-k')^2\over {k'}^2}\Big)
\nonumber\\
&&\hspace{-6mm}
=~-{1\over 16\pi^3}\Big({1\over Y^2}\ln{X^2\over\Delta^2}
-{1\over X^2}\ln{Y^2\over\Delta^2}\Big)\ln{X^2\over Y^2}
+{1\over 8\pi^3}\Bigg[{\pi^2\over 6X^2}+{\pi^2\over 6Y^2}
-{1\over X^2}\!\int_0^1\!{du\ln u \over u+{Y^2\over X^2-Y^2}}
-{1\over Y^2}\!\int_0^1\!{du\ln u \over u-{X^2\over X^2-Y^2}}\Bigg]
\label{term5}
\end{eqnarray}
\begin{eqnarray}
&&\hspace{-2mm}
\int\! \dhd^2 p\dhd^2l
~e^{i(p,\delta)+i(l,Y)}\!\int\!\dhd^2k'~\Bigg[
-{(p-k'-l,k')\over p^2{k'}^2(p-l-k')^2}\ln { (p-k')^2\over {k'}^2}
-{(p-k',k')\over (p-l)^2{k'}^2(p-k')^2}\ln { (p-l-k')^2\over {k'}^2}
\Bigg]
\nonumber\\
&&\hspace{-2mm}
=~{1\over 32\pi^3}\ln{X^2\over Y^2}\Big[{1\over Y^2}\ln{X^2\over \Delta^2}
-{1\over X^2}\ln{Y^2\over \Delta^2}\Big]
\label{term6}
\end{eqnarray}
\begin{eqnarray}
&&\hspace{-2mm}
\int\! \dhd^2 p~\dhd^2l
~e^{i(p,\Delta)+i(l,Y)}
\int\!\dhd^2k'~\Bigg[{4/p^2\over (p-k')^2-(p-k'-l)^2}
\ln { (p-k')^2\over (p-l-k')^2}-
{4\ln (p-l-k')^2/ {k'}^2\over p^2[(p-l-k')^2-{k'}^2]}
\label{term7}\\
&&\hspace{-2mm}
+~{4\over (p-l)^2[(p-k')^2-(p-k'-l)^2]}
\ln { (p-k')^2\over (p-l-k')^2}-
{4\ln (p-k')^2/ {k'}^2\over (p-l)^2[(p-k')^2-{k'}^2]}~
=~{1\over 4\pi^3Y^2}\ln{X^2\over\Delta^2}+ {1\over 4\pi^3X^2}\ln{Y^2\over\Delta^2}
\nonumber
\end{eqnarray}
\begin{eqnarray}
&&\hspace{-2mm}
\int\! \dhd^2 p\dhd^2l
~e^{i(p,\Delta)+i(l,Y)}
\int\!\dhd^2k'~\Bigg[{2(p,p-l-k')\over p^2(p-l-k')^2{k'}^2}\ln {(p-k')^2\over p^2}  
+{2(p-l,p-k')\over (p-k')^2{k'}^2(p-l)^2}\ln {(p-l-k')^2\over (p-l)^2}  
\Bigg]
\label{term8}\\
&&\hspace{-2mm}
=~
{i\kappa\over  16\pi^3X^2Y^2}\Bigg\{
2\!\int_0^1\! du\Big[
{\ln u\over u-{(X,\Delta)+i\kappa\over \Delta^2}}+{\ln u\over u+{(Y,\Delta)-i\kappa\over \Delta^2}}
-~c.c.\Big]
-\ln{X^2Y^2\over \Delta^4}\ln{(X,Y)+i\kappa\over  (X,Y)-i\kappa}
\Bigg\}
+{(X,Y)\over 16\pi^3X^2Y^2}\ln^2{X^2\over Y^2}
\nonumber
\end{eqnarray}
Adding the integrals (\ref{term2}) - (\ref{term6}) we obtain
\begin{eqnarray}
&&\hspace{-2mm}
\!\int\! \dhd^2 p~\dhd^2l~e^{i(p,\Delta)+i(l,Y)}F_2(p,l)
\label{f2contribution}\\
&&\hspace{-2mm}
=~{1\over 4\pi^3}\Big[{1\over X^2}\ln{Y^2\over \Delta^2}
+{1\over Y^2}\ln{X^2\over \Delta^2}\Big]  
+{(X+Y)^2\over 32\pi^3X^2Y^2}\ln{X^2\over\Delta^2}\ln{Y^2\over\Delta^2}
+{(X,Y)\over 8\pi^3X^2Y^2}\ln^2{X^2\over Y^2}
\nonumber\\
&&\hspace{-12mm}
+~{i\kappa\over 16\pi^3X^2Y^2}\Bigg\{\!\int_0^1\!du\Big[{\ln u\over u-{(\Delta,X)+i\kappa\over \Delta^2}}
+{\ln u\over u+{(\Delta,Y)-i\kappa\over \Delta^2}}-c.c.\Big]
-{1\over  2}\ln{X^2Y^2\over\Delta^4}\ln{(X,Y)+i\kappa\over (X,Y)-i\kappa} \Bigg\}
\nonumber
\end{eqnarray}
and therefore
\begin{eqnarray}
&&\hspace{-2mm}
\!\int\! \dhd^2 p~\dhd^2l~e^{i(p,\Delta)+i(l,Y)}[F_1^{\rm reg}(p,l)+F_2(p,l)]
~
\nonumber\\
&&\hspace{-2mm}=
~-{1\over 8\pi^3}{(X,Y)\over X^2Y^2}
\Big[{11\over 3}\ln{X^2Y^2\over \Delta^2}\mu^2+{67\over 9}-{\pi^2\over 3}\Big]
+{1\over 32\pi^3}{\Delta^2\over X^2Y^2}\ln{X^2\over\Delta^2}\ln{Y^2\over\Delta^2}
+{11\over 48\pi^3}\Big[{1\over X^2}\ln{Y^2\over\Delta^2}
+{1\over Y^2}\ln{X^2\over \Delta^2}\Big]
\nonumber\\
&&\hspace{-2mm}
+~{i\kappa\over 16\pi^3X^2Y^2}\Bigg\{\!\int_0^1\!du\Big[{\ln u\over u-{(\Delta,X)+i\kappa\over \Delta^2}}
+{\ln u\over u+{(\Delta,Y)-i\kappa\over \Delta^2}}-c.c.\Big]
-{1\over  2}\ln{X^2Y^2\over\Delta^4}\ln{(X,Y)+i\kappa\over (X,Y)-i\kappa} \Bigg\}
\label{f12}
\end{eqnarray}
Note that the r.h.s. of this equation is finite as $X\rightarrow Y$ (taken separately, the contributions of $F_1$ and $F_2$ are singular in this limit):  
\begin{eqnarray}
&&\hspace{-2mm}
\!\int\! \dhd^2 p~\dhd^2l~e^{i(l,X)}[F_1^{\rm reg}(p,l)+F_2(p,l)]
~=~-{1\over 8\pi^3}{1\over X^2}
\Big[{11\over 3}\ln X^2\mu^2+{67\over 9}-{\pi^2\over 3}\Big]
\label{f12xy}
\end{eqnarray}
Using Eqs. (\ref{f12}) and (\ref{f12xy}) we obtain
\begin{eqnarray}
&&\hspace{-2mm}
\!\int\! \dhd^2 p~\dhd^2l~e^{i(p,\Delta)+i(l,Y)}[F_1^{\rm reg}(p,l)+F_2(p,l)]
~(e^{i(p,X)}-e^{i(p,Y)})(e^{-i(p-l,X)}-e^{-i(p-l,Y)})
\nonumber\\
&&\hspace{-2mm}=
~-{1\over 8\pi^3}{\Delta^2\over X^2Y^2}
\Big[{11\over 3}\ln{X^2Y^2\over \Delta^2}\mu^2+{67\over 9}-{\pi^2\over 3}\Big]
-{1\over 16\pi^3}{\Delta^2\over X^2Y^2}\ln{X^2\over\Delta^2}\ln{Y^2\over\Delta^2}
\nonumber\\
&&\hspace{-2mm}
-~{i\kappa\over 8\pi^3X^2Y^2}\Bigg\{\!\int_0^1\!du\Big[{\ln u\over u-{(\Delta,X)+i\kappa\over \Delta^2}}
+{\ln u\over u+{(\Delta,Y)-i\kappa\over \Delta^2}}-c.c.\Big]
-{1\over  2}\ln{X^2Y^2\over\Delta^4}\ln{(X,Y)+i\kappa\over (X,Y)-i\kappa} \Bigg\}
\label{f12contribution}
\end{eqnarray}
Now we turn our attention to last  two terms in Eq. (\ref{a1}).
Using Fourier transformation
\begin{eqnarray}
&&\hspace{-2mm}
\int\! \dhd^2k_1\dhd^2k_2~e^{-i(k_1,x_1)-i(k_2,x_2)}
{k_{2i}\over (k_1+k_2)^2k_2^2}\ln{k_1^2\over k_2^2}
\label{fint1}\\
&&\hspace{-2mm}
=~{i\over 8\pi^2}\Big(x_{1i}-{(x_1,x_{12})\over x_{12}^2}x_{12i}\Big)
{1\over i\kappa_{12}}\Bigg\{
\int_0^1\! du \Big[{\ln u\over u-{(x_1,x_{12})-i\kappa\over x_1^2}}-c.c.\Big]
\nonumber\\
&&\hspace{72mm}+~{1\over 2}\ln{x_1^2\over x_{12}^2}
\ln{(x_2,x_{12})+i\kappa_{12}\over (x_2,x_{12})-i\kappa_{12}}\Bigg\}
+{ix_{12i}\over 16\pi^2x_{12}^2}\ln{x_1^2\over x_{12}^2}\ln{x_1^2\over x_2^2}
\nonumber
\end{eqnarray}
(where $x_{12}\equiv x_1-x_2$ and 
$\kappa_{12}\equiv \sqrt{x_1^2x_2^2-(x_1,x_2)^2}$) one easily obtains
\begin{eqnarray}
&&\hspace{-2mm}
\int\! \dhd^2p\dhd^2l\dhd^2k'
(e^{-i(p-l,X)}-e^{-i(p-l,Y)})(e^{i(p-k',X)+i(k',Y)}-e^{i(p-k',Y)+i(k',X)})
{(k',p-k')\over
(p-l)^2{k'}^2(p-l-k')^2}\ln{(p-l-k')^2\over {k'}^2}
\nonumber\\
&&\hspace{-2mm}
=~{i\kappa\over 16\pi^3X^2Y^2}\Bigg[
\int_0^1\! du \Big({\ln u\over u-{(X,Y)-i\kappa\over X^2}}
+{\ln u\over u-{(X,Y)-i\kappa\over Y^2}}-c.c.\Big)
+{1\over 2}\ln{X^2\over Y^2}
\ln{[(\Delta,X)+i\kappa][(\Delta,Y)+i\kappa]\over[(\Delta,X)-i\kappa][(\Delta,Y)-i\kappa]}
\Bigg]
\nonumber\\
&&\hspace{-2mm}
=~{i\kappa\over 16\pi^3X^2Y^2}\Bigg[-\!\int_0^1\!du\Big[{\ln u\over u-{(\Delta,X)+i\kappa\over \Delta^2}}
+{\ln u\over u+{(\Delta,Y)-i\kappa\over \Delta^2}}-c.c.\Big]
-{1\over  2}\ln{X^2Y^2\over\Delta^4}\ln{(X,Y)+i\kappa\over (X,Y)-i\kappa}\Bigg]
-{(X,Y)\over 32\pi^3 X^2Y^2}\ln^2{X^2\over Y^2}\nonumber\\
\label{1kycok}
\end{eqnarray}
Similarly,
\begin{eqnarray}
&&\hspace{-6mm}
\int\! \dhd^2k_1\dhd^2k_2~e^{-i(k_1,x_1)-i(k_2,x_2)}
{(k_1,k_2)k_{1i}\over (k_1+k_2)^2k_1^2k_2^2}\ln{k_1^2\over k_2^2}
\label{fint2}\\
&&\hspace{-6mm}
=~{i\over 16\pi^2}
\Big(x_{1i}-{(x_1,x_{12})\over x_{12}^2}x_{12i}\Big){1\over i\kappa}\Bigg[
\int_0^1\! du \Big[{\ln u\over u-{(x_1,x_{12})-i\kappa\over x_1^2}}-c.c.\Big]
+{1\over 2}\ln{x_1^2\over x_{12}^2}
\ln{(x_2,x_{12})+i\kappa\over (x_2,x_{12})-i\kappa}\Bigg]
\nonumber\\
&&\hspace{-6mm}
-~{i\over 16\pi^2}
\Big(x_{2i}-{(x_2,x_{12})\over x_{12}^2}x_{12i}\Big){1\over i\kappa}\Bigg[
\int_0^1\! du \Big[{\ln u\over u+{(x_2,x_{12})-i\kappa\over x_2^2}}-c.c.\Big]
+{1\over 2}\ln{x_2^2\over x_{12}^2}
\ln{(x_1,x_{12})+i\kappa\over (x_1,x_{12})-i\kappa}\Bigg]
+{ix_{12i}\over 32\pi^2x_{12}^2}\ln{x_1^2x_2^2\over x_{12}^4}\ln{x_1^2\over x_2^2}
\nonumber\\
&&\hspace{-6mm}
+~{i\over 16\pi^2}
\Big(x_{12i}-{(x_1,x_{12})\over x_1^2}x_{1i}\Big){1\over i\kappa}\Bigg[
\int_0^1\! du \Big[{\ln u\over u-{(x_1,x_{12})-i\kappa\over x_{12}^2}}-c.c.\Big]
-{1\over 2}\ln{x_1^2\over x_{12}^2}
\ln{(x_1,x_2)+i\kappa\over (x_1,x_2)-i\kappa}\Bigg]
-{ix_{1i}\over 32\pi^2x_1^2}\ln{x_1^2\over x_2^2}\ln{x_2^2\over x_{12}^2}
\nonumber
\end{eqnarray}
and therefore
\begin{eqnarray}
&&\hspace{-2mm}
\int\! \dhd^2p\dhd^2l\dhd^2k'
(e^{-i(p-l,X)}-e^{-i(p-l,Y)})(e^{i(p-k',X)+i(k',Y)}-e^{i(p-k',Y)+i(k',X)})
\nonumber\\
&&\hspace{32mm}
\times~{2(p-k',p-l-k')(k',p-l-k')\over
(p-l)^2(p-k')^2{k'}^2(p-l-k')^2}\ln{(p-l-k')^2\over {k'}^2}\nonumber\\
&&\hspace{-2mm}
=~{i\kappa Y^{-2}\over 16\pi^3X^2}\Bigg[
\int_0^1\! du \Big(
-2{\ln u\over u-{(\Delta,X)+i\kappa\over \Delta^2}}
-2{\ln u\over u+{(\Delta,Y)-i\kappa\over \Delta^2}}-c.c.\Big)
+~\ln{X^2Y^2\over \Delta^4}\ln{(X,Y)+i\kappa\over (X,Y)-i\kappa}\Bigg]
\nonumber\\
&&\hspace{-2mm}
-~{(X,Y)\over 32\pi^3 X^2Y^2}\ln{X^2\Delta^2\over Y^4}\ln{X^2\over\Delta^2}
-{(X,Y)\over 32\pi^3 X^2Y^2}\ln{Y^2\Delta^2\over X^4}\ln{Y^2\over\Delta^2}
-{1\over 32\pi^3 }\Big({1\over X^2}+{1\over Y^2}\Big)
\ln{X^2\over Y^2}\ln{Y^2\over\Delta^2}
\label{2kycok}
\end{eqnarray}
Adding the equations (\ref{1kycok}) and  (\ref{2kycok}) we obtain 
\begin{eqnarray}
&&\hspace{-6mm}
\int\! \dhd^2p\dhd^2l\dhd^2k'
(e^{-i(p-l,X)}-e^{-i(p-l,Y)})(e^{i(p-k',X)+i(k',Y)}-e^{i(p-k',Y)+i(k',X)})
\label{summa12}\\
&&\hspace{32mm}
\times~{(k',p-k')(p-k')^2-2(p-k',p-l-k')(k',p-l-k')\over
(p-l)^2(p-k')^2{k'}^2(p-l-k')^2}\ln{(p-l-k')^2\over {k'}^2}
\nonumber\\
&&\hspace{-6mm}
=~{i\kappa \over 16\pi^3X^2Y^2}\Bigg[
\int_0^1\! du \Big({\ln u\over u-{(\Delta,X)+i\kappa\over \Delta^2}}
+{\ln u\over u+{(\Delta,Y)-i\kappa\over \Delta^2}}-c.c.\Big)
-{1\over 2}\ln{X^2Y^2\over \Delta^4}\ln{(X,Y)+i\kappa\over (X,Y)-i\kappa}\Bigg]
+{\Delta^2\over 32\pi^3 X^2Y^2}\ln{X^2\over \Delta^2}\ln{Y^2\over\Delta^2}
\nonumber
\end{eqnarray}
It is easy to see that the contribution of the last term in Eq. (\ref{a1}) is equal to
(\ref{summa12}) so we get
\begin{eqnarray}
&&\hspace{-2mm}
\langle{\rm Tr}\{\hat{U}_x 
\hat{U}^\dagger_y\}\rangle_{\rm Fig. \ref{2cutdms}~z\rightarrow z'}~
=~
{g^4\over 8\pi^2}\!\int_0^\sigma\!{d\alpha\over\alpha}
\!\int d^2z~
[{N_c\over 2}{\rm Tr}\{U_xU^\dagger_z\}{\rm Tr}\{U_zU^\dagger_y\}
-{1\over 2}{\rm Tr}\{U_xU^\dagger_y\} ]
\nonumber\\
&&\hspace{-2mm}
\times~\Bigg[
\int\! \dhd^2 p\dhd^2l~[F_1^{\rm reg}(p,l)+F_2(p,l)]
~(e^{i(p,X)}-e^{i(p,Y)})(e^{-i(p-l,X)}-e^{-i(p-l,Y)})
\nonumber\\
&&\hspace{-2mm}
+~2\!\int\! \dhd^2 p\dhd^2l\dhd^2k'(e^{-i(p-l,X)}-e^{-i(p-l,Y)})(e^{i(p-k',X)+i(k',Y)}-e^{i(p-k',Y)+i(k',X)})
\nonumber\\
&&\hspace{32mm}
\times~{(k',p-k')(p-k')^2-2(p-k',p-l-k')(k',p-l-k')\over
(p-l)^2(p-k')^2{k'}^2(p-l-k')^2}\ln{(p-l-k')^2\over {k'}^2}\Bigg]
\nonumber\\
&&\hspace{-2mm}
=~~-{\alpha_s^2N_c\over 8\pi^3}
\!\int d^2z~
[{\rm Tr}\{U_xU^\dagger_z\}{\rm Tr}\{U_zU^\dagger_y\}
-{1\over N_c}{\rm Tr}\{U_xU^\dagger_y\} ]
{\Delta^2\over X^2Y^2}
\Big[{11\over 3}\ln{X^2Y^2\over \Delta^2}\mu^2+{67\over 9}-{\pi^2\over 3}\Big]
\label{appendixresult}
\end{eqnarray}
Note that the dilogarithms and products of logarithms have canceled. 
The simplicity of the final result  indicates that there should be a less
tedious derivation but we were not able to find it.

\section{Appendix B: Cutoff dependence of the NLO kernel.}
We will repeat the procedure from Sect. (\ref{sect:losubtraction}), this time using the cutoff by the slope.
\begin{eqnarray}
&&\hspace{-6mm}
 \langle K_{\rm NLO}{\rm Tr}\{\hat{U}_x\hat{U}^\dagger_y\}\rangle_{\rm shockwave}=
{\partial\over\partial\eta}\langle{\rm Tr}\{\hat{U}_x\hat{U}^\dagger_y\}\rangle_{\rm shockwave}-
\langle K_{\rm LO}{\rm Tr}\{\hat{U}_x\hat{U}^\dagger_y\}\rangle_{\rm shockwave}
\label{subtractionapp}
\end{eqnarray}
Instead of Eq. (\ref{selfenergy1}) we get
\begin{eqnarray}
&&\hspace{-6mm}g^4\!\int_0^{\infty}\!du\int^0_{-\infty} \!dv
\langle \hat{A}^a_{\bullet}(un+x_{\perp})\hat{A}^b_{\bullet}(vn+y_{\perp})\rangle 
\label{selfenergy1app}\\
&&\hspace{-6mm}
=~{1\over 2}
g^2{s^2\over 4}f^{anl}f^{bn'l'}\!\int\!
\dhd\alpha\dhd\alpha_1\dhd\beta\dhd\beta' \dhd\beta_1\dhd\beta'_1
\dhd\beta_2\dhd\beta'_2\!\int \!d^2z d^2z'
\!\int \!\dhd^2q_1\dhd^2q_2\dhd^2k_1\dhd^2k_2~e^{i(q_1+q_2,x)_{\perp}
-i(k_1+k_2,y)_{\perp}}
\nonumber\\ 
&&\hspace{-6mm}
{4\alpha_1(\alpha-\alpha_1) 
U^{nn'}_zU^{ll'}_{z'}e^{-i(q_1-k_1,z)_\perp-i(q_2-k_2,z')_\perp}
d_{\bullet\lambda}(\alpha p_1+\beta p_2+q_{1\perp}+k_{1\perp})
d_{\lambda'\bullet}(\alpha p_1+\beta' p_2+q_{2\perp}+k_{2\perp})
\over(\beta-\beta_1-\beta_2+i\epsilon)
(\beta'-\beta'_1-\beta'_2+i\epsilon)(\beta+\xi\alpha-i\epsilon)(\beta'+\xi\alpha'-i\epsilon)[ \alpha\beta
s-(q_1+q_2)_\perp^2+i\epsilon][\alpha\beta'
s-(k_1+k_2)_\perp^2+i\epsilon]}
\nonumber\\ 
&&\hspace{-6mm}
{d_{\mu\xi}(\alpha_1 p_1+\beta_1 p_2+q_{1\perp})
\over \alpha_1\beta_1 s-q_{1\perp}^2+i\epsilon}
{d^\xi_{~\mu'}(\alpha_1 p_1+\beta'_1 p_2+k_{1\perp})
\over \alpha_1\beta'_1 s-k_{1\perp}^2+i\epsilon}
{d_{\nu\eta}((\alpha-\alpha_1) p_1+\beta_2 p_2+q_{2\perp})
\over (\alpha-\alpha_1)\beta_2 s-q_{2\perp}^2+i\epsilon}
~{d^\eta_{~\nu'}((\alpha-\alpha_1)p_1+\beta'_2 p_2+k_{2\perp})
\over (\alpha-\alpha_1)\beta'_2 s-k_{2\perp}^2+i\epsilon}
\nonumber\\ 
&&\hspace{-6mm}
\Gamma^{\mu\nu\lambda}
(\alpha p_1+q_{1\perp},(\alpha-\alpha_1)p_1+q_{2\perp},-\alpha p_1-q_{1\perp}
-q_{2\perp})~
\Gamma^{\mu'\nu'\lambda'}
(\alpha p_1+k_{1\perp},(\alpha-\alpha_1)p_1
+k_{2\perp},-\alpha p_1-k_{1\perp}-k_{2\perp})
\nonumber
\end{eqnarray}
where $\xi=e^{-2\eta_1}$.
In this formula
${1\over \beta+\xi\alpha-i\epsilon}$ comes from the integration over $u$ parameter in the l.h.s. and  ${1\over \beta'+\xi\alpha' -i\epsilon}$ from the integration over $v$ parameter. 
 
Taking residues at $\beta=-\xi\alpha$ and $\beta'=-\xi\alpha'$ and $\beta_2=-\beta_1$, $\beta'_2=-\beta'_1$ 
we obtain
\begin{eqnarray}
&&\hspace{-6mm}\int_0^{\infty}du\int^0_{-\infty}dv 
\langle \hat{A}^a_{\bullet}(un+x_{\perp})\hat{A}^b_{\bullet}(vn+y_{\perp})\rangle
\label{selfenergy2app}\\
&&\hspace{-6mm}=~{1\over 2}
g^2{s^2\over 4}f^{anl}f^{bn'l'}\!\int\!
\dhd\alpha\dhd\alpha_1\dhd\beta_1\dhd\beta'_1\!\int d^2z d^2z'
\!\int \!\dhd^2q_1\dhd^2q_2\dhd^2k_1\dhd^2k_2
~e^{i(q_1+q_2,x)_{\perp}
-i(k_1+k_2,y)_{\perp}}
\nonumber\\ 
&&\hspace{-6mm}
4{\alpha_1(\alpha-\alpha_1)\over\alpha^2}
 U^{nn'}_zU^{ll'}_{z'}e^{-i(q_1-k_1)z-i(q_2-k_2)z'}
~{(q_{1\perp}+q_{2\perp})_\lambda\over (q_1+q_2)_\perp^2+\xi\alpha^2}
{(k_{1\perp}+k_{2\perp})_{\lambda'}\over (k_1+k_2)_\perp^2+\xi\alpha^2}
\nonumber\\ 
&&\hspace{-6mm}
{d_\mu^{~\xi}(\alpha_1 p_1+q_{1\perp})
\over \alpha_1\beta_1 s-q_{1\perp}^2+i\epsilon}
{d_{\xi\mu'}(\alpha_1 p_1+k_{1\perp})
\over \alpha_1\beta'_1 s-k_{1\perp}^2+i\epsilon}
{d^\eta_{~\eta}((\alpha-\alpha_1) p_1+q_{2\perp})
\over -(\alpha-\alpha_1)\beta_1 s-q_{2\perp}^2+i\epsilon}
~{d_{\eta\nu'}((\alpha-\alpha_1)p_1+k_{2\perp})
\over -(\alpha-\alpha_1)\beta'_1 s-k_{2\perp}^2+i\epsilon}
\nonumber\\ 
&&\hspace{-6mm}
\Gamma^{\mu\nu\lambda}
(\alpha_1 p_1+q_{1\perp},(\alpha-\alpha_1)p_1+q_{2\perp},-\alpha p_1-q_{1\perp}
-q_{2\perp})~
\Gamma^{\mu'\nu'\lambda'}
(\alpha_1 p_1+k_{1\perp},(\alpha-\alpha_1)p_1
+k_{2\perp},-\alpha p_1-k_{1\perp}-k_{2\perp})
\nonumber
\end{eqnarray}
which leads to (cf. Eq. (\ref{selfenergy4})
\begin{eqnarray}
&&\hspace{-2mm}
{d\over d\eta}\langle {\rm Tr}\{\hat{U}_x \hat{U}^\dagger_y\}\rangle_{\rm Fig. \ref{nlobk2}a}
~=~{g^4\over 4\pi^2}{\rm Tr}\{t^aU_xt^bU^\dagger_y\}f^{anl}f^{bn'l'}
\!\int\! d^2z d^2z'U_z^{nn'}U_{z'}^{ll'}
\label{selfenergy4app}\\
&&\hspace{-2mm}
\times~\xi{d\over d\xi}\!
\int_0^\infty\!\!
{d\alpha\over\alpha}\!\int_0^1\! du~ \bar{u}u
\!\int\! \dhd^2 q_1\dhd^2q_2\dhd^2k_1\dhd^2k_2~
{e^{i(q_1,X)_\perp+i(q_2,X')_\perp
-i(k_1,Y)_\perp-i(k_2,Y')_\perp}
\over [(q_1+q_2)^2+\xi\alpha^2][(k_1+k_2)^2+\xi\alpha^2]
(q_1^2\bar{u}+q_2^2u)(k_1^2\bar{u}+k_2^2u)}
\nonumber\\ 
&&\hspace{-2mm}\times~
\Big[(q_1^2-q_2^2)\delta_{ij}-{2\over u}q_{1i}(q_1+q_2)_j+
{2\over \bar{u}}(q_1+q_2)_iq_{2j}\Big]
\Big[(k_1^2-k_2^2)\delta_{ij}-{2\over u}k_{1i}(k_1+k_2)_j+
{2\over \bar{u}}(k_1+k_2)_ik_{2j}\Big]
\nonumber\\
&&\hspace{-2mm}
~=~{g^4\over 4\pi^2}{\rm Tr}\{t^aU_xt^bU^\dagger_y\}f^{anl}f^{bn'l'}
\!\int\! d^2z d^2z'U_z^{nn'}U_{z'}^{ll'}
\nonumber\\
&&\hspace{-2mm}
\times~\xi{d\over d\xi}\!
\int_0^\infty\!\!\!
d\alpha\!\int_0^\alpha\! \!\!d\alpha'
\!\int\! \dhd^2 q_1\dhd^2q_2\dhd^2k_1\dhd^2k_2~
{(\alpha-\alpha')\alpha'e^{i(q_1,x-z)_\perp+i(q_2,x-z')_\perp
-i(k_1,y-z)_\perp-i(k_2,y-z')_\perp}
\over [(q_1+q_2)^2+\xi\alpha^2][(k_1+k_2)^2+\xi\alpha^2]
(q_1^2(\alpha-\alpha')+q_2^2\alpha')(k_1^2(\alpha-\alpha')+k_2^2\alpha')}
\nonumber\\ 
&&\hspace{-2mm}\times~
\Big[{\delta_{ij}\over\alpha}(q_1^2-q_2^2)-{2\over \alpha'}q_{1i}(q_1+q_2)_j+
{2\over \alpha-\alpha'}(q_1+q_2)_iq_{2j}\Big]
\Big[{\delta_{ij}\over\alpha}(k_1^2-k_2^2)-{2\over \alpha'}k_{1i}(k_1+k_2)_j+
{2\over \alpha-\alpha'}(k_1+k_2)_ik_{2j}\Big]
\nonumber
\end{eqnarray}
(recall that ${d\over d\eta}=-2\xi{d\over d\xi}$). 
The contribution which is sensitive to the subtraction of (LO)$^2$ is
\begin{eqnarray}
&&\hspace{-4mm}
{d\over d\eta}\langle {\rm Tr}\{\hat{U}_x \hat{U}^\dagger_y\}\rangle_{\rm Fig. \ref{nlobk2}a}
~=~{g^4\over \pi^2}{\rm Tr}\{t^aU_xt^bU^\dagger_y\}f^{anl}f^{bn'l'}
\!\int\! d^2z d^2z'U_z^{nn'}U_{z'}^{ll'}
\nonumber\\
&&\hspace{-4mm}
\times~\xi{d\over d\xi}\!
\int_0^\infty\!\!\!
d\alpha\!\int_0^\alpha\!\! d\alpha'
\!\int\! \dhd^2 q_1\dhd^2q_2\dhd^2k_1\dhd^2k_2~
{e^{i(q_1,X)_\perp+i(q_2,X')_\perp
-i(k_1,Y)_\perp-i(k_2,Y')_\perp}
\over [(q_1+q_2)^2+\xi\alpha^2][(k_1+k_2)^2+\xi\alpha^2]
(q_1^2(\alpha-\alpha')+q_2^2\alpha')(k_1^2(\alpha-\alpha')+k_2^2\alpha')}
\nonumber\\ 
&&\hspace{-4mm}\times~
\Big[{\alpha\over\alpha'}(q_1,k_1)+{\alpha\over\alpha-\alpha'}(q_2,k_2)\Big](q_1+q_2,k_1+k_2)
\label{sensi}
\end{eqnarray}
The ``+''-prescription (\ref{pluscription})  leads to the subtraction
\begin{eqnarray}
&&\hspace{-2mm}
{d\over d\eta}\langle {\rm Tr}\{\hat{U}_x \hat{U}^\dagger_y\}\rangle_{\rm Fig. \ref{nlobk2}a}
\nonumber\\ 
&&\hspace{-2mm}=~\xi{d\over d\xi}{g^4\over \pi^2}{\rm Tr}\{t^aU_xt^bU^\dagger_y\}f^{anl}f^{bn'l'}
\!\int\! d^2z d^2z'U_z^{nn'}U_{z'}^{ll'}
\!\int_0^\infty\!
{d\alpha\over\alpha}
\!\int\! \dhd^2 q_1\dhd^2q_2\dhd^2k_1\dhd^2k_2~(q_1+q_2,k_1+k_2)
\nonumber\\ 
&&\hspace{-2mm}
\times~
{e^{i(q_1,X)_\perp+i(q_2,X')_\perp
-i(k_1,Y)_\perp-i(k_2,Y')_\perp}
\over [(q_1+q_2)^2+\xi\alpha^2][(k_1+k_2)^2+\xi\alpha^2]}\Bigg\{\!\int_0^\alpha\! {d\alpha'\over\alpha'}
\Big[
{\alpha^2(q_1,k_1)\over (q_1^2(\alpha-\alpha')+q_2^2\alpha']
[k_1^2(\alpha-\alpha')+k_2^2\alpha']}-
{(q_1,k_1)\over q_1^2k_1^2}\Big]
\nonumber\\ 
&&\hspace{42mm}
+~\!\int_0^\alpha\! {d\alpha'\over\alpha-\alpha'}
\Big[{\alpha^2(q_2,k_2)\over [q_1^2(\alpha-\alpha')+q_2^2\alpha']
[k_1^2(\alpha-\alpha')+k_2^2\alpha']}-
{(q_2,k_2)\over q_2^2k_2^2}\Big]\Bigg\}
\label{b7}
\end{eqnarray}
The details of the upper cutoff in $\alpha$ do not matter since they
correspond to changes in the impact factor which do not affect the evolution.
For example,
\begin{eqnarray}
&&\hspace{-2mm}
-2\xi{d\over d\xi}\!\int_0^\infty\!
{d\alpha\over\alpha}
{1\over [(q_1+q_2)^2+\xi\alpha^2][(k_1+k_2)^2+\xi\alpha^2]}
\!\int_0^\alpha\! {d\alpha'\over\alpha'}\Big[
{\alpha^2(q_1,k_1)\over (q_1^2(\alpha-\alpha')+q_2^2\alpha']
[k_1^2(\alpha-\alpha')+k_2^2\alpha']}-
{(q_1,k_1)\over q_1^2k_1^2}\Big]
\nonumber\\ 
&&\hspace{-2mm}
=~
{1\over (q_1+q_2)^2(k_1+k_2)^2}
\!\int_0^1 {du\over u}\Big[
{(q_1,k_1)\over (q_1^2\bar{u}+q_2^2u]
[k_1^2\bar{u}+k_2^2u]}-
{(q_1,k_1)\over q_1^2k_1^2}\Big]
\nonumber\\ 
&&\hspace{-2mm}
=~{d\over d\sigma}\!\int_0^\sigma\!
{d\alpha\over\alpha}
{1\over (q_1+q_2)^2(k_1+k_2)^2}
\!\int_0^\alpha\! {d\alpha'\over\alpha'}\Big[
{\alpha^2(q_1,k_1)\over (q_1^2(\alpha-\alpha')+q_2^2\alpha']
[k_1^2(\alpha-\alpha')+k_2^2\alpha']}-
{(q_1,k_1)\over q_1^2k_1^2}\Big]
\label{b8}
\end{eqnarray}
where the last line is exactly  our ``rigid cutoff'' with ``+''  subtraction (\ref{pluscription}).

On the contrary, the details of the upper cutoff in $\alpha'$ are essential for the evolution equation  (\ref{subtractionapp}).
The contribution to $\langle K_{\rm LO} {\rm Tr}\{\hat{U}_x \hat{U}^\dagger_y\}\rangle$ corresponding to the ``slope'' cutoff (\ref{defy})
has the form
\begin{eqnarray}
&&\hspace{-2mm}
\langle K_{\rm LO}{\rm Tr}\{\hat{U}_x \hat{U}^\dagger_y\}\rangle^{\rm slope}_{\rm Fig. \ref{nlobk2}a}
~
\nonumber\\ 
&&\hspace{-2mm}=~-{g^4\over 2\pi^2}{\rm Tr}\{t^aU_xt^bU^\dagger_y\}f^{anl}f^{bn'l'}
\!\int\! d^2z d^2z'U_z^{nn'}U_{z'}^{ll'}
\!\int\! \dhd^2 q_1\dhd^2q_2\dhd^2k_1\dhd^2k_2~{(q_1+q_2,k_1+k_2)\over (q_1+q_2)^2 (k_1+k_2)^2}
\nonumber\\ 
&&\hspace{-2mm}\times
~e^{i(q_1,X)_\perp+i(q_2,X')_\perp
-i(k_1,Y)_\perp-i(k_2,Y')_\perp}
\int_0^\infty\! {d\alpha'\over\alpha'}~
\Big[{(q_1,k_1)\over (q_1^2+\xi{\alpha'}^2)(k_1^2+\xi{\alpha'}^2)}
+
{(q_2,k_2)\over (q_2^2+\xi{\alpha'}^2)(k_2^2+\xi{\alpha'}^2)}
\Big]
\nonumber
\end{eqnarray}
and therefore the difference between the subtractions in ``rigid cutoff'' 
(\ref{b7}) and ``slope cutoff'' (\ref{b8}) prescriptions can be written as
\begin{eqnarray}
&&\hspace{-2mm}
\langle K_{\rm LO}{\rm Tr}\{\hat{U}_x \hat{U}^\dagger_y\}^{\rm rigid}-
K_{\rm LO}{\rm Tr}\{\hat{U}_x \hat{U}^\dagger_y\}^{\rm slope}
\rangle_{\rm Fig. \ref{nlobk2}a}
~
\nonumber\\ 
&&\hspace{-2mm}=~{g^4\over 2\pi^2}{\rm Tr}\{t^aU_xt^bU^\dagger_y\}f^{anl}f^{bn'l'}
\!\int\! d^2z d^2z'U_z^{nn'}U_{z'}^{ll'}
\!\int\! \dhd^2 q_1\dhd^2q_2\dhd^2k_1\dhd^2k_2~e^{i(q_1,X)_\perp+i(q_2,X')_\perp
-i(k_1,Y)_\perp-i(k_2,Y')_\perp}
\nonumber\\ 
&&\hspace{-2mm}
\times~\Bigg\{{(q_1+q_2,k_1+k_2)\over (q_1+q_2)^2 (k_1+k_2)^2}\!\int_0^\infty\! {d\alpha'\over\alpha'}~
\Big[{(q_1,k_1)\over (q_1^2+\xi{\alpha'}^2)(k_1^2+\xi{\alpha'}^2)}
+{(q_2,k_2)\over (q_2^2+\xi{\alpha'}^2)(k_2^2+\xi{\alpha'}^2)}
\Big]
\nonumber\\ 
&&\hspace{-2mm}
+~2\xi{d\over d\xi}\!\int_0^\infty\! {d\alpha\over\alpha}~{(q_1+q_2,k_1+k_2)\over [(q_1+q_2)^2+\xi\alpha^2][(k_1+k_2)^2+\xi\alpha^2]}
\Big[{(q_1,k_1)\over q_1^2k_1^2}+{(q_2,k_2)\over q_2^2k_2^2}\Big]
\!\int_0^\alpha\! {d\alpha'\over\alpha'}
\Bigg\}
\nonumber\\ 
&&\hspace{-2mm}
=~-{g^4\over 2\pi^2}{\rm Tr}\{t^aU_xt^bU^\dagger_y\}f^{anl}f^{bn'l'}
\!\int\! d^2z d^2z'U_z^{nn'}U_{z'}^{ll'}
\!\int\! \dhd^2 q_1\dhd^2q_2\dhd^2k_1\dhd^2k_2~e^{i(q_1,X)_\perp+i(q_2,X')_\perp
-i(k_1,Y)_\perp-i(k_2,Y')_\perp}
\nonumber\\ 
&&\hspace{-2mm}
\times~\!\int_0^\infty\! {d\alpha'\over\alpha'}~
\Bigg\{{(q_1+q_2,k_1+k_2)\over (q_1+q_2)^2 (k_1+k_2)^2}
\Big[{(q_1,k_1)\over (q_1^2+\xi{\alpha'}^2)(k_1^2+\xi{\alpha'}^2)}
+{(q_2,k_2)\over (q_2^2+\xi{\alpha'}^2)(k_2^2+\xi{\alpha'}^2)}
\Big]
\nonumber\\ 
&&\hspace{-2mm}
-~{(q_1+q_2,k_1+k_2)\over [(q_1+q_2)^2+\xi{\alpha'}^2][(k_1+k_2)^2+\xi{\alpha'}^2]}
\Big[{(q_1,k_1)\over q_1^2k_1^2}+{(q_2,k_2)\over q_2^2k_2^2}\Big]
\Bigg\}
\nonumber
\end{eqnarray}
It is instructive to rewrite this result in Schwinger's notations
\begin{eqnarray}
&&\hspace{-12mm}
\langle K_{\rm LO}{\rm Tr}\{\hat{U}_x \hat{U}^\dagger_y\}^{\rm rigid}-
K_{\rm LO}{\rm Tr}\{\hat{U}_x \hat{U}^\dagger_y\}^{\rm slope}
\rangle_{\rm Fig. \ref{nlobk2}a}
~
\nonumber\\ 
&&\hspace{-2mm}=~
{g^4\over 2\pi^2}{\rm Tr}\{t^aU_xt^bU^\dagger_y\}f^{anl}f^{bn'l'}
\!\int\! d^2z 
\int_0^\infty\!{d\alpha'\over\alpha'}\Big[(x|{p_i\over p^2+\xi{\alpha'}^2}|z)U_z^{nn'}
(z|{p_i\over p^2+\xi{\alpha'}^2}|y)(z|{p_j\over p^2}U^{ll'}{p_j\over p^2}|y) ~
\nonumber\\ 
&&\hspace{42mm}-~ (x|{p_i\over p^2}|z)U_z^{nn'}
(z|{p_i\over p^2}|y)(z|{p_j\over p^2+\xi{\alpha'}^2}U^{ll'}{p_j\over p^2+\xi{\alpha'}^2}|y)\Big]
\nonumber
\end{eqnarray}
We see now that the difference between
the two regularizations of the  longitudinal divergence is given by the difference of (LO)$^2$ contributions with cutoffs in $\alpha$ 
determined by the momenta on the first and on the second step of (LO)$^2$ evolution.

It is easy to see that for the sum of all diagrams this yields (see eq. (\ref{colors}))
\begin{eqnarray}
&&\hspace{-3mm}
\langle K_{\rm LO}{\rm Tr}\{\hat{U}_x \hat{U}^\dagger_y\}^{\rm rigid}-
K_{\rm LO}{\rm Tr}\{\hat{U}_x \hat{U}^\dagger_y\}^{\rm slope}
\rangle
\nonumber\\ 
&&\hspace{-3mm}
=~\alpha_s^2\!\int\! d^2zd^2z'~[{\rm Tr}\{U_xU^\dagger_z\}{\rm Tr}\{U_zU^\dagger_{z'}\}
{\rm Tr}\{U_{z'}U^\dagger_y\}-{\rm Tr}\{U_xU^\dagger_z U_{z'}U^\dagger_yU_zU^\dagger_{z'}\}
+(z\leftrightarrow z')]\!\int_0^\infty\!{dt\over t}
\nonumber\\
&&\hspace{-3mm}
\times~\Big\{\Big[(x|{p_i\over p^2+t}|z)-(y|{p_i\over p^2+t}|z)\Big]^2
\Big[(z|{p_i\over p^2}|z')-(y|{p_i\over p^2}|z')\Big]^2
-\Big[(x|{p_i\over p^2}|z)-(y|{p_i\over p^2}|z)\Big]^2
\Big[(z|{p_i\over p^2+t}|z')-(y|{p_i\over p^2+t}|z')\Big]^2
\nonumber\\
&&\hspace{-3mm}
+~\Big[(x|{p_i\over p^2+t}|z')-(y|{p_i\over p^2+t}|z')\Big]^2
\Big[(z'|{p_i\over p^2}|z)-(x|{p_i\over p^2}|z)\Big]^2
-\Big[(x|{p_i\over p^2}|z')-(y|{p_i\over p^2}|z')\Big]^2
\Big[(z'|{p_i\over p^2+t}|z)-(x|{p_i\over p^2+t}|z)\Big]^2\Big\}
\nonumber
\end{eqnarray}

Using the integral
\begin{eqnarray}
&&\hspace{-2mm}
\int_0^\infty\!{dt\over t^{1-\epsilon}}(x|{p_i\over p^2+t}|z)(y|{p_i\over p^2+t}|z)
~=~-{(X,Y)\over 4\pi^2}\!\int_0^1\! du~
{\Gamma(\epsilon)\Gamma(2-\epsilon)(4\bar{u}u)^{-\epsilon}\over (X^2\bar{u}+Y^2u)^{2-\epsilon} }
\nonumber\\\
&&\hspace{-3mm}
=~{(X,Y)\over 4\pi^2X^2Y^2}\Big(-{1\over\epsilon}-\ln 4
+{X^2\ln Y^2-Y^2\ln X^2\over X^2-Y^2}
-\ln X^2Y^2\Big)+O(\epsilon)
\nonumber
\end{eqnarray}
\begin{eqnarray}
&&\hspace{-3mm}
\langle K_{\rm LO}{\rm Tr}\{\hat{U}_x \hat{U}^\dagger_y\}^{\rm rigid}-
K_{\rm LO}{\rm Tr}\{\hat{U}_x \hat{U}^\dagger_y\}^{\rm slope}
\rangle
\label{slopedif}\\ 
&&\hspace{-3mm}
=~-{\alpha_s^2\over 16\pi^4}\!\int\! d^2zd^2z'~
{1\over(z-z')^2X^2Y^2}[{\rm Tr}\{U_xU^\dagger_z\}{\rm Tr}\{U_zU^\dagger_{z'}\}
{\rm Tr}\{U_{z'}U^\dagger_y\}-{\rm Tr}\{U_xU^\dagger_z U_{z'}U^\dagger_yU_zU^\dagger_{z'}\}
+z\leftrightarrow z']
\nonumber\\
&&\hspace{-12mm}
\times~\Bigg(
{Y^2\over {Y'}^2}\Big\{(X,Y)\Big[{X^2+Y^2\over X^2-Y^2}\ln {X^2\over Y^2}+2\Big]+(\Delta,Y)\ln X^2-(\Delta,X)\ln Y^2\Big\}
\nonumber\\
&&\hspace{-12mm}
-~{\Delta^2\over {Y'}^2}
\Big\{(z-z',Y')\Big[{(z-z')^2+{Y'}^2\over (z-z')^2-{Y'}^2}\ln {(z-z')^2\over {Y'}^2}
+2\Big]-(Y,Y')\ln (z-z')^2+(Y,z-z')\ln {Y'}^2\Big\}+x\leftrightarrow y\Bigg)
\nonumber
\end{eqnarray}
The NLO kernel for the evolution of color dipoles with respect to the slope is the sum 
of Eq. (\ref{nlobk}) and the correction (\ref{slopedif}). Note that the correction term  (\ref{slopedif})
is not conformally invariant (cf. Ref. \cite{baba}).
This is hardly surprising since  the non-light-like Wilson line turns into a circle under the inversion $x_\mu\rightarrow x_\mu/x^2$.

 

\section*{References}

\vspace{-5mm}
\begin{thebibliography}{99}

\bibitem{mobzor}
I. Balitsky, {\it ``High-Energy QCD and Wilson Lines''}, 
In *Shifman, M. (ed.): At the frontier of particle 
physics, vol. 2*, p. 1237-1342  (World Scientific, Singapore,2001)
[hep-ph/0101042] 

\bibitem{mu94}
A.H. Mueller, 
{\it Nucl. Phys.}  {\bf B415}, 373 (1994);
 A.H. Mueller and Bimal Patel, 
{\it Nucl. Phys.}  {\bf B425}, 471 (1994).

\bibitem{nnn}
N.N. Nikolaev and B.G. Zakharov,
{\it Phys. Lett.} {\bf B 332}, 184 (1994);
{\it Z. Phys.}  {\bf C64}, 631 (1994);
N.N. Nikolaev B.G. Zakharov, and V.R. Zoller,
{\it JETP Letters} {\bf 59}, 6 (1994).


\bibitem{npb96}
I. Balitsky, 
{\it Nucl. Phys.}  {\bf B463}, 99 (1996);
{\it ``Operator expansion for diffractive high-energy scattering''},
[hep-ph/9706411]; 


\bibitem{yura}
Yu.V. Kovchegov,  
{\it Phys. Rev.} {\bf D60}, 034008 (1999);
{\it Phys. Rev.} {\bf D61},074018 (2000).

\bibitem{bfkl}
V.S. Fadin, E.A. Kuraev, and L.N. Lipatov,
{\it Phys. Lett.} {\bf B 60}, 50 (1975);
I. Balitsky and L.N. Lipatov,
{\it Sov. Journ. Nucl. Phys.} 
{\bf 28}, 822 (1978).


\bibitem{saturation}
L.V. Gribov, E.M. Levin, and M.G. Ryskin, 
{\it Phys. Rept.} {\bf 100}, 1 (1983),
A.H. Mueller and J.W. Qiu,
{\it Nucl. Phys.}  {\bf B268}, 427 (1986);
A.H. Mueller, 
{\it Nucl. Phys.}  {\bf B335}, 115 (1990).

\bibitem{satreviews}
E. Iancu and  R. Venugopalan ,
In *Hwa, R.C. (ed.) et al.: Quark gluon plasma* 249-3363,
[e-Print: hep-ph/0303204];\\
H. Weigert ,
{\it Prog.Part.Nucl.Phys.}{\bf 55}, 461(2005);\\
J. Jalilian-Marian and Yu.V. Kovchegov,
  {\it Prog.Part.Nucl.Phys.}{\bf 56}, 104(2006).



\bibitem{prd75}
I. Balitsky, 
{\it  Phys.Rev.}{\bf D75},014001(2007).

\bibitem{kw1}
Yu. V. Kovchegov and H. Weigert,
{\it Nucl. Phys.}  {\bf A784}, 188 (2007),

\bibitem{balbel}
I. Balitsky and A.V. Belitsky, 
{\it Nucl. Phys.}  {\bf B629}, 290 (2002).
 

 \bibitem{renormalons}
 M. Beneke,
{\it Phys.Rept.}{\bf 317},1(1999);
M. Beneke and V.M. Braun,
 {\it ``Renormalons and power corrections.''}, 
In *Shifman, M. (ed.): At the frontier of particle 
physics, vol. 3*, p. 1719-1773  (World Scientific, Singapore,2001)
[hep-ph/0010208] 

\bibitem{nlobfkl}
V.S. Fadin and L.N. Lipatov,
  {\it  Phys. Lett.}{\bf B429}, 127 (1998);
G. Camici and M. Ciafaloni,
  {\it  Phys. Lett.}{\bf B430}, 349 (1998).

\bibitem{prd99}
I. Balitsky, 
{\it Phys. Rev.} {\bf D60}, 014020 (1999).


\bibitem{physlett01}
I. Balitsky, 
{\it Phys.Lett.}{\bf B518}, 235(2001).


\bibitem{integral}
V.S. Fadin , M.I. Kotsky, and L.N. Lipatov
{\it ``Gluon pair production in the quasimulti - Regge kinematics''},
[hep-ph/9704267]. 

 \bibitem{kw2}
Yu.V. Kovchegov and H. Weigert
{\it Nucl.Phys.}{\bf A789}, 260(2007).

\bibitem{lipkot00}
 A.V. Kotikov and L.N. Lipatov, 
{\it Nucl. Phys.}  {\bf B582}, 19 (2000).

\bibitem{3loops}
A. Vogt, S. Moch, and J.A.M. Vermaseren,
{\it  Nucl.Phys.}{\bf B691}, 129 (2004)
 
\bibitem{fadin07}
V. S. Fadin, R. Fiore, A.V. Grabovsky, and A. Papa, 
{\it  Nucl.Phys.}{\bf B784}, 49(2007).

 \bibitem{nfnlobfkl}
V.S. Fadin and R. Fiore, 
 {\it Phys. Rev.} {\bf D72}, 014018 (2005).


 \bibitem{runcon}
E. Gardi, J. Kuokkanen, K. Rummukainen, and H. Weigert, 
 {\it Nucl.Phys.}{\bf A784}, 282(2007);
 J.L. Albacete and Yu.V. Kovchegov, 
{\it Phys.Rev.}{\bf D75},125021 (2007).


\bibitem{effaction}
A. Kovner and M. Lublinsky,
{\it Phys.Rev.}{\bf D71}, 085004(2005);
{\it Phys.Rev.Lett.}{\bf 94}, 181603(2005);
 {\it JHEP}, 0503:001(2005);
Y. Hatta, E. Iancu,  L. McLerran,  A. Stasto and D.N. Triantafyllopoulos,
  {\it Nucl.Phys.}{\bf A764}, 423 (2006);
I. Balitsky, 
{\it Phys. Rev.} {\bf D72}, 074027 (2005):
 A.H. Mueller, A.I. Shoshi, and S.M.H. Wong,
  {\it Nucl.Phys.}{\bf B715}, 440(2005).

\bibitem{baba}
A. Babansky and I. Balitsky, 
{\it Phys. Rev.} {\bf D67}, 054026 (2003).

\end{thebibliography}
\end{document}